\documentclass[a4paper,11pt]{article}
\usepackage{jheppub}
\usepackage{amsmath,multirow,graphicx,tabularx,amssymb,braket}
\usepackage{hyperref}
\usepackage{slashed}

\def\tablename{Table}
\def\figurename{Figure}

\newcommand\one{\leavevmode\hbox{\small1\normalsize\kern-.33em1}}

\newcommand{\lag}{\mathcal{L}}

\newcommand{\qqquad}{\qquad \qquad}
\newcommand{\qqqquad}{\qquad \qquad \qquad}

\newcommand{\CP}{CP}

\newcommand{\gev}{{\ensuremath\rm GeV}}

\def\slashchar#1{\setbox0=\hbox{$#1$}           
   \dimen0=\wd0                                 
   \setbox1=\hbox{/} \dimen1=\wd1               
   \ifdim\dimen0>\dimen1                        
      \rlap{\hbox to \dimen0{\hfil/\hfil}}      
      #1                                        
   \else                                        
      \rlap{\hbox to \dimen1{\hfil$#1$\hfil}}   
      /                                         
   \fi}

\def\eg{{\sl e.g.} \,}
\def\ie{{\sl i.e.} \,}

\newcommand{\be}{\begin{eqnarray*}}
\newcommand{\ee}{\end{eqnarray*}}

\newcommand{\bee}{\begin{eqnarray}}
\newcommand{\eee}{\end{eqnarray}}
\newcommand{\beeq}{\begin{equation}}
\newcommand{\eeeq}{\end{equation}}

\newcommand{\hzero}{h^0}
\newcommand{\Hzero}{H^0}
\newcommand{\Azero}{A^0}
\newcommand{\PHiggs}{H}
\newcommand{\PW}{W}
\newcommand{\PZ}{Z}

\newcommand{\sw}{\ensuremath{s_w}}
\newcommand{\cw}{\ensuremath{c_w}}
\newcommand{\swd}{\ensuremath{s^2_w}}
\newcommand{\cwd}{\ensuremath{c^2_w}}

\newcommand{\mhhd}{\ensuremath{m^2_{\Hzero}}}

\newcommand{\mlhd}{\ensuremath{m^2_{\hzero}}}

\newcommand{\mad}{\ensuremath{m^2_{\Azero}}}
\newcommand{\mhpd}{\ensuremath{m^2_{\PHiggs^{\pm}}}}
\newcommand{\mhp}{\ensuremath{m_{\PHiggs^{\pm}}}}

 \newcommand{\sa}{\ensuremath{\sin\alpha}}
 \newcommand{\ca}{\ensuremath{\cos\alpha}}
 \newcommand{\cad}{\ensuremath{\cos^2\alpha}}
 \newcommand{\sad}{\ensuremath{\sin^2\alpha}}
 \newcommand{\sbd}{\ensuremath{\sin^2\beta}}
 \newcommand{\cbd}{\ensuremath{\cos^2\beta}}
 \newcommand{\cb}{\ensuremath{\cos\beta}}
 \renewcommand{\sb}{\ensuremath{\sin\beta}}

 \newcommand{\tanb}{\ensuremath{\tan\beta}}
 
 \newcommand{\cotb}{\ensuremath{\cot\beta}}

\title{Measuring extended Higgs sectors \\
       as a consistent free couplings model}

\author[a,*]{David L\'opez-Val,\note{Corresponding author.}}
\author[a]{Tilman Plehn,}
\author[b]{Michael Rauch}

\affiliation[a]{Institut f\"ur Theoretische Physik, Universit\"at Heidelberg, Germany}
\affiliation[b]{Institute for Theoretical Physics, Karlsruhe Institute of Technology (KIT), Germany}

\emailAdd{lopez@thphys.uni-heidelberg.de}
\emailAdd{plehn@uni-heidelberg.de}
\emailAdd{michael.rauch@kit.edu}

\abstract{Extended Higgs sectors appear in many models for physics
  beyond the Standard Model. Current Higgs measurements at the LHC are
  starting to significantly constrain them. We study their Higgs
  coupling patterns at tree level as well as including quantum
  corrections. Our benchmarks include a dark singlet-doublet extension
  and several two-doublet setups. Using SFitter we translate the
  current Higgs coupling measurements for one light Higgs state into
  their respective parameter spaces.  Finally, we show how
  two--Higgs--doublet models can serve as a consistent ultraviolet
  completion of an assumed single Standard--Model--like Higgs boson
  with free couplings.}

\begin{document} 

\maketitle

\newpage

\section{Introduction}
\label{sec:intro}

In the Standard Model of particle physics electroweak symmetry
breaking is described by the Higgs mechanism~\cite{higgs}. It assumes
the existence of a CP-even scalar field whose non-vanishing
expectation value breaks the $SU(2)_L \times U(1)_Y$ gauge group down
to the electromagnetic $U(1)$ gauge group. The Standard Model predicts
all properties of the Higgs boson, except for its mass~\cite{lecture}.
Recently, a Higgs boson with mass around 126~GeV has been discovered
by ATLAS~\cite{atlas} and CMS~\cite{cms}.  Its decays to $\gamma
\gamma$, $ZZ^*$, $WW^*$, and $\tau \tau$~\cite{higgschannels_ex} have been
established and the observed event numbers in each of these final
states agree with the Standard Model
predictions~\cite{sfitter_higgs,Azatov:2012bz,others,carmi,fits_ex}.\bigskip

While some alternative theoretical descriptions of such a scalar,
linked to electroweak symmetry breaking or not, predict significant
deviations from the Standard Model operator structure and coupling
strengths, many perturbative extensions of the Standard Model also
predict a light Higgs boson with very similar properties as in the
Standard Model. Reasons for this tendency are general decoupling
patterns as well as experimental constraints in the electroweak
precision and flavor sectors. This defines two ways to look at
extended Higgs sectors including a light Higgs state with a mass of
126~GeV: on the one hand we might interpret the recently discovered
Higgs boson as just one of several Higgs states. On the other hand, we
can use extended Higgs sectors as possible ultraviolet completions of
a light Higgs boson with couplings deviating from the Standard Model
predictions.\bigskip

Higgs couplings are defined as prefactors of the respective Lagrangian
terms coupling the Higgs field to other particles. To leading order in
perturbation theory this is obvious, while including higher orders we
need to add a proper definition of the counter terms.  Defining a set
of couplings to be compared to experimental measurements always relies
on a hypothetical Lagrangian. Given the current experimental knowledge
we start from the renormalizable Lagrangian of the Standard Model and
measure each coupling to a Standard Model particle
$x$~\cite{sfitter_higgs,Duhrssen:2005sp,Bechtle:2013xfa},
\begin{alignat}{9}
g_{xxH} &\equiv g_x  = 
\left( 1 + \Delta_x \right) \; g_x^\text{SM} \; .
\label{eq:delta}
\end{alignat}
The loop-induced Higgs coupling to photons then reads
\begin{alignat}{9}
g_{\gamma\gamma H} &\equiv g_{\gamma}  = 
\left( 1 + \Delta_\gamma^\text{SM} + \Delta_\gamma \right) \;
g_\gamma^\text{SM} \; .
\label{eq:deltagamma}
\end{alignat}
Any modification of the underlying tree-level couplings to Standard
Model particles inside the loop induces $\Delta_\gamma^\text{SM}$. The
remaining $\Delta_\gamma$ characterizes genuine non-Standard Model
contributions. The total coupling shift, for example appearing in the
measured signal strength, reads $\Delta_\gamma^\text{tot} \equiv
\Delta_\gamma^\text{SM} + \Delta_\gamma$. The same setup
applies to the loop--induced Higgs coupling to gluons. Equivalent parameters
$\kappa_x \equiv 1+\Delta_x$ are used in Ref.~\cite{HiggsXS}. 

The one problem with this model independent approach is that such a
theory with free Higgs couplings has poor ultraviolet properties. It
may violate unitarity just above LHC energy scales and is not
renormalizable. This means that once LHC observables reach a precision
where electroweak corrections become relevant the approach of
Eq.\eqref{eq:delta} is not well defined~\cite{passarino}. We need to
also define an ultraviolet completion which at low energies reduces to
a single light Higgs boson and which is renormalizable. This
limitation becomes particularly important when one quantifies the
potential of a linear collider for probing new physics effects in the
Higgs sector~\cite{sfitter_ilc}.

A prime candidate for such a theory is a perturbative extended Higgs
sector, which includes additional singlet, doublet, or triplet
fields. It clearly allows for precision calculations in electroweak
perturbation theory. One of the key questions which we will answer in
this paper is if such models give us enough freedom for a full set of
independent Higgs couplings differing from the Standard Model. In
addition to modified Higgs couplings to all the Standard Model
particles, extended Higgs sectors provide candidate particles that may
induce invisible Higgs decays.\bigskip

Extended Higgs sectors have attractive features.  Adding for instance
a singlet to the minimal Higgs sector of the Standard Model is one of
the few ways to link the Standard Model to a new physics sector in a
renormalizable way. Such a Higgs portal makes distinctive predictions
for the LHC, including Higgs decays to invisible
states~\cite{singlet-portal, portal-collider}.

Adding a second Higgs doublet to the Standard Model is a natural
extension of the very minimal Higgs sector of the Standard Model,
where one is no longer forced to rely on $\Phi$ and $\Phi^\dag$ to give masses to up-type and
down-type fermions. For example in supersymmetric models the
appearance of the conjugate Higgs field is not allowed, so we need two
Higgs doublets to give mass to up-type and down-type fermions. With
two vacuum expectation values this setup can lead to key signatures at
the LHC as well as to new effects in flavor physics.

The Glashow--Weinberg--Salam group can be embedded in different grand
unified theories, like the Pati--Salam group~\cite{pati},
$E_6$~\cite{E6}, $SO(10)$~\cite{SO10}, or the trinification
group~\cite{trini,Stech:2012zr}. The subgroups relevant for the Higgs
sector can be $SU(3)_L \times SU(3)_R$ or $SU(2)_L \times SU(2)_R$,
which means they include two or more doublet representations of scalar
fields. One of them can have couplings to gauge bosons and fermions
proportional to the Standard Model couplings while the other misses
Yukawa couplings completely~\cite{Stech:2008wd}. The LHC could then
see two Higgs fields with suppressed coupling patterns. It can even
occur that the Higgs resonance is a nearly degenerate state composed
of two orthogonal states of this
type~\cite{Stech:2012zr,Stech:2013pda}.

As we will see, each of these extensions predicts distinct features in
the set of light Higgs couplings observed at the LHC. This allows us
to significantly constrain the available parameter space in the more
constrained extended Higgs sectors. Obviously, the most general
extended models essentially allow for a free variation of all measured
couplings of the lightest discovered Higgs state.\bigskip

This paper consists of two main parts. First, we introduce different
extensions of the Standard Model Higgs sector and study their impact
on the light Higgs couplings to leading order and to next-to-leading
order. The latter includes a multitude of new qualitative and
quantitative results. In the second part we define a set of benchmark
models for extended Higgs sectors and determine their parameters from
current ATLAS and CMS results. This includes an update of the
\textsc{SFitter}~\cite{sfitter_higgs} coupling measurement of the
Standard Model Higgs boson, including the Moriond/Aspen results from
the 2012 run. In both parts we will comment on how suitable extended
Higgs sectors are as renormalizable, ultraviolet--consistent
extensions of a Standard Model Higgs Lagrangian with free couplings.
Many details about the benchmark parameterizations we give in the
appendices.

\section{Extended Higgs sectors}
\label{sec:extensions}

A brief overview of extended Higgs sector models in this section
serves two purposes: first, we determine which of these models give us
enough flexibility to be considered as an ultraviolet completion of a model
with a single light Higgs boson and variable couplings differing from
the Standard Model at the $20\%-50\%$ level. Second, we need to define
benchmark models which we can compare to current ATLAS and CMS data.

Deviations in the Higgs couplings in a consistent framework can occur
in two ways: first, new particles with direct or indirect couplings to
the Higgs can contribute to the loop-induced $hgg$ and $h\gamma\gamma$
vertices. Second, new scalar multiplets can give rise to additional
neutral and charged states, in turn leading to mixing
effects~\cite{Killick:2013mya}. Beyond leading order, the two avenues
are no longer separated; the new scalar degrees of freedom will
contribute to quantum effects and may exhibit a non-decoupling
behavior discussed in Section~\ref{sec:quantum_higgs}.

In the mixing approach we consider models with complex $SU(2)$
doublets $\Phi_i$ and singlets $S_j$,
\begin{equation}
\Phi_i = \begin{pmatrix} h^+_i \\[1mm] \dfrac{v_i + h_i^0 + i a_i^0}{\sqrt{2}}
         \end{pmatrix} 
\qqqquad 
S_j = \cfrac{v_j + s_j}{\sqrt{2}} \; .
\label{eq:fields-def}
\end{equation}
The relation between the fields appearing in the Lagrangian and the
physical mass eigenstates is spelled out in the appendix.  The weak
doublets couple to the gauge bosons via the $SU(2)_L \times U(1)_Y$
covariant derivatives.  Fermion masses can be related to the Yukawa
couplings through a VEV of the respective Higgs field, $m_f =
y_f\,v_f/\sqrt{2}$.  The extension to more general flavor patterns is
straightforward.\bigskip

Any extension of the electroweak sector of the Standard Model is
severely constrained by electroweak precision data. The $\rho$ or $T$
parameter constraints point to a global custodial symmetry protecting
tree-level relations.  For example, a number of Higgs fields with
weak isospins $T_i$, hypercharge $Y_i$, and vacuum expectation values
$v_i$ gives~\cite{Langacker:1980js}
\begin{equation}
\frac{m^2_W}{m^2_Z \cwd} 
= \frac{\sum_i\left[ T_i(T_i+1) - \cfrac{1}{4}Y_i^2\right]v^2_i}{\cfrac{1}{2}\sum_i Y_i^2\,v^2_i} 
\stackrel{\text{doublets}}{=} 
  \frac{\sum_i\left[ \dfrac{3}{4} - \cfrac{1}{4} \right]v^2_i}{\cfrac{1}{2}\sum_i v^2_i} 
= 1 \; .
 \label{eq:rho-tree}
\end{equation}
Any number of singlets and doublets respects custodial symmetry at tree
level; loop-induced contributions naturally remain small.  In
contrast, higher isospin representations violate custodial symmetry at
tree level.  This means that phenomenologically viable extensions
beyond singlets and doublets must
either~\cite{Barger:2009me}
\begin{enumerate}
\item obey stringent tree-level constraints, like type-II see-saw
  models with one additional Higgs triplet. Its
  VEV is strongly constrained, as described in Section~\ref{sec:triplet},
\item carefully align different Higgs fields, like in the Georgi-Machacek
  model~\cite{triplets},
\item or combine exotic representations, like a Higgs doublet combined
  with a septet~\cite{Hisano:2013sn}, such that custodial symmetry is
  accidentally preserved.
\end{enumerate}
None of these options is particularly appealing if we are looking for
a simple extension of the Standard Model Higgs sector. Thus, we
limit ourselves to singlet and doublet extensions whenever possible.

\subsection{Adding one singlet}
\label{sec:singlet}

An additional real $SU(2)_L$ singlet field $S_i$ is the simplest
extension of the minimal Higgs sector.  For example, a Higgs
unparticle or a Randall--Sundrum radion can mix with the Standard
Model Higgs boson, leading to a universal depletion of all Higgs
couplings.  All branching ratios of the lightest Higgs mass eigenstate
stay the same as in the Standard Model while all production rates are
reduced by the universal mixing factor.  If the new scalar is light,
$m_S < m_{\PHiggs}/2 \simeq 63$~GeV, and not coupled to any other
Standard Model particle, it will lead to invisible Higgs
decays~\cite{portal-collider}.  Implications of LHC data on singlet
extensions of the Standard Model have been analyzed thoroughly in
Ref.~\cite{Chpoi:2013wga}.  In our search for a consistent model of a
light Higgs with variable couplings an additional singlet will be a
first step towards a more general layout.\bigskip

The scalar potential of the minimal doublet-plus-singlet model can be
written
as~\cite{Silveira:1985rk,Pruna:2013bma}
\begin{alignat}{5}
V(\Phi,S) = 
  \mu^2_1\,(\Phi^\dagger\,\Phi) 
+ \lambda_1\,|\Phi^{\dagger}\Phi|^2 
+ \mu^2_2\,S^2 
+ \kappa S^3 
+ \lambda_2\,S^4 
+ \lambda_3\,|\Phi^{\dagger}\,\Phi|S^2 \; ,
\label{eq:singlet-potential}
\end{alignat}
where $\mu_1^2$ and $\lambda_1$ form the Standard Model
potential. Doublet--singlet mixing is induced by $\lambda_3$ and gives
rise to a light and a heavy Higgs boson mass-eigenstate, $h^0$ and
$H^0$, and a mixing angle $\theta$ defined in Appendix~\ref{app:para}.
Strong mixing with one singlet or equivalently mixing with a large
number of additional fields translates into a too large depletion
factor, inconsistent with the experimental limit $\cos\theta \lesssim
0.7$~\cite{sfitter_higgs,carmi}. Additional theoretical (electroweak
precision, perturbativity and unitarity) and experimental constraints
(mainly direct exclusion) turn out to be special cases of the doublet
extension and will be discussed in Section~\ref{sec:doublet}.\bigskip

An interesting setup avoiding a second VEV introduces a global $Z_2$
parity under which $S$ is odd. This predicts a SM-like Higgs sector
alongside with a dark matter candidate whose decays are precluded by
this $Z_2$ parity~\cite{singlet-portal}.  Following
Eq.\eqref{eq:singlet-potential} this WIMP singlet model is defined in
terms of the singlet mass term $\mu^2_2$, its portal coupling to the
doublet $\lambda_3$, and the singlet self-interactions $\lambda_2$.

Interactions of the singlet with Standard Model fields are mediated by Higgs
exchange. This way the quartic coupling $\lambda_3$ determines both
WIMP-nucleon scattering and WIMP-WIMP annihilation and can be constrained
experimentally.  Depending on its mass the singlet annihilates into Standard Model
fermions, gauge bosons and Higgs bosons via an $s$-channel Higgs
propagator. For $m_S \lesssim 1$~TeV the present cosmic abundance
constrains the quartic coupling to $\lambda_3 =
\mathcal{O}(0.01-1)$, irrespective of the singlet self-interactions.
WIMP-nucleon scattering is sensitive to the singlet mass and the
singlet-doublet mixing.  Values of $\lambda_3$ giving the correct
relic density yield direct detection rates around $\sigma_\text{NS}
= \mathcal{O}(10^{-43})\, \text{cm}^{-3}$, within reach of
ongoing experiments. Last but not least, indirect searches in gamma
rays test the model through singlet annihilation into photon pairs via
an $s$-channel Higgs subsequently decaying into $\gamma\gamma$.

A light singlet with $m_S < m_{\PHiggs}/2$ contributes to an invisible
Higgs width~\cite{dm-invisible}.  While first direct searches for invisible
decays have recently been performed in the Higgs-strahlung processes 
at the LHC~\cite{LHC-invisible}, model dependent indirect limits can also be
derived and yield stronger constraints~\cite{Belanger:2013kya}. 
They rule out wide domains in the
$\lambda_3 - \mu_2$ plane corresponding to light
$\mathcal{O}{(10)}$~GeV WIMPs which otherwise agree with astrophysical
observations.  A global fit including the WMAP
results~\cite{Komatsu:2010fb}, XENON-100 direct detection
data~\cite{Aprile:2012nq}, the FERMI-LAT di-photon
spectrum~\cite{Ackermann:2011wa}, and LHC constraints singles out a
best--fit point. It lies close to the resonance $m_S = 63$~GeV, and is
correlated with a small quartic coupling $\lambda_3 =
\mathcal{O}(10^{-2})$ to reconcile a large annihilation rate with the
observed relic density~\cite{Cheung:2012xb}. A comprehensive update
has been made available very recently~\cite{Cline:2013gha}.\bigskip

The real singlet model can be extended to a complex singlet, with
largely unchanged phenomenological patterns.  If the complex singlet
remains inert we obtain a two--component dark matter model. When the
scalar potential is expanded by a soft $U(1)$-breaking term the
imaginary part of $S$ gives rise to a massive pseudo-Goldstone
boson. In the presence of additional vector-like matter, models with a
complex singlet are capable to describe the current LHC search
results~\cite{Batell:2012mj}.  Interesting implications have also been
highlighted in models with multiple singlet
fields~\cite{Drozd:2011aa}.

\subsection{Adding one doublet}
\label{sec:doublet}

The two Higgs doublet model (2HDM)~\cite{Branco:2011iw} adds a second
$SU(2)$ doublet with weak hypercharge one.  It provides a low-energy
description for a broad ensemble of TeV-scale models, such as the
Minimal Supersymmetric Standard Model (MSSM)~\cite{Carena:2002es},
GUTs~\cite{pati,E6,SO10,trini,Stech:2012zr}, composite Higgs
models~\cite{composite}, and little Higgs models~\cite{little}.  It
allows for CP violation as well as a complex vacuum
structure~\cite{Gunion:2005ja} and addresses problems like neutrino
mass generation~\cite{Ma:2006km}, electroweak
baryogenesis~\cite{baryogenesis}, or dark
matter~\cite{Gong:2012ri}. Because the 2HDM includes two fields which
can couple to fermions and gauge bosons independently, it is a
candidate for an ultraviolet completion of a light Higgs model with
variable couplings. The main question is how strongly such variable
couplings are correlated by construction or through experimental
constraints.

Numerous studies have explored signatures of the 2HDM at the
LHC~\cite{2hdm-collider} and attempted to interpret the LHC Higgs
discovery in this framework.  Owing to the SM-like Higgs observation
any 2HDM explanation tends to translate into parameter constraints.
Early attempts ascribed the new resonance to either the
lighter~\cite{Ferreira:2011aa} or the
heavier~\cite{Ferreira:2012my}
neutral CP-even states, as well as to their CP-odd
companion~\cite{Burdman:2011ki}. Updates cover all major requirements
for flavor structures: natural flavor
conservation~\cite{Azatov:2012wq,fits-nfc,Chen:2013kt},
Yukawa alignment~\cite{fits-alignment,Celis:2013rcs}, and minimal
flavor violation~\cite{Altmannshofer:2012ar}.\bigskip

The 2HDM Higgs sector contains four complex scalar fields defined in
Eq.\eqref{eq:fields-def}.  The most general gauge and CP-invariant
renormalizable potential reads
\begin{alignat}{5}
 V(\Phi_1,\Phi_2) 
&= m^2_{11}\,\Phi_1^\dagger\Phi_1
 + m^2_{22}\,\Phi_2^\dagger\Phi_2
 - \left[ m^2_{12}\,\Phi_1^\dagger\Phi_2 + \text{h.c.} \right] \notag \\
&+ \frac{\lambda_1}{2} \, (\Phi_1^\dagger\Phi_1)^2
 + \frac{\lambda_2}{2} \, (\Phi_2^\dagger\Phi_2)^2
 + \lambda_3 \, (\Phi_1^\dagger\Phi_1)\,(\Phi_2^\dagger\Phi_2) 
 + \lambda_4 \, |\Phi_1^\dagger\,\Phi_2|^2 \notag \\
&+ \left[ \frac{\lambda_5}{2} \, (\Phi_1^\dagger\Phi_2)^2 
        + \lambda_6 \, (\Phi_1^\dagger\Phi_1) \, (\Phi_1^\dagger\Phi_2)
        + \lambda_7 \, (\Phi_2^\dagger\Phi_2)\,(\Phi_1^\dagger\Phi_2) + \text{h.c.} 
   \right] \; ,
\label{eq:2hdmpotential}
\end{alignat}
where the mass terms $m^2_{ij}$ and the dimensionless self-couplings $\lambda_i$ are
real parameters and $v_j = \sqrt{2} \braket{\Phi_j^0}$.
Their ratio we denote as $\tan\beta = v_2/v_1$.  Electroweak symmetry
breaking requires $v_1^2 + v_2^2 = (246~\gev)^2$.  The physical
spectrum entails five mass-eigenstates: two neutral CP-even scalars
$\hzero,\Hzero$, one neutral CP-odd scalar $\Azero$, and a set of
charged scalars $\PHiggs^\pm$. As described in Appendix~\ref{app:para}
the two mass eigenstates $H^0$ and $h^0$ arise from a rotation by the
angle $\alpha$.  Throughout this study we interpret the observed
Higgs scalar as the lighter $h^0$ state.  At tree level, custodial
symmetry ensures that the couplings to the weak gauge bosons $V=W,Z$
scale with the same factor
\begin{equation}
g_V = \sin (\beta - \alpha) \, g_V^\text{SM} \; .
\label{eq:vvh_2hdm}
\end{equation}
The pattern of the Yukawa couplings depends on the 2HDM setup. In
general, two of them can be modified independently in terms of
$\alpha$ and $\beta$, correlated with $g_{VV\hzero}$. Quantum
corrections may change this picture and lead to non-universal coupling
shifts $\Delta g_Z \neq \Delta g_W$ or coupling enhancements $\Delta_V
> 0$. We will discuss them in Secs.~\ref{sec:quantum_higgs} and
\ref{sec:quantum_fermion}.\bigskip

A fully flexible spectrum described by Eq.\eqref{eq:2hdmpotential}
allows for different patterns:
\begin{enumerate}
\item compressed masses $m_{\hzero} \simeq m_{\Hzero}$,
\item twisted masses $m_{\Azero} < m_{\hzero,\Hzero}$,
\item single mass hierarchy $m_{\hzero} \ll m_{\Hzero,\Azero,\PHiggs^\pm}$,
\item or multiple mass hierarchies $m_{\hzero} \ll m_{\Hzero} \ll m_{\Azero,\PHiggs^\pm}$. 
\end{enumerate}
The Higgs sector of the MSSM is one example of a constrained 2HDM
Higgs sector descending from a more general UV
completion~\cite{Carena:2002es}. Another example are dark portal or
inert doublet
models~\cite{Ma:2006km,inert-first,inert-cosmo}. They
follow when we enforce a $Z_2$ symmetry in Eq.\eqref{eq:2hdmpotential}
and require one of the Higgs doublets to transform as $\Phi_1 \to
-\Phi_1$, such that $\braket{\Phi_1} = 0$. This doublet does not
participate in electroweak symmetry breaking, and therefore it does
not interact with weak gauge bosons or fermions. Because the SM-like
Higgs field is exclusively linked to $\Phi_2$, without any admixture
of $\Phi_1$, the resulting Higgs couplings do not depart from the
Standard Model, except for charged Higgs contributions to
$g_\gamma$.\bigskip

An extensive set of bounds restricts the phenomenologically viable
regions of the 2HDM parameter
space~\cite{elkaffas,bounds-flavor,Azatov:2012bz,Azatov:2012wq,Chen:2013kt}.
Accidental tree-level symmetries in the Higgs sector play a big role
facing these constraints.  We need a novel class of possible
symmetries which relate the different potential terms.  They can be
classified as Higgs family (HF) symmetries, linking $\Phi_1$ and
$\Phi_2$ via a unitary transformation, and generalized CP (GCP)
transformations~\cite{Neufeld:1987wa}, linking $\Phi_1$ and $\Phi_2^*$
via a unitary transformation~\cite{Branco:2011iw}.  No matter what
combination of HF and/or GCP symmetries are enforced one always ends
up with one out of six distinct classes of Higgs potentials described
in Refs.~\cite{Ivanov:2006yq,Branco:2011iw}.  We will
focus on the flavor sector, while keeping the general potential of
Eq.\eqref{eq:2hdmpotential}. Three types of symmetries allow us to
write down 2HDM potentials in agreement with experimental constraints:

\paragraph{Custodial symmetry:} an accidental global $SU(2)_{L+R}$ symmetry in the
Standard Model protects the relation $m_W = m_Z\cw$. It is broken by
different up-type and down-type fermion masses and by virtual Higgs
exchange at one loop~\cite{Veltman:1976rt}.  Correspondingly, mass
terms involving the additional 2HDM fields are strongly
constrained. This effectively reduces the number of free parameters
and leads to a larger predictive power.  Barring fine-tuned mixing
angle choices we distinguish two scenarios for a phenomenologically
viable 2HDM: a compressed mass spectrum with only moderately split
Higgs masses or a light SM-like Higgs with mass-degenerate heavy
companions $\Hzero, \Azero, \PHiggs^\pm$.  In the latter all Higgs
states fall into the singlet and triplet representations of
$SU(2)_{L+R}$
\begin{alignat}{7}
\Phi_1 \supset& 
\begin{pmatrix} \PHiggs^+ \\ \Azero \\ \PHiggs^- \end{pmatrix} 
\oplus \Hzero \quad  \text{or} \quad
\Phi_1 \supset 
\begin{pmatrix} \PHiggs^+ \\ \Hzero \\ \PHiggs^- \end{pmatrix}
 \oplus \Azero
\qqqquad 
\Phi_2\supset& 
\begin{pmatrix} G^+ \\ G^0 \\ G^- \end{pmatrix} 
\oplus \frac{v+ \hzero}{\sqrt{2}} 
 \; .
\label{eq:spectrum-custodial}
\end{alignat}
This structure nicely illustrates the mechanism that protects the
custodial symmetry: while $\Phi_2$ accommodates a light, SM-like Higgs
boson all other mass eigenstates belong to $\Phi_1$.  We can integrate
them out simultaneously and retrieve an effective field theory
description in terms of $\Phi_2$ only.\bigskip

With $m_{\PHiggs} = 126$~GeV electroweak precision
data~\cite{Peskin:1990zt,oblique-2hdm} requires $S = 0.03 \pm 0.01$,
$T = 0.05 \pm 0.12$, and $U = 0.03 \pm 0.10$~\cite{pdg,globalew}.  The
genuine 2HDM contributions to the dominant constraint read
\begin{alignat}{5}
T
&=-\frac{\sqrt{2}G_F}{16\pi^2 \alpha_{em}}
\Bigg[ 
m_{\PHiggs^\pm}^2 \left( 1-\frac{m_{\Azero}^2}{m_{\PHiggs^\pm}^2-m_{{\Azero}}^2}\,
                   \log\frac{m_{\PHiggs^\pm}^2}{m_{\Azero}^2}
          \right) \notag \\
&\qqqquad +\cos^2(\beta-\alpha)\,
m_{\hzero}^2 \left( \frac{m_{\Azero}^2}{m_{\Azero}^2-m_{\hzero}^2}\, \log\frac{m_{\Azero}^2}{m_{\hzero}^2} 
               -\frac{m_{\PHiggs^\pm}^2}{m_{\PHiggs^\pm}^2-m_{\hzero}^2}\, \log\frac{m_{\PHiggs^\pm}^2}{m_{\hzero}^2}
         \right) \notag \\
&\qqqquad +\sin^2(\beta-\alpha)\,
m_{\Hzero}^2 \left( \frac{m_{\Azero}^2}{m_{\Azero}^2-m_{\Hzero}^2}\, \log\frac{m_{\Azero}^2}{m_{\Hzero}^2}
               -\frac{m_{\PHiggs^\pm}^2}{m_{\PHiggs^\pm}^2-m_{\Hzero}^2}\, \log\frac{m_{\PHiggs^\pm}^2}{m_{\Hzero}^2}
         \right) 
\Bigg] \; ,
\label{eq:tparam}
\end{alignat}
For degenerate masses $m_{\Hzero} \simeq m_{\Azero} \simeq
m_{\PHiggs^\pm}$ we find $S = T = 0$. In contrast, in the decoupling
limit $\sin(\beta-\alpha) \simeq 1$ with $m_{\Hzero} \simeq
m_{\Azero}$, the corrections read
\begin{alignat}{5}
S \simeq -\cfrac{1}{12\pi}\, \log \cfrac{m^2_{\PHiggs^\pm}}{m^2_{\Azero}}
 \qqqquad
T \simeq  \frac{\sqrt{2} G_F}{16\pi^2 \alpha_{ew}}\,\left(m^2_{\PHiggs^{\pm}}-m^2_{\Azero}\right)  \; .
\label{eq:stlimits}
\end{alignat}
Moderate mass splittings are hence directly related to custodial
symmetry. This is a key ingredient to understanding the deviations in
the $W$ and $Z$ interactions to the light Higgs boson.

\paragraph{Flavor symmetry:} the symmetry group $U(3)_{Q_L} \times
U(3)_{U_R} \times U(3)_{D_R}$ acting on three different types of
quarks leaves the CKM matrix invariant and guarantees the absence of
tree-level FCNCs through GIM suppression.  It is broken by large
Yukawas which give rise to loop-induced FCNC interactions.
This picture does not hold in the presence of a second Higgs doublet.
Traditionally, natural flavor conservation imposes a global,
flavor-blind, $Z_2$ discrete symmetry $\Phi_{1,2} \to \mp \Phi_{1,2}$ and
demands any fermion family to couple to only one Higgs doublet. It
satisfies the Glashow-Weinberg theorem~\cite{Glashow:1976nt} and
defines the canonical 2HDM setups:
\begin{itemize}
\item[--] type-I, where all fermions couple to just one Higgs
  doublet, namely $\Phi_2$;
\item[--] type-II, where up-type (down-type) fermions couple
  exclusively to $\Phi_2$ ($\Phi_1$);
\item[--] lepton--specific, with a type-I quark sector and a type-II
  lepton sector;
\item[--] flipped, with a type-II quark sector and a type-I lepton
  sector.
\end{itemize}
The formulas for all these couplings are provided in
Appendix~\ref{app:para}. The discrete $Z_2$ symmetry $\Phi_{1,2} \to
\mp \Phi_{1,2}$ can be understood as a single-parameter HF symmetry. It
forbids the mass term $m^2_{12}$ and the self interactions
$\lambda_{6,7}$.  If extended to the Yukawa sector it automatically
removes all sources of tree-level FCNC interactions.  The continuous
$U(1)$ transformation is $\Phi_{1,2} \to \Phi_{1,2} \, \exp(\mp
i\theta)$ with a real phase $\theta$. It is a genuine Peccei-Quinn
(PQ) symmetry.  The field combination $\hzero_j + i a_j^0$ has a
non-vanishing PQ charge, explaining why the mass splitting between the
mass eigenstates $\Hzero,\Azero$ is controlled by the size of the
PQ-breaking terms $m^2_{12}$ and $\lambda_{6,7}$.

Yukawa structures featuring natural flavor conservation constitute one
case of a broader class of models, in which tree-level FCNCs are
largely suppressed, albeit not fully absent. This is ensured via
minimal flavor violation~\cite{D'Ambrosio:2002ex,Dery:2013aba}. This
mechanism is based on promoting the Yukawa couplings to auxiliary
fields or spurions, and imposes a link between the ($3\times3$) Yukawa
matrices $y_1,y_2$ which couple to the corresponding Higgs doublets.
All flavor transitions are controlled by CKM matrix elements, so that
they become naturally small.

One implementation of minimal flavor violation are aligned
models~\cite{Pich:2009sp}.  There, the fermions couple to both Higgs
doublets with \emph{aligned} Yukawa matrices, \ie linked to one
another by linear shifts $\epsilon_f$,
\begin{equation}
y_{u1} = \epsilon_u \; y_{u2} \qqquad  
y_{d2} = \epsilon_d \; y_{d1} \qqquad  
y_{l2} = \epsilon_{\tau} \; y_{l1} \; ,
\label{eq:alignment}
\end{equation}
with $y_{f,i} = \sqrt{2} m_f/v_i$.  This way the fermion masses and
Yukawa matrices diagonalize simultaneously.  The construction of the
aligned 2HDM relies on basis invariance
properties~\cite{Davidson:2005cw} to absorb the $\epsilon_f$ into a
field re-definition and rotate one of them away (usually
$\epsilon_u$).  In practice, we parameterize the dependence on the
different doublets by introducing angles $\gamma_{b,\tau}$, by which
we may write the bottom and tau Yukawas
\begin{alignat}{5}
\frac{m_{b,\tau}}{v}
= y_{b,\tau}\braket{\Phi_1}\cos\gamma_{b,\tau}
+ y_{b,\tau}\,\braket{\Phi_2}\,\sin\gamma_{b,\tau} 
\qquad  \Rightarrow  \qquad 
y_{b,\tau}\,\cos(\beta-\gamma_{b,\tau}) = \cfrac{\sqrt{2}m_{b,\tau}}{v} \; .
\label{eq:alignment2}
\end{alignat}
The aligned setup includes natural flavor conservation, but covers all
regimes with absent tree-level FCNC interactions. 

\paragraph{CP invariance:} while hermiticity of the Higgs potential
Eq.\eqref{eq:2hdmpotential} requires $m^2_{11},m^2_{22}$ and
$\lambda_{1,2,3,4}$ to be real, the parameters $m^2_{12},
\lambda_{5,6,7}$ may be complex. In that case not only the CP-even but
also the CP-odd neutral Higgs fields will mix and define physical
fields which are no longer CP-eigenstates. New CP-violating phases can
also appear in the Higgs Yukawa matrices.  We will omit CP-violating
terms in the Higgs potential.\bigskip

Additional experimental constraints from $B$-mesons and
taus~\cite{Aubert:2007my,bounds-flavor} we account for assuming no
cancellation of 2HDM effects and for example other supersymmetric
contributions.  The new heavy scalars may also contribute to the muon
anomalous magnetic moment $(g-2)_\mu$~\cite{gm2}, even though their
effect gives rise to mass constraints which are milder than those from
custodial symmetry. Last but not least, any Higgs mass spectrum ought
to satisfy all current limits from direct searches at LEP, Tevatron
and LHC~\cite{direct-searches}. Notice that systematic searches for
heavy Higgs bosons at the LHC have been so far restricted mainly to
the MSSM, \eg ruling out $\tan\beta \gtrsim 7.58$ for $m_{\Azero}
\simeq 300$~GeV.  We implement all the above mentioned constraints
using \textsc{2HDMC}~\cite{Eriksson:2009ws},
\textsc{HiggsBounds}~\cite{higgsbounds} and
\textsc{SuperIso}~\cite{superiso} together with a number of in-house
routines.

\subsection{Adding one doublet and one singlet}
\label{sec:singletdoublet}

Adding singlets to the 2HDM setup hardly changes its main features.
Following Eq.\eqref{eq:rho-tree} such models respect custodial
symmetry at tree level, which means that the Higgs couplings to the
$W$ and $Z$ bosons are linked. In the light of an ultraviolet
completion of a model with variable Higgs couplings the main advantage
of the additional singlet is a consistent modelling of invisible Higgs
decays.\bigskip

Usually, the main interest in these scenarios comes from a less strict
upper limit on the lightest Higgs mass in its supersymmetric version,
the NMSSM~\cite{Ellwanger:2009dp}.  The main virtue of the NMSSM is
that it explains the mass term $\mu \Phi_u \Phi_d$ of the order of the
soft SUSY-breaking scale $M_\text{SUSY}$ by generating it through a
singlet VEV and adding
\begin{equation}
 W_\text{NMSSM} \supset \lambda S \hat{H}_u\,\hat{H}_d + \frac{k}{3}\,\hat{S}^3
\label{eq:nmssm-potential}
\end{equation}
to the Higgs superpotential. The particle spectrum now includes three
neutral CP-even states, all linear combinations of
$\hat{H}_u$,$\hat{H}_d$ and $\hat{S}$, and two neutral CP-odd
scalars. The higgsinos mix with the usual neutralinos.  A 126~GeV
SM-like Higgs boson with slightly enhanced $g_\gamma$ can be realized
for large $\lambda$ and small $\tan\beta$ values. In these instances,
the two lightest mass eigenstates $\hzero_{1,2}$ typically have
nearby masses and very weak doublet mixing.\bigskip

If, in the spirit of the NMSSM we assume that the additional singlet
develops a VEV, we find a light Higgs state $h^0 =
\cos\theta\,(\cos\alpha\,h_2^0 - \sin\alpha\,h_1^0 ) + \sin\theta\,s$
with couplings
\begin{alignat}{5}
\frac{g_{VV\hzero}}{ g_V^\text{SM}} = \cos\theta\,\sin(\beta-\alpha)
\qqqquad 
\frac{g_{ff\hzero}}{g_f^\text{SM}} = \cos\theta\,\cfrac{\cos\alpha}{\sin\beta}  \; ,
\label{eq:param7}
\end{alignat}
where we assume a type-I 2HDM structure.  The 2HDM interaction pattern
is therefore simply rescaled.  

Given that an additional singlet with a VEV does not structurally
enhance our ability to describe a light Higgs state with free
couplings we can focus on an additional inert Higgs singlet as a
source of invisible Higgs decays. The mixing parameter $\eta$ is
defined by
\begin{alignat}{5}
 \lag \supset \eta\,\left( \Phi_1^\dagger\Phi_1 + \Phi_2^\dagger\Phi_2  \right)\,S^2
\label{eq:lag-singlet-doublet}.
\end{alignat}
The corresponding invisible width for the SM-like Higgs boson
$\Gamma^\text{inv}_{\hzero}$ reads
\begin{alignat}{5}
 \eta^2 = \frac{32\,\pi\,m_{\hzero}}{v^2\,\sin^2(\beta-\alpha)} 
          \left( 1- \frac{4m_s^2}{m^2_{\hzero}} \right)^{-1/2}
\,\Gamma_{\hzero}^\text{inv} \; .
\label{eq:invwidth}
\end{alignat}

\subsection{Adding one triplet}
\label{sec:triplet}

Adding a triplet to the Standard Model Higgs sector allows to
separately vary the light Higgs couplings to the $W$ and $Z$
bosons. The model has enough degrees of freedom to accommodate a wide
range of independent variations in all Higgs couplings.  The
phenomenological motivations for Higgs triplet models are based, among
others, on left-right symmetries~\cite{triplet-orig} or type-II
see-saw models for neutrino mass generation.  Interesting implications
have been highlighted in the context of non-minimal SUSY
extensions~\cite{Delgado:2013zfa}.  A study of the model at the
quantum level is available in Ref.~\cite{Aoki:2012jj}.\bigskip

The Higgs sector consists of an isospin doublet $\Phi$ with
hypercharge +1 and an $SU(2)_L$ triplet $\Delta =(\vec{T} \cdot
\vec{\tau})$, also with hypercharge +1. 
%
In terms of the doublet and triplet VEVs the tree-level gauge boson
masses are
\begin{equation}
m_W^2 = \frac{g^2}{4}\, \left( v^2_\Phi +2v_\Delta^2 \right)
\qqqquad
m_Z^2 = \frac{g^2}{4 \cwd}\, \left( v^2_\Phi + 4v_\Delta^2 \right) \; ,
\label{eq:gaugeboson3}
\end{equation}
manifesting the breaking of the custodial symmetry already at the
tree-level.  The physical spectrum includes two CP-even scalars
$\hzero, \Hzero$, one CP-odd scalar $\Azero$, one singly charged
scalar $\PHiggs^\pm$, and one doubly-charged state $\PHiggs^{\pm
  \pm}$.  Barring marginal corners of the parameter space, one of the
CP-even scalars will behave like a Standard Model Higgs boson. All
remaining states can be described by an almost-decoupled $SU(2)$
triplet -- minimizing the tree-level breaking of the custodial
symmetry.\bigskip

As alluded to above, adding a triplet is the obvious choice when we
look for an extended Higgs sector with maximum flexibility in the
couplings of the lightest state. However, because of the
phenomenological problems of these models we will stick to additional
singlets and doublets and enhance their flexibility by fully
exploiting not only tree-level but also loop effects.

\subsection{Degenerate spectrum} 
\label{sec:degenerate-description}

In extensions of the Standard Model often two or more representations
of scalar fields are necessary. It can happen that fields in one
representation have couplings to gauge bosons and fermions similar to
the Standard Model couplings, while fields in the other representation
only couple to gauge bosons.  Such models can arise in generalizations
of the Glashow--Weinberg--Salam group to left-right symmetric groups
like $E_6$, its subgroup $SU(3)_L \times SU(3)_R \times SU(3)_C$ ~\cite{Stech:2008wd,Stech:2010gf,Stech:2012zr}, or
even smaller subgroups.

In these cases the first representation can account for all fermion
couplings. Nonetheless, the corresponding fields are not sufficient to
provide the correct low-scale chirality properties of the Standard
Model. For instance, when the left-right symmetry is not broken by the
VEVs one needs fields of a second representation performing this
task. Such fields will then not directly couple to fermions, but only
via the gauge vector bosons. 

Interestingly, in such models, the Higgs candidate found at the LHC
might no longer be a single resonance. Instead, it can be a mass
degenerate state, consisting of two orthogonal components; each of
them is an admixture of a conventional, SM-like scalar with the new
fermiophobic counterparts~\cite{Stech:2013pda}.  Mass degenerate
scalar states of a different type have been considered in connection
with singlet extensions~\cite{Heikinheimo:2013cua}, the NMSSM
approach~\cite{Gunion:2012he} and also within conventional
2HDM~\cite{Ferreira:2012nv}. In order to discover or disprove such
near degeneracies with mass separations smaller than the experimental
mass resolution, dedicated strategies are
necessary~\cite{Gunion:2012gc}.  Although such models in general
contain more $SU(2)_L$ doublets, in Section~\ref{sec:fit_degenerate}
we concentrate on the case where the observed 126~GeV resonance is
composed of two scalar fields, with only one of them giving masses to
all of the fermions in a type-I Yukawa structure.

\subsection{Ultraviolet structure}
\label{sec:uv}

If we want to establish extended Higgs sectors as a consistent
perturbative framework for a light Higgs boson with variable couplings
we need to carefully study the high energy behavior of our theory. The
Standard Model Higgs sector with the observed mass of $m_H = 126$~GeV
navigates between two strong high-scale constraints: first, the
triviality bound forbids a significant enhancement of the relevant
Higgs self-couplings~\cite{triviality,trivial-fixedpoint}. Second, for
$m_H \lesssim 130$~GeV the electroweak vacuum may become metastable
and eventually decay into the global low-lying vacuum via thermal or
quantum-induced tunneling~\cite{vacuum-tree,vacuum-loop,
  metastable-decay}. While these constraints might be avoidable
through an appropriate ultraviolet embedding, we still consider them
for the study of these effective extended models.\bigskip

At high energy scales the running Higgs self-coupling develops a
Landau pole, which has to lie outside the range of validity of our
fundamental theory.  In the same spirit, but numerically more
relevant, the size of the Higgs self-couplings is limited by
unitarity~\cite{Lee:1977eg}.  To avoid this the partial waves of the
scalar--scalar, gauge boson--gauge boson and scalar--gauge boson
scattering processes must be limited from above.  Following the
Goldstone equivalence theorem the high energy behavior of all these
processes is determined by the scalar sector. The leading
contributions are governed by the scalar self-couplings $\lambda_i$ in
the Higgs potential in Eq.\eqref{eq:2hdmpotential}.  We can in general
apply unitarity bounds for an additional singlet, doublet and triplet.
In the 2HDM the different channels force the absolute value of each
combination of scalar
self-couplings~\cite{Casalbuoni:1986hy,Pruna:2013bma,unitarity,Gunion:2002zf}
\begin{alignat}{5} 
a_{\pm} &= 
\frac{1}{16\pi}\,\left[3(\lambda_1+\lambda_2)\, \pm \sqrt{9(\lambda_1-\lambda_2)^2 + 4(2\lambda_3+\lambda_4)^2} \right] 
\qqquad 
&f_1 &= f_2 = \frac{1}{8\pi}\,\left(\lambda_3 + \lambda_4\right) \notag \\
b_{\pm} &= 
\frac{1}{16\pi}\,\left[(\lambda_1+\lambda_2)\, \pm \sqrt{(\lambda_1-\lambda_2)^2 + 4\lambda_4^2} \right] 
&f_+ &= \frac{1}{8\pi}\,\left(\lambda_3 + 2\lambda_4 + 3\lambda_5\right) \notag \\
c_{\pm} &= 
\frac{1}{16\pi}\,\left[(\lambda_1+\lambda_2)\, \pm \sqrt{(\lambda_1-\lambda_2)^2 + 4\lambda_5^2} \right] 
&f_- &= \frac{1}{8\pi}\,\left(\lambda_3 + \lambda_5\right) \notag \\
e_1 &= \frac{1}{8\pi}\left(\lambda_3 + 2\lambda_4 - 3\lambda_5\right) 
&p_1 &= \frac{1}{8\pi}\left(\lambda_3 - \lambda_4\right) \notag \\
e_2 &= \frac{1}{8\pi}\,\left(\lambda_3 -\lambda_5\right); 
\label{eq:2hdm-unitarity1}
\end{alignat}
to be smaller than unity. In addition, tree-level unitarity in the
usual 2HDM-like mixing pattern imposes a set of sum rules for the
Higgs couplings to Standard Model particles~\cite{general-unitarity}
\begin{alignat}{5}
g_{VV \hzero}^2 + g_{VV \Hzero}^2 &=  \left( g_V^\text{SM} \right)^2 
\qqqquad 
&g^2_{ff \hzero} + g^2_{ff \Hzero} + g^2_{ff \Azero}&= \frac{m_f^2}{v^2} \equiv \left( g_f^\text{SM} \right)^2 
 \notag \\
g^2_{\hzero\Azero\PZ^0} +  g^2_{\Hzero\Azero\PZ^0} &= \frac{g^2}{4\cwd} 
\qqqquad 
&g^2_{\phi ZZ} + 4 m_Z^2 g^2_{\phi \Azero\PZ^0} &= \frac{g^2 m_Z^2}{\cwd} \notag \\ 
g_{ff \hzero}\,g_{VV \hzero} + g_{ff \Hzero}\,g_{VV \Hzero} &= g_{f}^\text{SM} \,g_{V}^\text{SM}\, ,
\label{eq:sumrules}
\end{alignat}
with $\phi = \hzero,\Hzero$ and $V = W,Z$. The weak coupling is $g =
e/\sw$.  These tree--level sum rules have obvious implications: first,
the Higgs couplings to massive gauge bosons can be at most as strong
as in the Standard Model; second, the related sum rules for $W$ and
$Z$ bosons suggest a universal modification to both couplings
$\Delta_W \simeq \Delta_Z$, reflecting custodial invariance; third,
and owing to the structure of the gauge derivative, all vertices
containing at least one gauge boson and exactly one non-standard Higgs
field ($\Hzero,\Azero,\PHiggs^\pm$) will be proportional to
$\cos(\beta-\alpha)$.  Consequently, these heavy fields will decouple
from the Standard Model dynamics in the limit $\alpha \to \beta -
\pi/2$~\cite{Gunion:2002zf}.\bigskip

The second condition on ultraviolet models is the stability of the
vacuum.  For a 2HDM with the most general potential the non-linear
nature of the vacuum conditions $\partial V/\partial \Phi = 0$ as a
function of the VEVs may lead to different vacua featuring spontaneous
CP or $U(1)$ breaking.  Vacuum stability can be
ensured by requiring~\cite{Gunion:2002zf,vacuum-tree,Kanemura:1999xf}
\begin{equation} 
\lambda_1 > 0 \qqquad 
\lambda_2 > 0 \qqqquad 
\sqrt{\lambda_1\lambda_2} + \lambda_3 
    + \text{min}\left( 0, \lambda_4 - |\lambda_5| \right) > 0 \; .
\label{eq:vacuum}
\end{equation}
Using the renormalization group equations for the 2HDM~\cite{RG} we
can demand these conditions to hold for running self-couplings up to
any arbitrarily high scale~\cite{Kanemura:1999xf,Ferreira:2009jb}.  A
few generic properties govern the outcome: very small weak--scale
self-couplings typically lead to unbounded high--scale potentials;
very large weak--scale self-coupling values would hit a low-lying
Landau pole; all constraints are very sensitive to additional
symmetries in the 2HDM potential like a global $Z_2$
parity~\cite{Branco:2011iw}.  In practice, we envision our 2HDM setup
as an effective Higgs sector parameterization of a generic TeV-scale
UV-completion. We thus allow for new physics entering around
$\mathcal{O}$(1 - 10)~TeV. Therefore tree-level bounds provide a
suitable vacuum stability prescription.\bigskip

Finally, to ensure that the entire modified Higgs sector remains
weakly interacting all Yukawas should be sufficiently small
at the weak scale, $y_{f}/\sqrt{2} < \sqrt{4 \pi}$.  This translates
into $\tan\beta > 0.28$ for all natural flavor conservation 2HDM
models, $\tan\beta < 140 $ for type-II and the flipped models, and
$\tan\beta < 350$ in the lepton--specific case~\cite{Chen:2013kt}.

\section{Coupling patterns}
\label{sec:patterns}

If we assume that the observed light Higgs resonance is part of an
extended Higgs sector the key question is what we can say about the
structure of such an extended Higgs sector when looking at patterns in
the light Higgs couplings.  The challenge in this section is to
identify model features which allow light Higgs particles with
flexible couplings $\Delta_x =
(g_x-g_x^\text{SM})/g_x^\text{SM}$~\cite{sfitter_higgs} and an
invisible decay width.  

Turning this argument around we can ask if a sufficiently general
extended Higgs sector can serve as a consistent renormalizable
framework to describe free Higgs couplings in the Standard Model Higgs
Lagrangian.  We will attempt to build such models based on the
extended Higgs sectors summarized in Section~\ref{sec:extensions}.  For
these models we can compute electroweak quantum corrections to Higgs
observables without assuming a Standard Model
structure~\cite{passarino}. After largely reviewing the tree--level
patterns of extended Higgs sectors we will give a comprehensive
discussion of quantum effects from this perspective.

\subsection{Tree-level couplings}
\label{sec:patterns_tree}

\begin{table}[t]
\begin{center} \begin{small}
\begin{tabular}{|l|l|ll|ll|} \hline
\multicolumn{2}{|c}{ } & \multicolumn{4}{|c|}{$hVV$} \\ \hline
\multirow{2}{1.8cm}{extension} & \multirow{2}{1.8cm}{model} &
\multicolumn{2}{c|}{universal} & \multicolumn{2}{c|}{non-universal}
\\ & & \multicolumn{2}{|c|}{rescaling} &
\multicolumn{2}{c|}{rescaling} \\ \hline \multirow{2}{1.8cm}{singlet}
& inert ($v_S = 0$) & & & & \\ \cline{2-6} & EWSB ($v_S \neq 0$) &
$\theta$ & $\Delta_V< 0$ & & \\ \hline
\multirow{5}{1.8cm}{doublet} & inert ($v_d = 0$) & & & &
\\ \cline{2-6}
& type-I & $\alpha-\beta$ & $\Delta_V < 0$ & $\mathcal{O}(y_f,\lambda_H)$ 
& $\Delta_V \gtrless 0$ 
\\ \cline{2-6}
& type-II-IV & $\alpha-\beta$ & $\Delta_V< 0$&   $\mathcal{O}(y_f,\lambda_H)$ & $\Delta_V\gtrless 0$ \\ \cline{2-6}
 &aligned, MFV & $\alpha-\beta$ & $\Delta_V< 0$ &  $\mathcal{O}(y_f,\lambda_H)$ & $\Delta_V\gtrless 0$  \\\hline 
singlet+doublet &  
& $\alpha-\beta,\theta$ & $\Delta_V< 0$ &  $\mathcal{O}(y_f,\lambda_H)$ & $\Delta_V\gtrless 0$  
\\\hline 
triplet &   &  & & $\alpha,\beta_n,\beta_c$ & 
$\Delta_V\gtrless 0$ 
\\ \hline
\end{tabular}
\end{small} \end{center}
\caption{Interaction patterns for a light Higgs boson to weak gauge
  bosons ($V= W^\pm,Z$), allowing for universal or non-universal
  departures from the Standard Model interactions.  We indicate the
  relevant model parameters defined in Appendix~\ref{app:para} and the
  possibility of coupling enhancement vs suppression.}
\label{tab:structure-gauge}
\end{table}

In models with additional singlets and doublets the light Higgs
interactions to massive gauge bosons show a systematic suppression
\begin{alignat}{5}
\frac{g_V}{g^\text{SM}_V} 
\stackrel{\text{singlet}}{=} 
 \cos \theta
\qqquad \text{and} \qqquad 
\frac{g_V}{g^\text{SM}_V} 
\stackrel{\text{2HDM}}{=} 
 \sin(\beta-\alpha) \; ,
\label{eq:couptogauge}
\end{alignat}
arising from tree-level mixing in the light Higgs mass eigenstate.
The Standard Model coupling $g_V^\text{SM}$ is fixed by unitarity and
renormalizability. Assuming the singlets and doublets all develop a
non-zero VEV, it is composed of the different \CP-even scalar
components. In the 2HDM the mixing factors $\cos (\beta-\alpha)$ and
$\sin (\beta - \alpha)$ realize the unitarity sum rule in
Eq.\eqref{eq:sumrules}.  These couplings appear in the covariant
derivative, so the suppression is universal for $g_Z$ and $g_W$,
unless we introduce a Higgs triplet.  As we will see in
Sections~\ref{sec:quantum_higgs} and \ref{sec:quantum_fermion} quantum
effects modify this coupling pattern slightly.  We document this
simple structure of the couplings of the lightest Higgs boson in
Table~\ref{tab:structure-gauge}. The notation
$\mathcal{O}(y_f,\lambda_H)$ stands for fermion--mediated or
Higgs--mediated loop contributions which we will discuss
below.\bigskip

\begin{table}[b!]
\begin{center} \begin{small}
\begin{tabular}{|l|l|ll|ll|} \hline
\multicolumn{2}{|c|}{ } & \multicolumn{4}{c|}{$hf\bar{f}$} \\ \hline
\multirow{2}{1.8cm}{extension} & \multirow{2}{1.8cm}{model} & \multicolumn{2}{c|}{universal} & \multicolumn{2}{c|}{non-universal} \\ 
 & & \multicolumn{2}{c|}{rescaling} & \multicolumn{2}{c|}{rescaling} \\ \hline 
\multirow{2}{1.8cm}{singlet} & inert ($v_S = 0$) & & &  & \\ \cline{2-6} 
 & EWSB ($v_S \neq 0$) & $\theta$ & $\Delta_f< 0$ & &  \\ \hline
\multirow{5}{1.8cm}{doublet} & inert ($v_d = 0$) & & &  & \\ \cline{2-6}
 & type-I & $\alpha-\beta$ & $\Delta_f\gtrless 0$ & $\mathcal{O}(y_f,\lambda_H)$ 
& $\Delta_f\gtrless 0$ 
\\ \cline{2-6}
  & type-II &  & &  $\alpha-\beta$, $\mathcal{O}(y_f,\lambda_H)$ & $\Delta_f\gtrless 0$ \\ \cline{2-6}
 &aligned/MFV & $y_f$, & $\Delta_f\gtrless 0$ & $y_{f}, 
\mathcal{O}(y_f,\lambda_H)$ & $\Delta_f\gtrless 0$  \\\hline 
singlet+doublet &  &  $y_f,\theta$ & $\Delta_f\gtrless 0$  & $y_f, 
\mathcal{O}(y_f,\lambda_H)$& 
$\Delta_f\gtrless 0$  
\\ \hline
triplet &   & $\beta_n$ & $\Delta_f \gtrless 0$&  $\mathcal{O}(y_f,\lambda_H)$& 
$\Delta_f\gtrless 0$ 
\\ \hline
\end{tabular}
\end{small} \end{center} 
\caption{Interaction patterns for a light Higgs boson to fermions,
  allowing for universal or non-universal departures from the Standard
  Model interactions.  We indicate the relevant model parameters
  defined in Appendix~\ref{app:para} and the possibility of coupling
  enhancement vs suppression.}
\label{tab:structure-fermions}
\end{table}

\begin{table}[t]
\begin{center} \begin{small}
\begin{tabular}{|l|l|ll|ll|l|} \hline
\multicolumn{2}{|c|}{ } &  \multicolumn{2}{c|}{$h\gamma\gamma$}&\multicolumn{2}{c|}{$hgg$} & $\Gamma_\text{inv}$ \\ \hline
\multirow{2}{1.8cm}{extension} & \multirow{2}{1.8cm}{model}  
& \multicolumn{2}{c|}{ } & \multicolumn{2}{c|}{ } &  \\ 
 & & \multicolumn{2}{c|}{ } &\multicolumn{2}{c|}{ } & \\
 \hline 
\multirow{2}{1.8cm}{singlet} & inert ($v_S = 0$) & & & & & $\lambda_{hSS}$ \\ \cline{2-7} 
 & EWSB ($v_S \neq 0$)  & $\theta$ & $\Delta_\gamma^\text{tot} < 0$ &  $\theta$&   $\Delta_g^\text{tot} <  0$ &  \\ \hline
\multirow{5}{1.8cm}{doublet} & inert ($v_d = 0$)  & $\lambda_{hH^+H^-}$ & $\Delta_\gamma^\text{tot}\gtrless 0$ & & & $\lambda_H$ \\ \cline{2-7}
 & type-I &
 $\alpha-\beta,\lambda_{hH^+H^-}$ & $\Delta_\gamma^\text{tot}\gtrless 0$ & $\alpha-\beta$ & $\Delta_g^\text{tot}\gtrless 0$ &  \\ \cline{2-7}
  & type-II-IV  &
 $\alpha-\beta,\lambda_{hH^+H^-}$ & $\Delta_\gamma^\text{tot}\gtrless 0$ & $\alpha-\beta$ & $\Delta_g^\text{tot}\gtrless 0$ &  \\ \cline{2-7}
 &aligned, MFV  &
 $\alpha-\beta,\lambda_{hH^+H^-}$ & $\Delta_\gamma^\text{tot}\gtrless 0$ & $y_{f}$& $\Delta_g^\text{tot}\gtrless 0$& \\\hline
singlet + doublet &  
 &  
 $\alpha-\beta,\theta,\lambda_{hH^+H^-}$ & $\Delta_\gamma^\text{tot}\gtrless 0$ & $y_{f}, \theta$& $\Delta_g^\text{tot}\gtrless 0$
&$\lambda_{hSS}$ 
\\ \hline
 \multirow{3}{1.8cm}{triplet} &    & $\alpha,\beta_n,\beta_c, $ 
 & \multirow{3}{1.8cm}{$\Delta_\gamma^\text{tot}\gtrless 0$} 
 & \multirow{3}{1.8cm}{$\beta_n$} 
 & \multirow{3}{1.8cm}{$\Delta_g^\text{tot} \gtrless 0$} &  \\ 
& & $\lambda_{hH^+H^-}, $      & & & & \\
& & $\lambda_{hH^{++}H^{--}}$  & & & & 
\\ \hline
\end{tabular}
\end{small} \end{center}
\caption{Interaction patterns for a light Higgs boson to photons,
  gluons and invisible states, allowing for universal or non-universal
  departures from the Standard Model interactions.  We indicate the
  relevant model parameters defined in Appendix~\ref{app:para} and the
  possibility of coupling enhancement vs suppression.}
\label{tab:structure-loopinduced}
\end{table}

Departures from the Standard Model Higgs couplings to fermions can
have two origins: first, the fermionic 2HDM mixing structure includes
a \CP-odd gauge boson $\Azero$.  This coupling comes with an
additional factor $i \gamma_5$, which means that $g^2_{f f \Azero} <
0$ in the sum rule quoted in Eq.\eqref{eq:sumrules} allows for at
least one of the \CP-even couplings to lie above
$g^\text{SM}_f$. Second, multi-doublet structures typically allow for
independent variations for the up-type and down-type fermions.
Combining both effects, positive and negative tree-level non-universal
shifts $\Delta_f$ are attainable, as shown in
Table~\ref{tab:structure-fermions}. A serious limitation of the 2HDM
setup is that we only have two parameters to describe the leading
tree-level effects in $g_V$ and $g_{t,b,\tau}$. If ($\beta - \alpha$)
is fixed by $g_V$ all three Yukawas are described by $\tan
\beta$. Only Yukawa alignment gives rise to a more flexible pattern,
unlinking the bottom and tau Yukawas via the independent angles
$\gamma_{b,\tau}$.\bigskip

Higgs interactions to photons and gluons are generated by loops of all
relevant particles in a given model.  For the gluon case this requires
colored states, so we can immediately apply the modified quark Yukawa
patterns of an extended Higgs sector. In the Standard Model the bottom
contribution to these Higgs coupling loops is negligible, so we can
generate a non-trivial scaling of $g_g$ with respect to $g_t$ by
significantly increasing the bottom Yukawa, as we will see later.  The
photon coupling depends on the three heavy Yukawas and on $g_W$, but
will also receive corrections due to new charged scalars in the Higgs
sector.  The effect of additional states is relatively enhanced as it
overlays to the destructive interference between the leading top and
$W$ contributions in the Standard
Model~\cite{Kribs:2007nz}. Electroweak corrections to the two channels
can have similarly enhanced effects.  A charged Higgs loop is governed
by the Higgs potential~\cite{self}.  Deviations from
$g_\gamma^\text{SM}$ can manifest themselves both as an increase and a
reduction, depending on the size and sign of the $\hzero
\PHiggs^+\PHiggs^-$ coupling.

\begin{figure}[t]
\begin{center}
\includegraphics[width=0.8\textwidth]{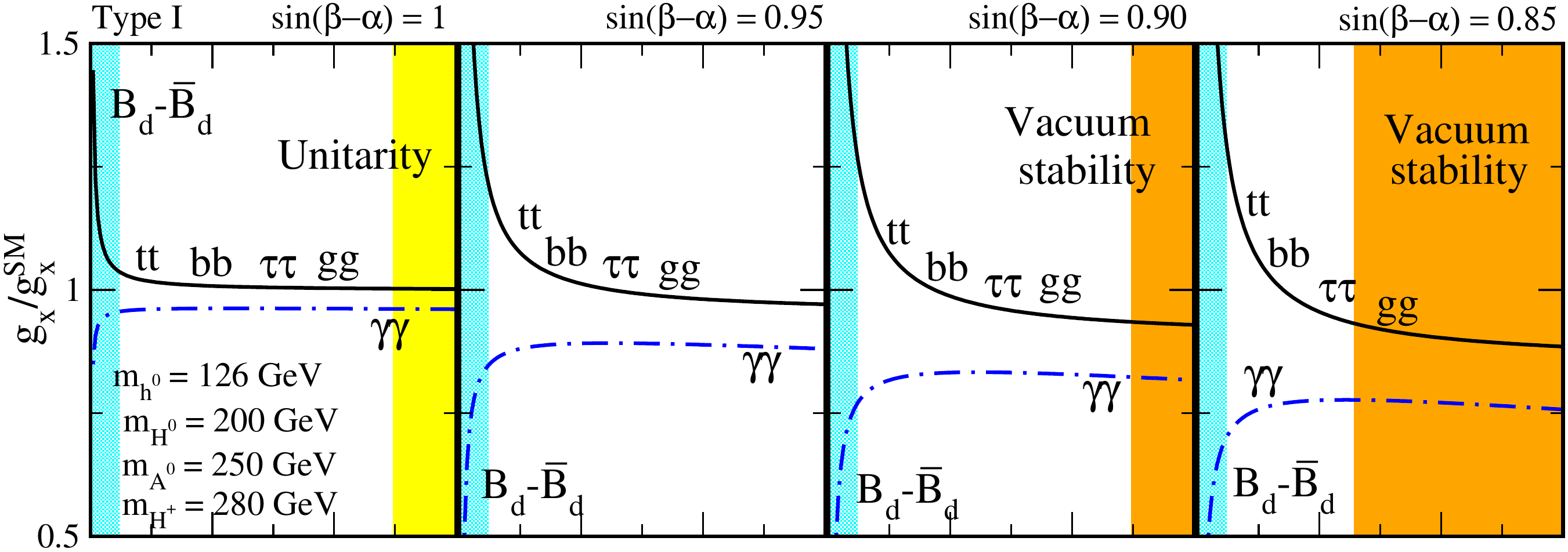} \\
\includegraphics[width=0.8\textwidth]{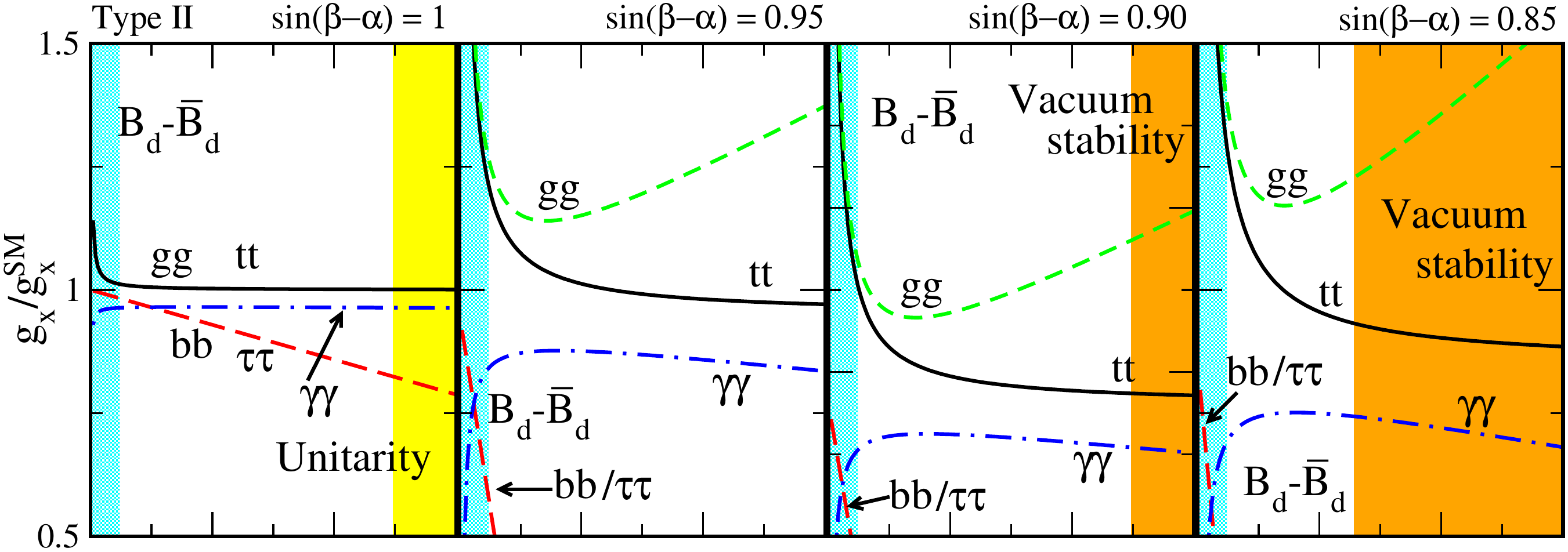} \\
\includegraphics[width=0.8\textwidth]{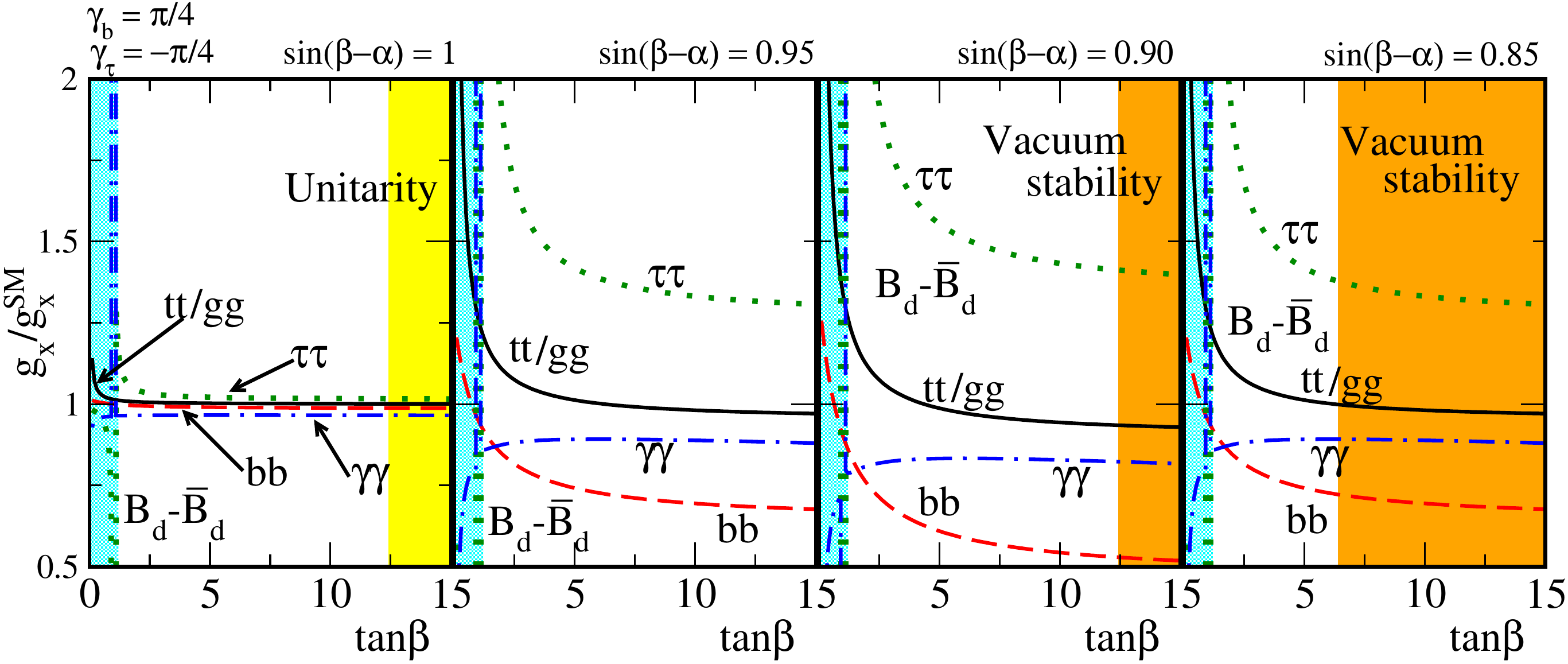} \\
\end{center}
\caption{Dependence of the Higgs couplings $g_x/g_x^\text{SM} =
  1+\Delta_x$ on $\tan\beta$. The couplings to weak gauge bosons are
  fixed by the values of $\sin(\beta-\alpha)$. For type-I (top),
  type-II (center) and one specific aligned configuration with
  $\gamma_{b,\tau}= \pm \pi/4$ (bottom) we illustrate four choices of
  mixing angles $\sin(\beta-\alpha)$. The shaded regions are
  excluded by the leading constraints.}
\label{fig:treelevel}
\end{figure}

Finally, if one of the extra Higgs multiplets does not develop a VEV
the Higgs boson can decay into a pair of inert light scalars. This
will be observed as an invisible Higgs width $\Gamma_\text{inv}$.
Again, this decay will depend on the self-coupling structure of the
potential.  The patterns of such modifications are illustrated in
Table~\ref{tab:structure-loopinduced}.\bigskip

We can then easily summarize the general Higgs coupling structure at
tree level: in the absence of a Higgs triplet any modification of
$g_{W,Z}$ will be negative and fully correlated. Changes in the
Yukawas $g_{b,t,\tau}$ can follow a wide range of model--specific
patterns.  The effective coupling $g_g$ will only be affected through
the modified Yukawas, while $g_\gamma$ as well as invisible Higgs
decays can be generated through self-couplings in the Higgs
potential. Adding a singlet will always give a further universal
reduction of all Higgs couplings, as shown in Eq.\eqref{eq:param7}, so
we only resort to such a singlet to describe invisible Higgs
decays.\bigskip

Each panel in Figure~\ref{fig:treelevel} shows different Higgs
interactions as a function of $\tan\beta$ for a fixed value of
$\sin(\beta-\alpha) = 0.85...1$, \ie implying a reduction of $g_V$ by
up to 15\%.  to soften the impact of the large-$\tan\beta$ constraints
by unitarity and vacuum stability we allow for a moderate
$Z_2$-breaking term $m_{12} = 50...60$~GeV in
Eq.\eqref{eq:2hdmpotential}. If we ignore these constraints we find
that towards small values of $\tan \beta$ the top Yukawa is no longer
perturbative.  Note that we do not attempt to fully account for all
constraints because extended Higgs sectors as consistent descriptions
of single-Higgs models with variable Higgs couplings are not meant to
be realistic models describing all available data in high energy
physics.

By definition, in a type-I 2HDM all Yukawa couplings vary
simultaneously, as shown in the upper panels of
Figure~\ref{fig:treelevel}.  As long as we remain close to the
Standard Model limit $\alpha - \beta = - \pi/2+ \delta$ the largest
deviations are found at small $\tan\beta$. Following
Eq.\eqref{eq:param7} the corrections then scale like
$\cos\alpha/\sin\beta \simeq 1 + \sin\delta/\tan\beta$.  This means
that for all 2HDM setups the couplings barely deviate from the
Standard Model for $\sin(\beta-\alpha) \simeq 1$.  Once we allow for a
stronger suppression of $g_V$ we find sizeable modifications in the
Yukawa couplings. Shifts in $g_g$ and $g_\gamma$ are inherited from
the tree-level couplings, with an additional contribution to
$g_\gamma$ from new charged Higgs states in the loop.

In a type-II model up-type and down-type Yukawas can vary separately,
scaling like
\begin{alignat}{5}
\frac{g_{b,\tau}}{g_{b,\tau}^\text{SM}} &= 1+\Delta_{b,\tau} 
&=& - \frac{\sin\alpha}{\cos\beta}
&=& \cos\delta-\sin\delta \tan\beta 
= 1 - \delta \tan\beta - \cfrac{\delta^2}{2} + \mathcal{O}(\delta^3)
\notag \\
\frac{g_t}{g_t^\text{SM}} &= 1+\Delta_t 
&=& + \frac{\cos \alpha}{\sin \beta} 
&=& 1+ \cfrac{\sin\delta}{\tan\beta}\, = 1+ \cfrac{\delta}{\tan\beta} - \frac{\delta^2}{2}   + \mathcal{O}(\delta^3) \; . 
\label{eq:yukawa-delta}
\end{alignat}
If $\sin\delta \simeq \delta > 0$ the down-type couplings decrease
with large $\tan\beta$ while the top Yukawa shows an increase at small
$\tan \beta$ values, and vice versa. The lepton--specific and flipped
setups combine a type-I pattern for the quark Yukawas with a type-II
pattern for the lepton Yukawas, and vice versa.

Once we depart from natural flavor conservation more possibilities
arise. Aligned models allow for an independent variation of the three
Yukawa interactions and accommodate enhancements, suppressions and
even sign flips~\cite{Celis:2013rcs}. The lower panels of
Figure~\ref{fig:treelevel} illustrate this for $\gamma_{b,\tau} = \pm
\pi/4$, as defined in Eq.\eqref{eq:alignment} and in
Appendix~\ref{app:para}.  We identify two distinct patterns for
$g_{b}$ and $g_\tau$.  The bottom Yukawa is suppressed with respect to
the Standard Model, with growing deviations for larger
$\tan\beta$. Instead, the tau lepton is increasingly enhanced as we
lower $\tan\beta$.  Coupling shifts of $\mathcal{O}(\pm 30\%)$ are
possible within the allowed $\tan\beta$ range.\bigskip

Building on these tree-level patterns we can construct a model which
gives us maximal flexibility in the light Higgs coupling
variations. We consider an aligned 2HDM with one additional inert
singlet. The relevant light Higgs couplings $g_x$ are determined by
six model parameters, with the only constraint $\Delta_W = \Delta_Z =
\Delta_V$ due to custodial symmetry,
\begin{alignat}{5}
\frac{g_V}{g_V^\text{SM}} &= 1 + \Delta_V(\alpha,\tan\beta)  
\notag \\
\frac{g_t}{g_t^\text{SM}} &= 1 + \Delta_t(\alpha,\tan\beta) \qqquad 
\frac{g_{b,\tau}}{g_{b,\tau}^\text{SM}} = 1 + \Delta_{b,\tau}(\alpha,\tan\beta,\gamma_{b,\tau}) \; 
\notag \\
\frac{g_\gamma}{g_\gamma^\text{SM}} &= 1 + 
\Delta_\gamma^\text{SM}(\alpha,\tan\beta,\gamma_{b,\tau}) +
\Delta_\gamma(\alpha,\tan\beta,m^2_{12},m^2_{\PHiggs^{\pm}}) 
\qqquad 
\frac{g_g}{g_g^\text{SM}} &= 1 + 
\Delta_g^\text{SM}(\Delta_t,\Delta_b) \; .
\label{eq:coup-dependence}
\end{alignat}
The loop-induced Higgs coupling to photons receives a contribution
from a charged Higgs loop, which means it depends on the trilinear
self-interactions. This self--interaction we can trade for the Higgs
boson masses and the PQ-breaking scale $m_{12}$. Additional
non-minimal doublet mixing can be introduced through non-vanishing
$Z_2$-breaking quartic couplings $\lambda_{6,7}$.

In the case of an inert or purely dark singlet, none of the above
couplings carries information on the singlet-doublet portal
interaction $\lambda_3 (\Phi^\dagger\Phi)\,S^2$ of
Eq.\eqref{eq:singlet-potential}.  It can only be determined from
invisible Higgs decays. With the exception of $\Delta_W = \Delta_Z <
0$ and the parametric dependence of $\Delta_g$ the setup in
Eq.\eqref{eq:coup-dependence} indeed describes a set of completely
independent Higgs couplings~\cite{sfitter_higgs}.

\subsection{Quantum effects: Higgs and gauge sector}
\label{sec:quantum_higgs}

Beyond the simple tree-level patterns described in the last section an
extended Higgs sector will affect all Higgs couplings at the quantum
level. These loop effects do not have as simple patterns and depend on
the heavy Standard Model states as well as on the entire Higgs
spectrum. They significantly increase the number of degrees of freedom
which we can use to for example modify the couplings of the lightest
Higgs state to all Standard Model particles.  Typical weak corrections
are too small to account for experimentally relevant coupling shifts
in the $20\%$ range. However, non-decoupling effects can have the
desired strength. Such effects can be linked to the Higgs self
couplings or to very large values of $\tan \beta$ or $\cot \beta$. In
this section we discuss the first kind, the latter will follow in
Section~\ref{sec:quantum_fermion}.\bigskip

\begin{figure}[b!]
\begin{center}
\includegraphics[width=0.20\textwidth]{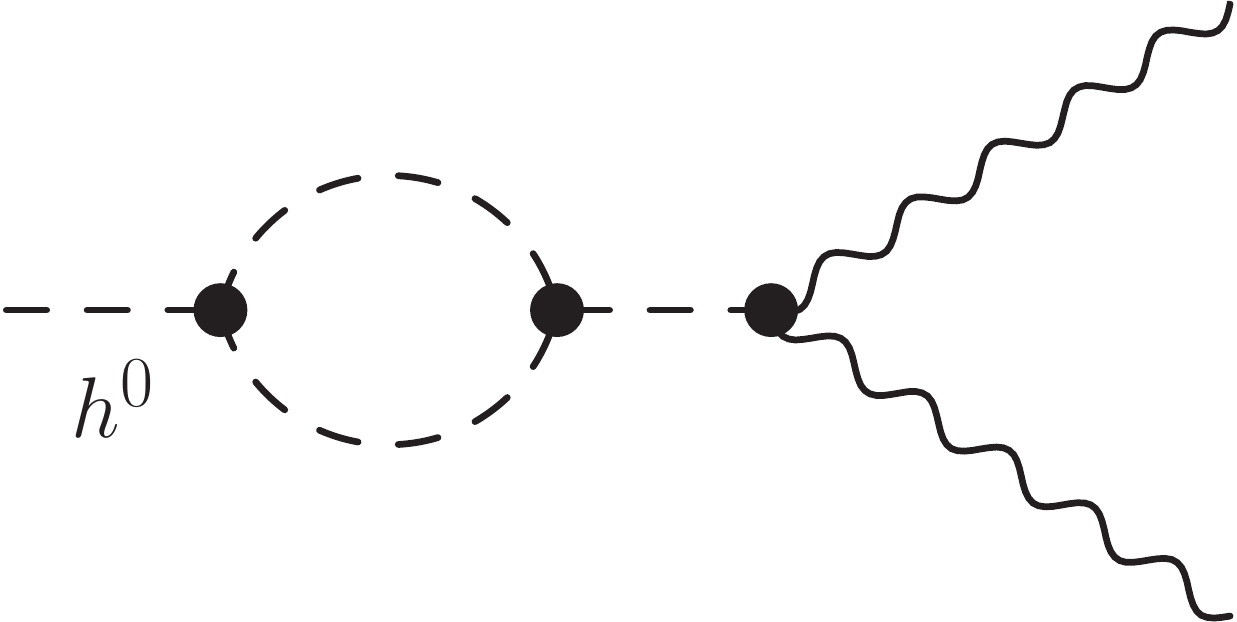} 
\hspace*{0.05\textwidth}
\includegraphics[width=0.20\textwidth]{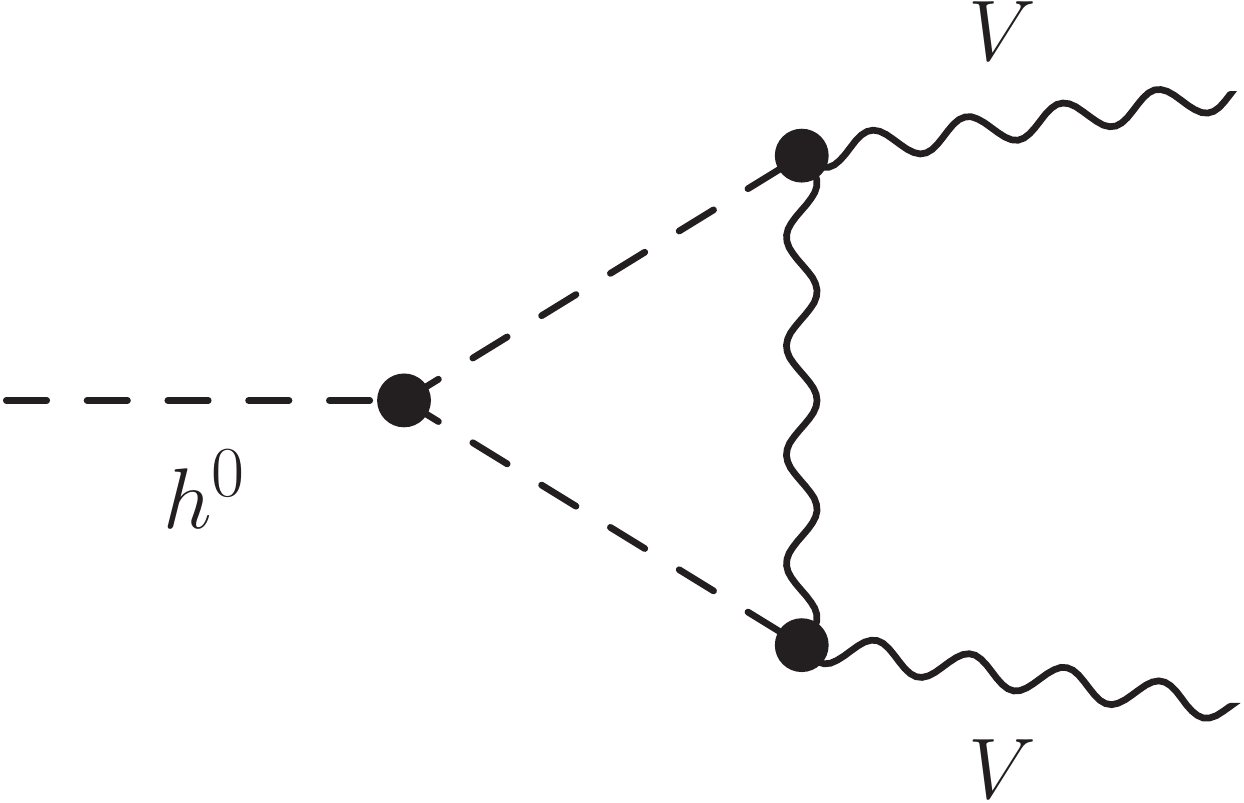} 
\hspace*{0.05\textwidth}
\includegraphics[width=0.20\textwidth]{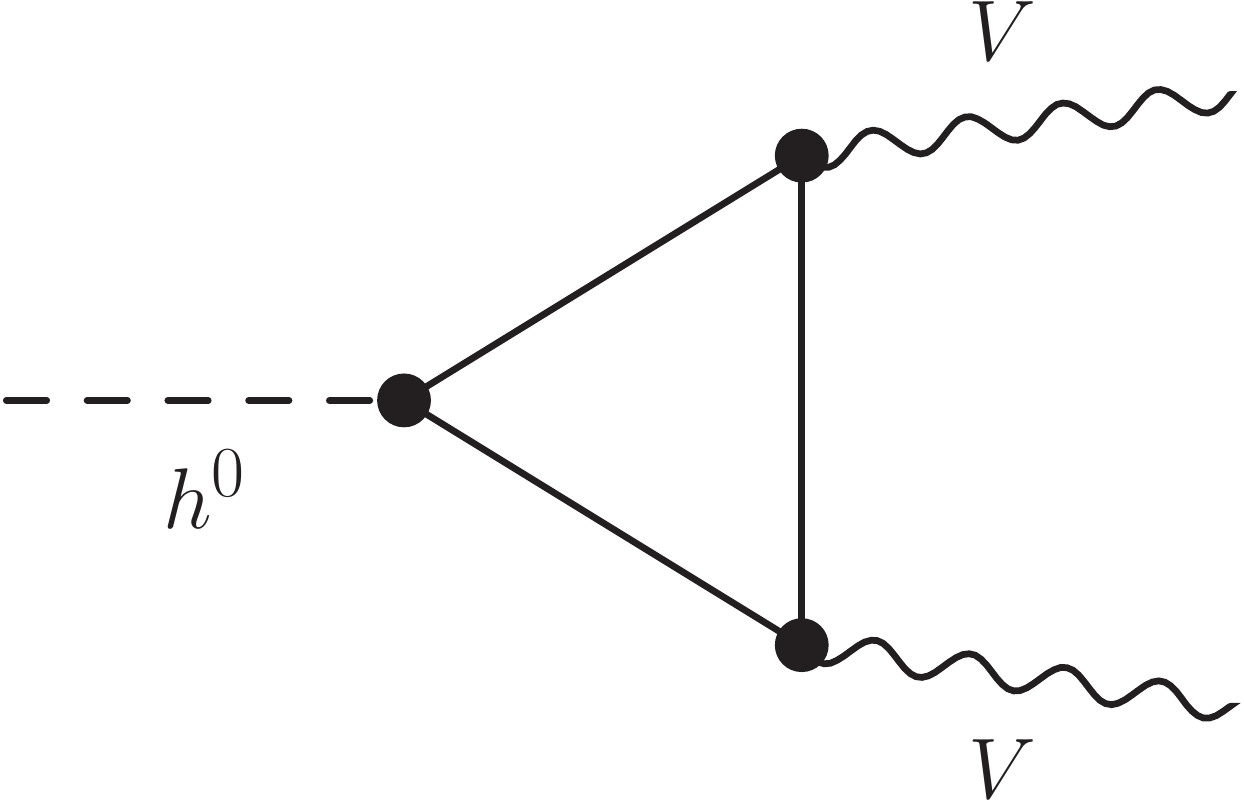} 
\end{center}
\caption{Sample Feynman diagrams accounting for the electroweak
  corrections to the Higgs couplings. In terms of self-couplings they
  scale with the square (left), linearly (center) and flat (right).}
\label{fig:loopdiagrams}
\end{figure}

One source of decoupling effects is the mass of a new state unrelated
to the Higgs mechanism. For example in the singlet extension of
Eq.\eqref{eq:singlet-potential} the heavier neutral CP-even state
obtains its mass $ m^2_{\Hzero} \simeq \mu_2^2 + \lambda_3\,v_1^2/2$
from the singlet dimension-2 operator $\mu_2^2$ as well as from the
Higgs VEV via the quartic interaction $\lambda_3$.  The first term is
unrelated to electroweak symmetry breaking and decouples. The second
term cannot be detached from the light Higgs dynamics and leads to
non-decoupling.  This is in agreement with the Appelquist-Carrazzone
theorem~\cite{Appelquist:1974tg} which is based on three conditions:
renormalizability, no Yukawa couplings, and no spontaneous symmetry
breaking. If $\lambda^2_3 v^2_1 \gg \mu^2_2$, heavy Higgs effects can
be large and for example useful in stabilizing the vacuum through a
large shift in the Higgs quartic coupling from threshold
corrections~\cite{EliasMiro:2012ay}.

The same effects occur in the 2HDM. The decoupling of the heavy states
can be described in terms of the small parameter $\cos(\beta -
\alpha)$, as shown in Appendix~\ref{app:hierarchy}. Just like in the
singlet case we can relate the heavy masses $M^2_\text{heavy} =
m^2_{\Hzero,\Azero,\PHiggs^\pm}$ to the mass terms and self-couplings
in the potential of Eq.\eqref{eq:2hdmpotential}. This gives us
\begin{alignat}{5}
m^2_{\Hzero,\Azero,\PHiggs^\pm} = 
\frac{m_{12}^2}{\sb\cb} 
+ \mathcal{O}(\lambda_i\,v^2) 
+ \mathcal{O} \left( \frac{\lambda_i v^4}{M_\text{heavy}^2} \right) \; .
\end{alignat}
A non-decoupling behavior manifests itself as power-like contributions
from the heavy masses and may shift the light Higgs couplings to weak
gauge bosons and heavy quarks by up to $10\%$~\cite{nondecoupling}.
For the trilinear Higgs self-interaction such effects can reach
$\mathcal{O}(100\%)$.\bigskip

Independent of the origin of the heavy Higgs masses the
self-interactions in the 2HDM potential Eq.\eqref{eq:2hdmpotential}
can induce sizeable quantum effects. We show the corresponding Feynman
diagrams in Figure~\ref{fig:loopdiagrams}.  This is because
$\mathcal{O}(\lambda_i v^2)$ contributions are not limited by any
underlying symmetry and only a subset of them is related to physical
masses and constrained indirectly.  While large self-interactions are
often identified with heavy masses this does not hold in the general
2HDM~\cite{LopezVal:2009qy,self}.  Usually, we select $m_{12}$ and
$\lambda_{6,7}$ as model parameters which are only related to
independent self-couplings.  From Section~\ref{sec:uv} we know that
perturbativity, vacuum stability and unitarity slightly tame these
effects in realistic scenarios.\bigskip

\begin{figure}[t]
\begin{center}
\includegraphics[width=0.8\textwidth]{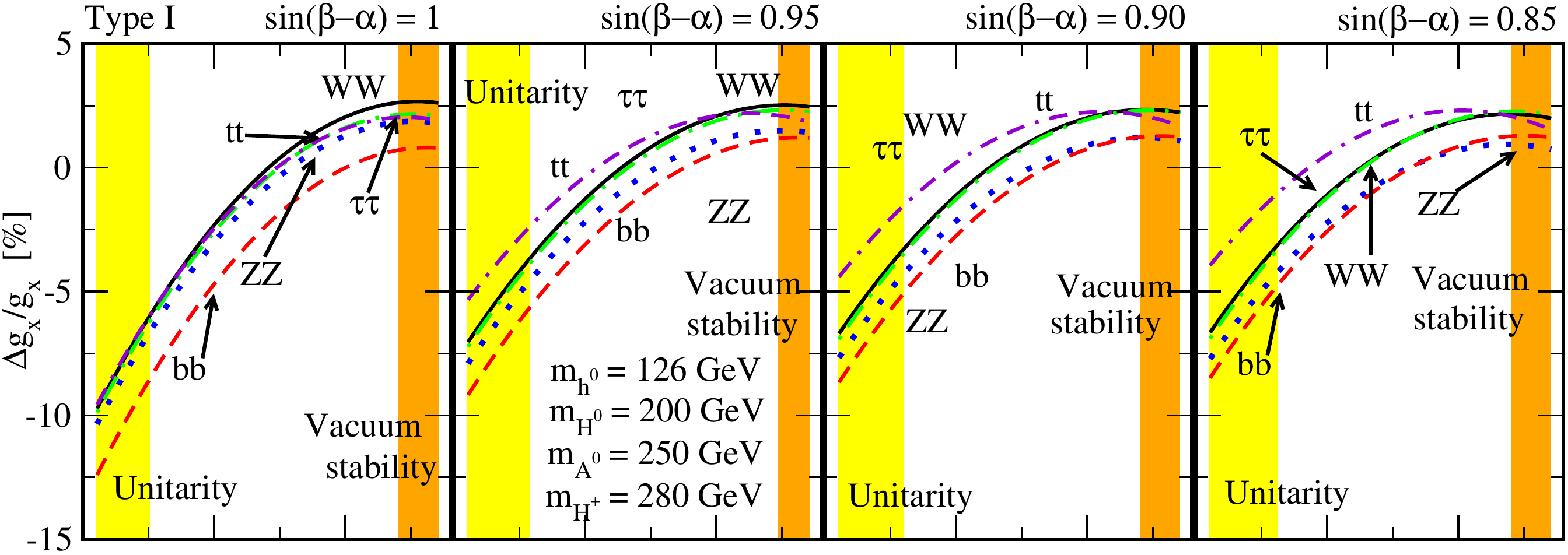} \\ 
\includegraphics[width=0.8\textwidth]{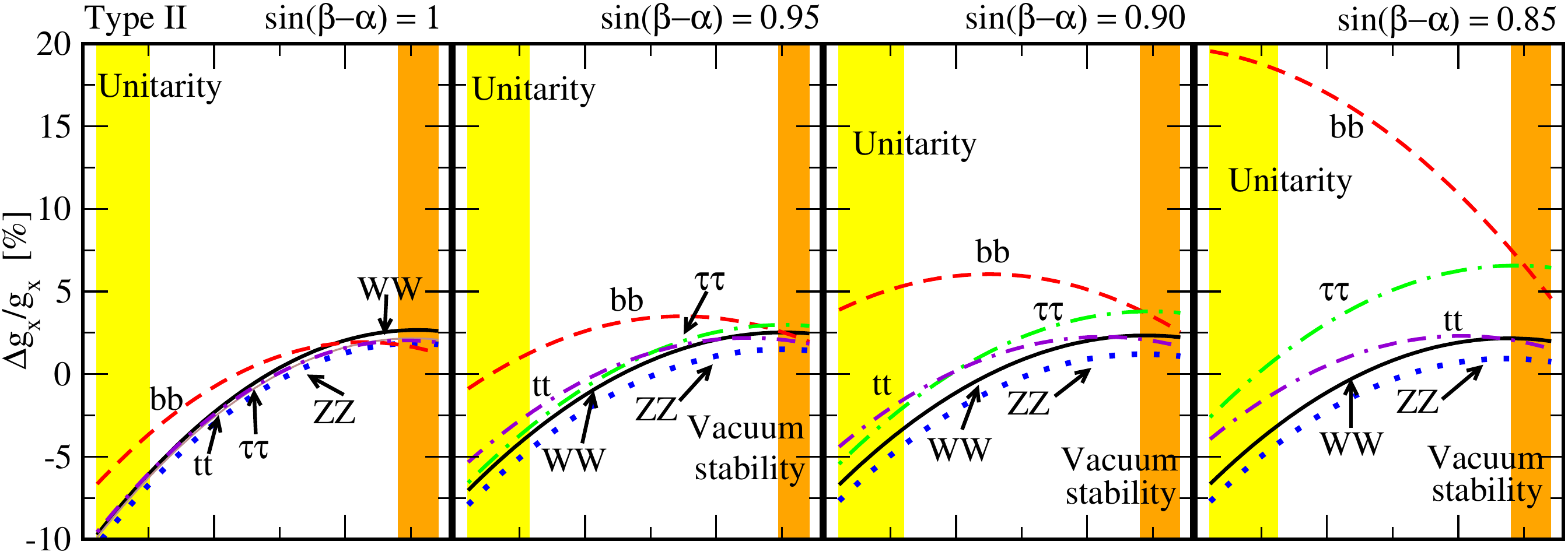} \\ 
\includegraphics[width=0.8\textwidth]{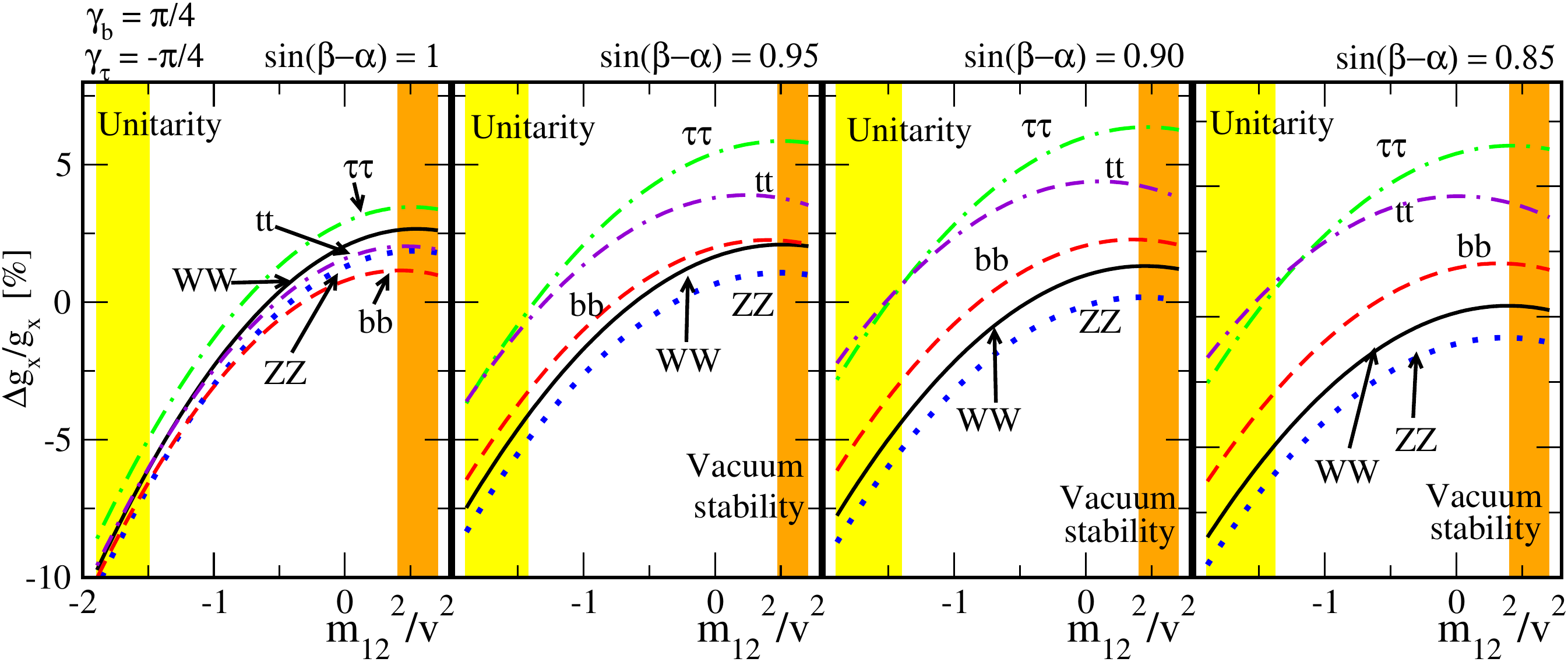} \\ 
\end{center}
\caption{Loop corrections to the light Higgs couplings $\Delta
  g_x/g_x$ (in \%) for type-I (top), type-II(center), and aligned
  configurations (bottom) as a function of $m^2_{12}$. Just like in
  Figure~\ref{fig:treelevel} the latter model is defined with
  $\gamma_{b,\tau}= \pm \pi/4$.  The shaded regions are excluded by
  unitarity and vacuum stability as discussed in
  Section~\ref{sec:uv}.}
\label{fig:quantum_higgs}
\end{figure}

In Figure~\ref{fig:quantum_higgs} we show the electroweak corrections
to the light Higgs couplings to heavy quarks and to massive gauge
bosons as a function of the PQ soft-breaking scale $m^2_{12}$.  We
consider type-I, type-II, and aligned setups, just like for the
tree-level analysis in Figure~\ref{fig:treelevel}.  The difference is
that we now fix $\tan\beta = 1.4$ and vary $m^2_{12}$.  The values for
$m_{12}^2$ are constrained by unitarity and vacuum stability.  The
calculation is based on \textsc{FeynArts}, \textsc{FormCalc} and
\textsc{LoopTools}~\cite{feynarts} with an appropriate
renormalization~\cite{LopezVal:2009qy}.  Technically, we relate the
relative coupling shift to the corresponding loop-corrected decay
rate, $\Delta\,g_x/g_x = 1/2\,\Delta\Gamma(H\to xx)/\Gamma(H \to
xx)$.\bigskip

\begin{figure}[t]
\includegraphics[width=0.31\textwidth]{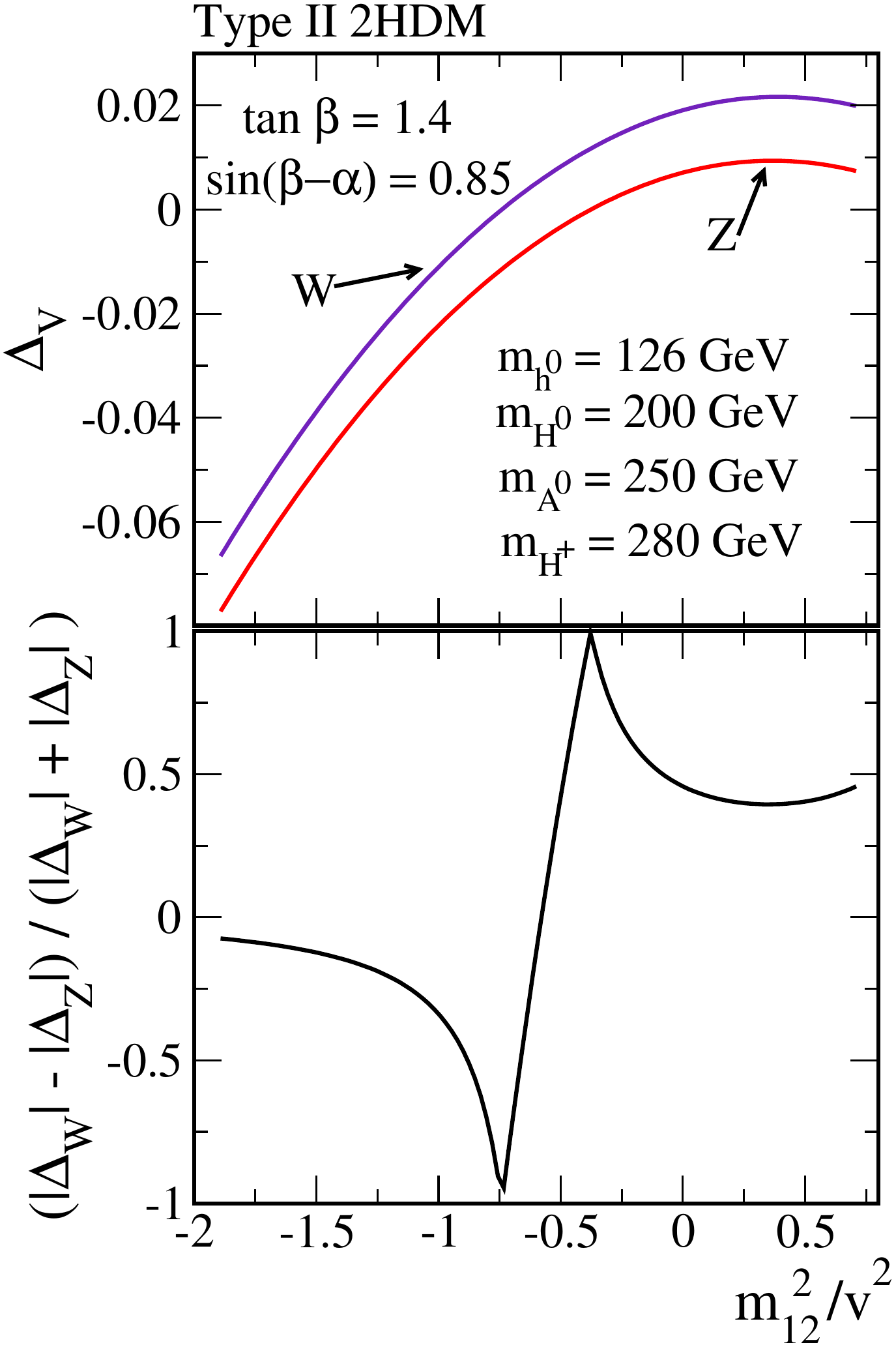} 
\hspace*{0.02\textwidth}
\includegraphics[width=0.31\textwidth]{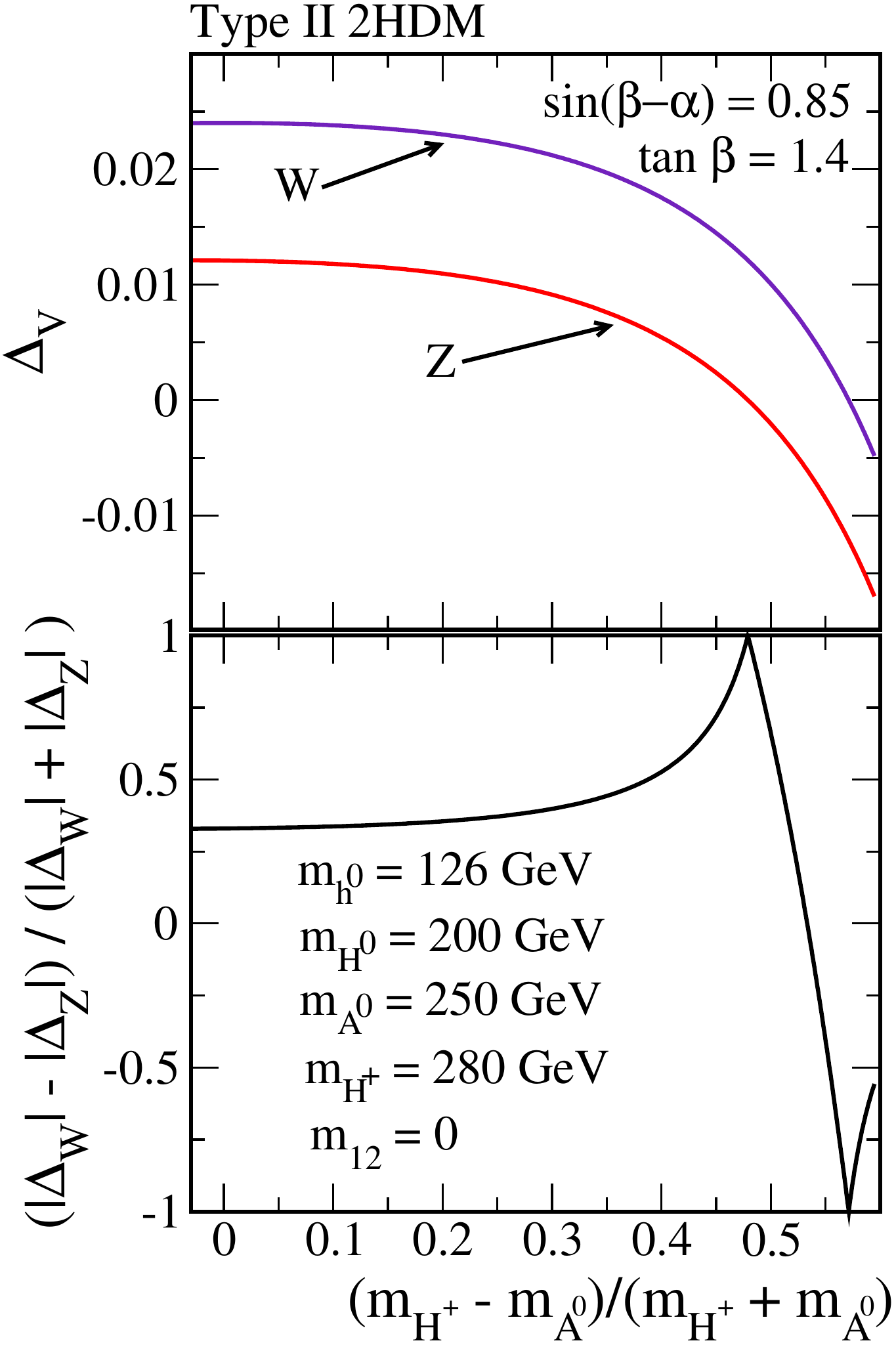} 
\hspace*{0.02\textwidth}
\includegraphics[width=0.31\textwidth]{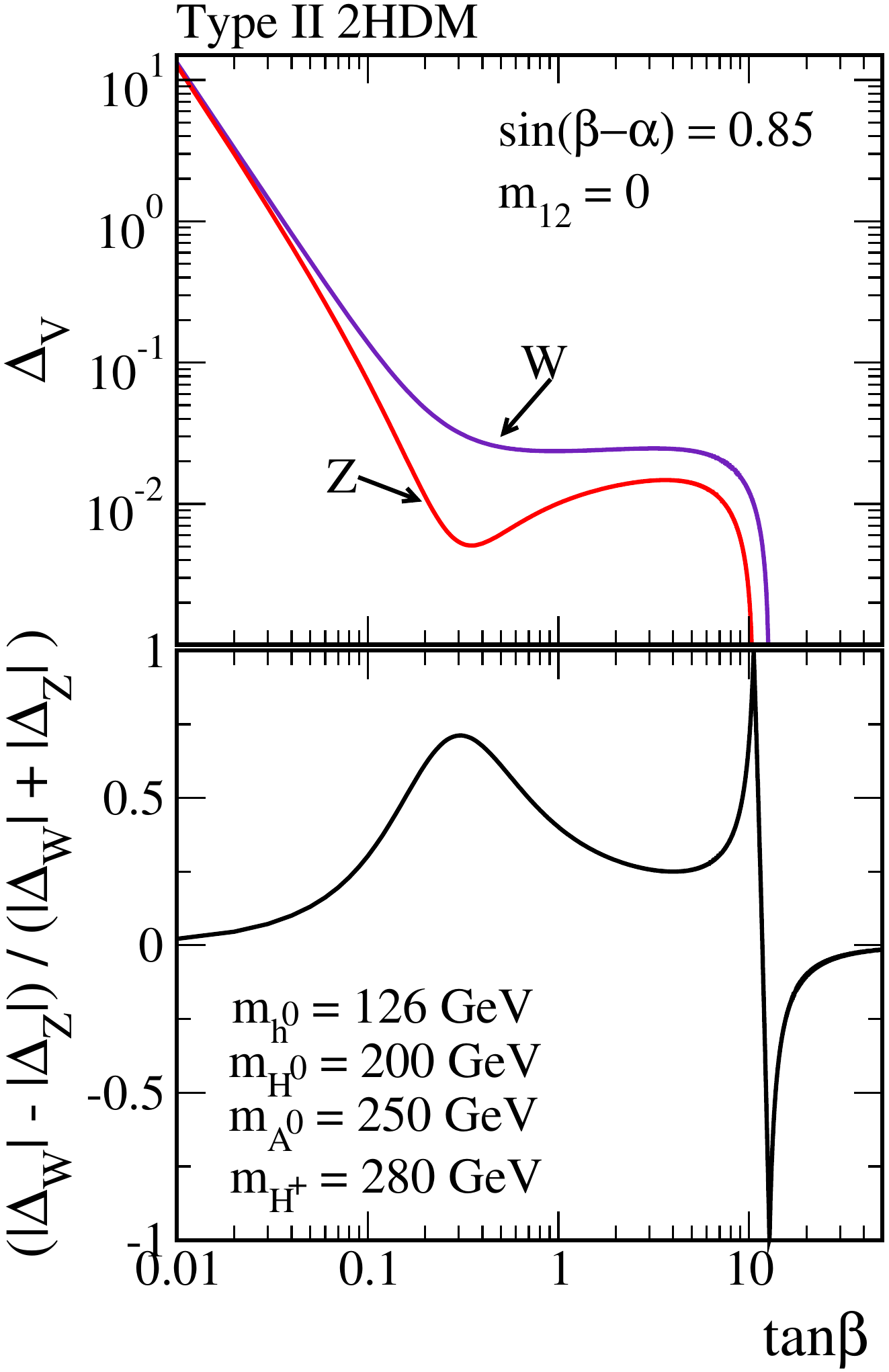} 
\caption{Loop corrections to $\Delta_{W,Z}$ and to the asymmetry
  $A_{W/Z}$ as a function of $m_{12}^2/v^2$ (left), $x_m$ (right), and
  $\tan \beta$ (right). We only show typical results for a type-II
  2HDM with light additional Higgs states. All curves include the full
  set of Higgs/gauge and fermion loops.}
\label{fig:wz_asymmetry}
\end{figure}

Some basic features are common to all couplings, with the exception of
the more model dependent case of the bottom quark.  Quantum effects
give rise to a systematic $\mathcal{O}(10\%)$ depletion, mostly
stemming from the Higgs-mediated finite wave-function corrections to
the light Higgs boson. The leading contributions are governed by two
trilinear Higgs self-interactions, shown in
Figure~\ref{fig:loopdiagrams}. In combination they scale as
$|m^2_{12}|^2$. The key question is if such corrections can either
separate $\Delta_W$ from $\Delta_Z$ or lead to positive corrections
$\Delta_V > 0$. From the leading Feynman diagram we know that this is
unlikely to happen: large corrections induced by Higgs self-couplings
will affect the wave function renormalization of the light Higgs and
the renormalization of mixing angles in the Higgs sector. Both are
negative and universal. Because the leading corrections to the Higgs
wave function renormalization are proportional to the self coupling
squared, signs or phases in the Higgs potential do not affect this
conclusion.

It turns our that the situation is even worse: from
Eq.\eqref{eq:tparam} we know that custodial symmetry can be broken by
non-degenerate masses in the Higgs sector. This effect we can trace
through the asymmetry
\begin{alignat}{5}
x_m = \frac{\mhp - m_{\Azero}}{\mhp + m_{\Azero}}
\qqquad \text{and} \qqquad
A_{W/Z}(x_m) = \frac{|\Delta_W| - |\Delta_Z|}{|\Delta_W| + |\Delta_Z|} \; .
\label{eq:asymmetry}
\end{alignat}
Note that $A_{W/Z}$ is defined on the loop-induced shift in the $W$
and $Z$ couplings, not on their coupling values.  Effects of this kind
arise for example from the central Higgs vertex diagram in
Figure~\ref{fig:loopdiagrams}.  However, these mass--induced
corrections are tied to the gauge couplings and unrelated to
potentially enhanced scalar self-interactions. In practice, the
$A_{W/Z}$ variation due to universal non-decoupling effects is
reduced.\bigskip

In the left panel of Figure~\ref{fig:wz_asymmetry} we show the
behavior of the loop corrections to the Higgs couplings to massive
gauge bosons as a function of $m^2_{12}$. As expected, the weak loop
corrections can be large. For negative values $m_{12}^2 \simeq -2 v^2$
they shift $\Delta_W \simeq \Delta_Z$ down by 6-7\%. On the other hand,
positive contributions to the $W$ and $Z$ couplings are at the
per-cent level and cannot compensate the tree-level reduction by 15\%
for our assumed parameter point. Similarly, the split $A_{W/Z}$ is
small wherever the absolute size of the loop corrections is large,
reflecting again the universal character of the leading loop-induced
deviations.  Around $m_{12}^2 = -v^2/2$ the loop corrections both
cross zero. The large value of $A_{W/Z}$ at large positive values of
$m_{12}^2$ is due to vanishing loop-induced $\Delta_Z$.

In the central panel of Figure~\ref{fig:wz_asymmetry} we show the size
of the coupling shifts as a function of $x_m$ defined in
Eq.\eqref{eq:asymmetry}. Wherever the second diagram in
Figure~\ref{fig:loopdiagrams} dominates the asymmetry roughly scales
like
\begin{alignat}{5}
A_{W/Z} \sim 
\dfrac{\left| \dfrac{m^2_{\PHiggs^{\pm}}}{m_{\hzero}^2-m^2_{\PHiggs^{\pm}}}
              \log \dfrac{m^2_{\hzero}}{m^2_{\PHiggs^{\pm}}} \right|
     - \left| \dfrac{m^2_{\Azero}}{m_{\hzero}^2-m^2_{\Azero}}
              \log \dfrac{m^2_{\hzero}}{m^2_{\Azero}} \right|}
      {\left| \dfrac{m^2_{\PHiggs^{\pm}}}{m_{\hzero}^2-m^2_{\PHiggs^{\pm}}}
              \log \dfrac{m^2_{\hzero}}{m^2_{\PHiggs^{\pm}}} \right|
     + \left| \dfrac{m^2_{\Azero}}{m_{\hzero}^2-m^2_{\Azero}}
              \log \dfrac{m^2_{\hzero}}{m^2_{\Azero}} \right|} \; .
\label{eq:deltaWZ-scaling}
\end{alignat}
Comparing to the $T$ contributions shown in Eq.\eqref{eq:tparam} the
asymmetry $A_{W/Z}$ will indeed behave similarly and therefore be
strongly constrained by electroweak precision data. Numerically, the
corrections to $\Delta_{W,Z}$ are indeed small and degenerate.

A richer range of patterns only appears for the bottom Yukawa.  There,
quantum effects are strongly model dependent and may eventually yield
large (and positive) $20\%$ corrections, including sizable $
m_t/\tan\beta$ pieces from charged Higgs corrections, which can be
further boosted by the trilinear $\hzero\PHiggs^+\PHiggs^-$
coupling. 

However, to overcome the tree-level limitations of
Eq.\eqref{eq:coup-dependence} loop effects from the gauge and Higgs
sector are generally of limited use. They hardly predict structurally
new positive or non-degenerate $\Delta_{W,Z}$ contributions. The question
is if fermion-induced loops lead to the desired effects.

\subsection{Quantum effects: fermions}
\label{sec:quantum_fermion}

At tree level, Figure~\ref{fig:treelevel} shows that the variation of
the modified light Higgs couplings in the aligned model the range of
patterns is rich and non-universal.  However, the couplings to weak
gauge bosons show no dependence on $\tan \beta$ for fixed
$(\alpha-\beta)$. Obviously, this simple picture will not hold once we
include fermion loop effects.

\begin{figure}[t]
\begin{center}
\includegraphics[width=0.8\textwidth]{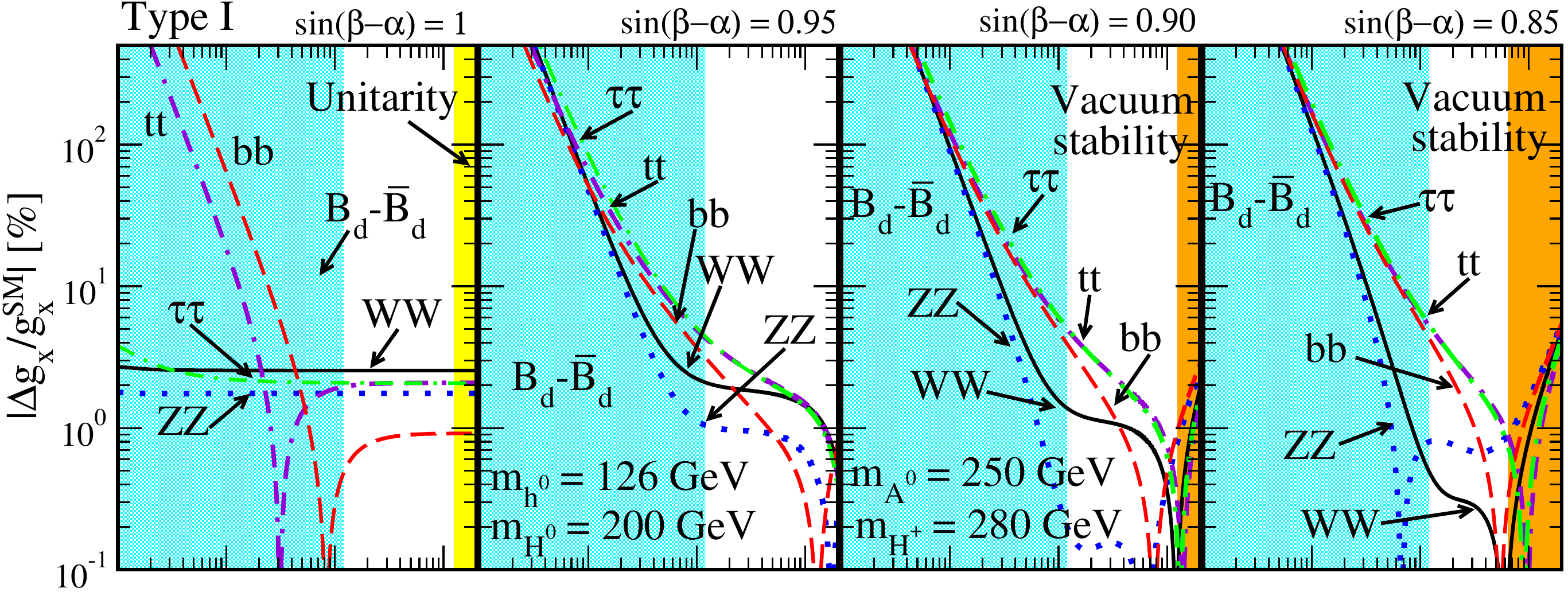} \\
\includegraphics[width=0.8\textwidth]{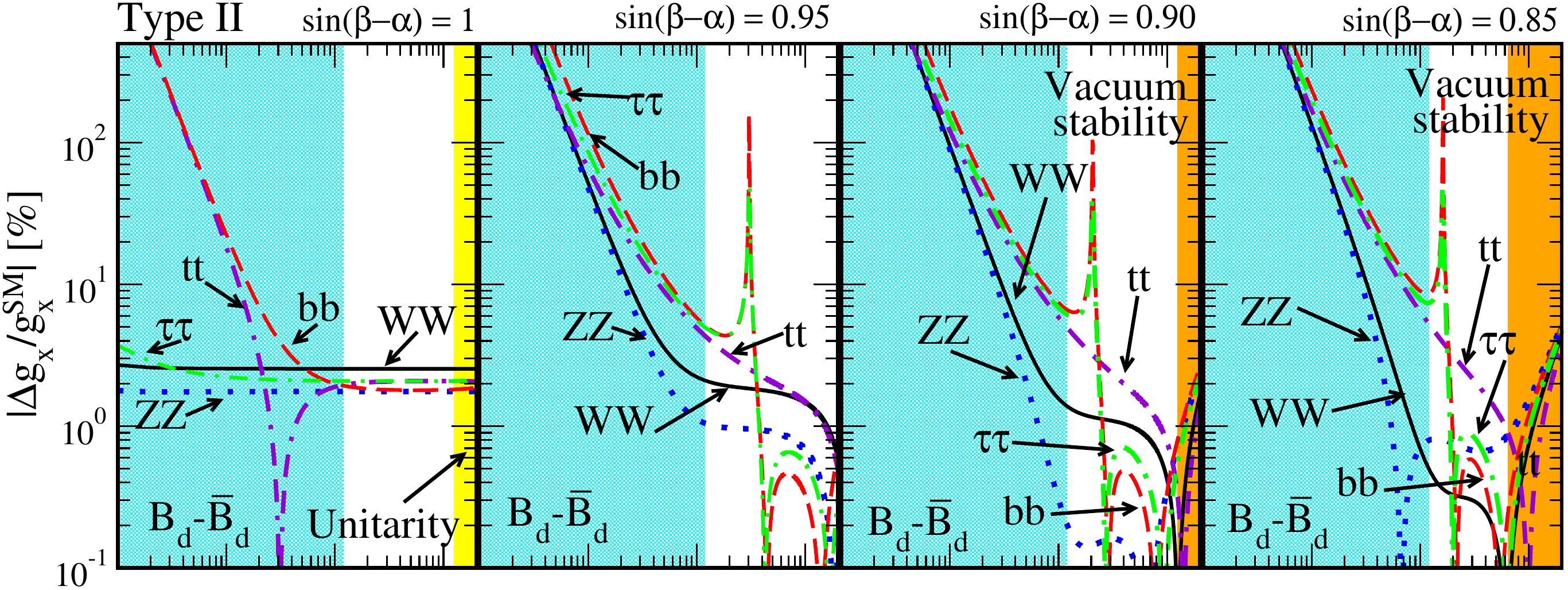} \\
\includegraphics[width=0.8\textwidth]{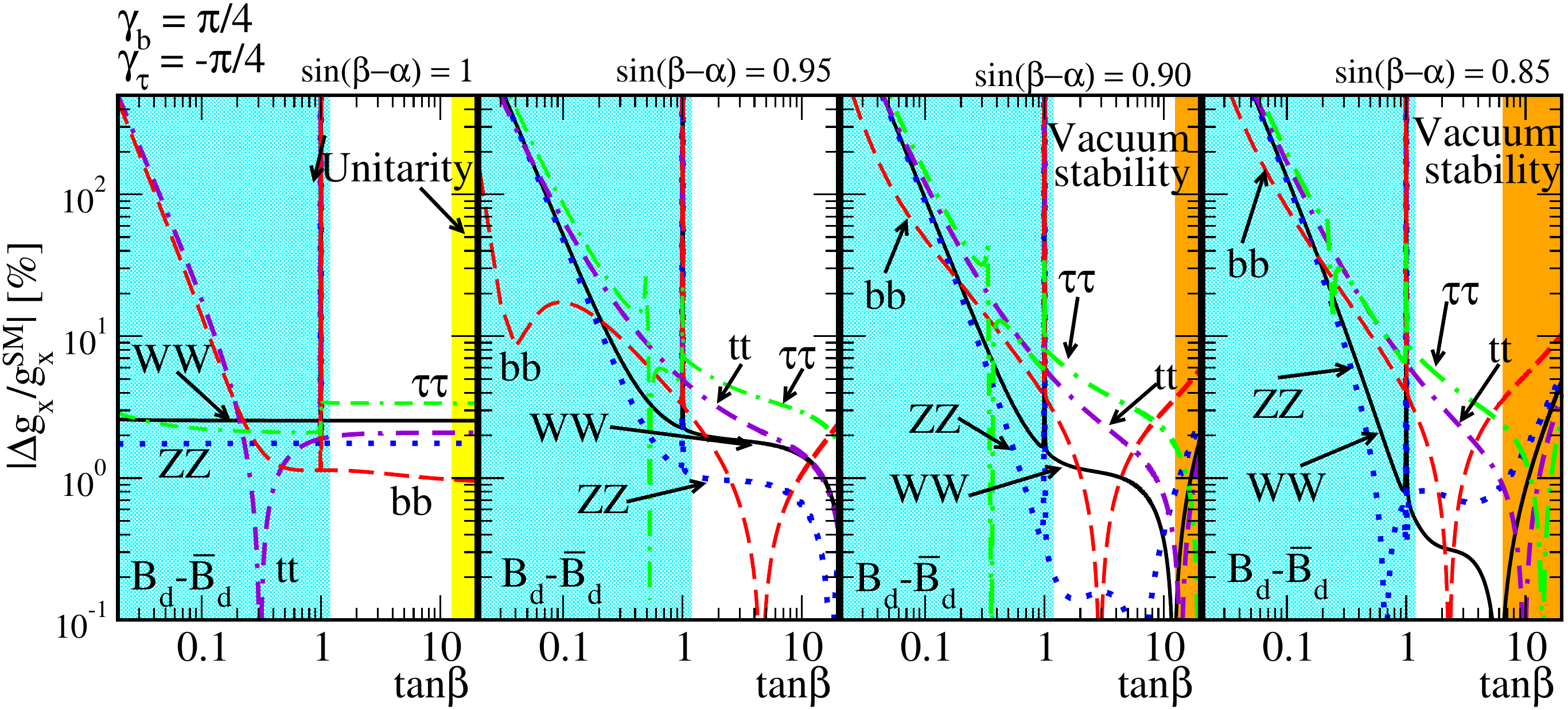} \\
\end{center}
\caption{Loop corrections to the light Higgs couplings for type-I
  (top), type-II(center), and aligned configurations (bottom) as a
  function of $\tan\beta$.  Just like in Figure~\ref{fig:treelevel}
  the latter is defined with $\gamma_{b,\tau}= \pm \pi/4$.  The shaded
  regions are excluded by unitarity and vacuum stability as discussed
  in Section~\ref{sec:uv}. The dashed area at low $\tan\beta$ is ruled
  out by the flavor bounds from $B_d^0-\bar{B}_d^0$ mixing.}
\label{fig:quantum_fermions}
\end{figure}

The trademark feature for top-mediated effects in all conventional
2HDMs with natural flavor conservation is a modified Yukawa coupling
scaling like $m_t/\tan\beta$ for $\tan\beta \ll 1$. For type-II models
a complementary $\tan\beta$ enhancement of down-type Yukawas is given
in Eq.\eqref{eq:yukawa-delta}.  In Figure~\ref{fig:treelevel} we see
that small $\tan \beta$ enhances the top Yukawa
enough to overshoot perturbativity constraints. Experimentally, this
low $\tan\beta$ region is in strong tension with the
$B^0_d$--$\bar{B}^0_d$ mixing rate data, a constraint we will ignore
for the structural argument in this section. The question is then how
this richer Higgs coupling pattern to fermions may be imprinted on the
loop-induced effects.\bigskip

For small $\tan \beta$ the right panel of
Figure~\ref{fig:wz_asymmetry} indicates for the first time sizeable
and non-universal quantum contributions to $\Delta_W$ and $\Delta_Z$.
They arise because towards small $\tan \beta$ the fermionic and the
Higgs-mediated contributions to the wave function renormalization
shown in Figure~\ref{fig:loopdiagrams} become similar in size and
carry opposite signs.  As long as this partial compensation holds, the
one-loop contributions are dominated by the $1/\tan\beta$-enhanced
top-mediated vertex corrections. For $\tan \beta \simeq 0.1$ the
positive contributions to $\Delta_{W,Z}$ exceed 10\% and become of
similar size as the tree level contributions.  Because the top and
bottom couplings are different for $W$ and $Z$ bosons this induces a
violation of custodial symmetry $\Delta_Z \neq \Delta_W$ with an
asymmetry $A_{WZ} \simeq 0.5$. This scan over $\tan\beta$ we extend to
the full set of relevant couplings in
Figure~\ref{fig:quantum_fermions}, similar to the Higgs and
gauge--induced loop corrections in Figure~\ref{fig:quantum_higgs}.  We
examine type-I (top), type-II (center) and aligned (bottom) 2HDM
configurations. The different $(\alpha-\beta)$ choices cover tree
level $g_V/g_V^\text{SM}$ suppressions to $-15\%$.

Once we continue beyond $\tan\beta \simeq 0.1$ the top Yukawa grows much
faster than the scalar self-interactions. This gives rise to strongly
enhanced but universal shifts in $g_{W,Z}$.  Moreover, in this regime
the top Yukawa is extremely large already at the weak scale.

We see that parameter regions with positive non-universal
$\Delta_{W,Z}$ values are severely limited by many constraints on the
low-$\tan\beta$ regime. On the other hand, this structural pattern is
attainable within the 2HDM, as long as we assume that experimental
constraints should only be placed on the full and unknown ultraviolet
model.  The only structural constraint is a perturbative top Yukawa
coupling.\bigskip

An alternative model inducing the desired effects on $\Delta_{W,Z}$ is
based on new vector-like fermions coupling to the Higgs and eventually
to weak gauge bosons. Their contribution to the Higgs wave-function
and to the Higgs-gauge boson vertex corrections yield coupling shifts
$\Delta_W \neq \Delta_Z > 0$.  The relevant interactions we can model
as Higgs and vector portal-like operators~\cite{vector-portal}
\begin{alignat}{5}
\lag \supset 
  c_{HF} \,\frac{\psi^c\,\Phi\,\psi\,(\Phi^\dagger\Phi)}{\Lambda^2} 
+ c_{WF} \,\frac{\bar{\psi}\,\sigma^{\mu\nu}\,W_{\mu\nu}\psi}{\Lambda} 
+ c_{BF} \,\frac{\bar{\psi}\,\sigma^{\mu\nu}\,B_{\mu\nu}\psi}{\Lambda} \; ,
\label{eq:eff-fermions}
\end{alignat}
with the usual definition for the field-strength tensors and
$\sigma^{\mu\nu} \equiv i[\gamma^\mu,\gamma^\nu]/2$.  The scale
$\Lambda$ determines the mass range for a specific ultraviolet
completion. Realizations of these nature have been proposed in the
literature~\cite{newfermions}.  As long as the new fermions do not
carry color they do not affect the Higgs--gluon coupling and are
therefore not constrained by the experimental limits on a chiral 4th
generation~\cite{g4}.\bigskip

In passing we remark that in the supersymmetric 2HDM the situation is
very different. There, the self-couplings are gauge couplings and
scale like $g^2 v^2$. Consequently, heavy MSSM Higgs bosons
decouple. Non-decoupling effects may arise 
when we break the strict type-II structure of the Yukawa
couplings via enhanced quantum effects at large $\tan\beta$ ~\cite{Hall:1993gn,Carena:1999py}. 
In the non-supersymmetric
2HDM such $\tan\beta$-enhanced couplings appear as well, but are
numerically not as relevant.

\section{LHC data}
\label{sec:fits}

In this second part of the paper we compare some of the extended Higgs
sectors discussed in Section~\ref{sec:extensions} to the available
ATLAS and CMS measurements. For details on the corresponding Standard
Model analysis and the channels included we refer to
Ref.~\cite{sfitter_higgs}.  The coupling extraction is performed in
the fully correlated \textsc{SFitter}
framework~\cite{sfitter,sfitter_higgs,sfitter_ilc} which gives us
two--dimensional correlations in the log-likelihood distributions.
The difference to the best fit value defined as $-2 ( \log \mathcal{L}
- \log \mathcal{L}_\text{best}) < \{ 1,4,9 \}$ would in the Gaussian
approximation mark the $\{ 1,2,3 \} \, \sigma$ confidence level
regions. Note that in computing error bars on the model parameters
\textsc{SFitter} does not use a Gaussian approximation. Correlations
between the theory uncertainties are not included because the analyses
in the different channels are hardly comparable and because theory
errors are sub-leading at this point in time. First \textsc{SFitter}
studies show that the 2011-2012 coupling fit will not significantly
change when we fully correlate the theory uncertainties on the
production and decay sides.\bigskip

After very briefly reviewing and updating the fully general Higgs
coupling fit we will turn to a set of extended Higgs sectors defined
according to Eq.\eqref{eq:coup-dependence}. Quantum corrections to the
Higgs couplings are not taken into account, with the exception of a
charged Higgs loop in the effective Higgs--photon coupling. For all
2HDM scenarios considered this ansatz differs from the most general
Higgs fit only in that $\Delta_W = \Delta_Z \equiv \Delta_V$, \ie we
only consider extended Higgs sectors with negligible violation of
custodial symmetry. The quantum corrections analyzed in
Sections~\ref{sec:quantum_higgs} and \ref{sec:quantum_fermion} are not
included in the fit because the current LHC data is not sensitive to
such small effects. This has to be kept in mind when the naive
tree--level constraints $\Delta_W = \Delta_Z$ and $\Delta_W < 0$ are
reflected in the results. Moreover, when deriving Higgs couplings
from extended sectors we do not enforce the usual \textsc{SFitter}
condition $\Delta_W > -1$ and deal with ambiguities from over-all
re-rotations of the Higgs field individually.\bigskip

Because the full parameter space for extended models is vast and the
number of measurements is limited we adopt a number of assumptions:
the lightest Higgs mass is fixed at $m_{\hzero} = 126$~GeV, possibly
accompanied by a single heavy mass scale for additional Higgs states
$m_{\Hzero} \simeq m_{\Azero} \simeq m_{\PHiggs^{\pm}}\equiv
M_\text{heavy}$. In the Higgs potential of Eq.\eqref{eq:2hdmpotential}
we assume vanishing quartic couplings $\lambda_{6,7} = 0$ and limit PQ
soft-breaking contributions to the bilinear term $m_{12}$.  For
notational convenience we trade $m_{12}$ for the self-coupling it
induces after minimizing the potential,
\begin{alignat}{5}
 \tilde{\lambda} = \cfrac{2m^2_{12}}{\sin\beta\cos\beta\,v^2} \; .
\label{eq:ltilde}
\end{alignat}
We define a few representative benchmark points with the parameter
ranges given in Table~\ref{tab:parameters}.  Electroweak and flavor
physics constraints, unitarity, and vacuum stability will be discussed
separately.

\begin{table}[t]
 \begin{center} \begin{small}
  \begin{tabular}{l|ll}
\hline
& parameters \\ \hline\hline
dark singlet 
&$m_s = ( 0 \to 65)$~GeV  &$\lambda_3 =0 \to 1$ \\ \hline
hierarchical 2HDM
&$\tan\beta = 1 \to 50$      &$\xi = 0.0 \to 1.0$ \\ \hline
\multirow{2}{*}{general 2HDM}
&$\tan\beta = 0.01 \to 50$      &$\sin\alpha = -1.0 \to 1.0$ \\ 
&$M_\text{heavy} = (200 \to 1000)$~GeV  &$\tilde{\lambda}= -10 \to 5$ \\ \hline
\multirow{3}{*}{Yukawa--aligned 2HDM}
&$\tan\beta = 0.01 \to 50$      &$\sin\alpha = -1.0 \to 1.0$ \\ 
&$M_\text{heavy} = (200 \to 1000)$~GeV       &$\tilde{\lambda} = -10 \to 5$ \\
&$\gamma_{b,\tau} = 0 \to 2\pi$ \\ \hline
\multirow{3}{*}{degenerate spectrum}
&$\tan\beta = 0.01 \to 50$      &$\sin\alpha = -1.0 \to 1.0$ \\ 
&$M_\text{heavy} = (200 \to 1000)$~GeV       &$\tilde{\lambda} = -10 \to 5$ \\
&$\gamma_{b,\tau} = 0 \to 2\pi$ \\ \hline
  \end{tabular}
\end{small} \end{center}
\label{tab:parameters}
\caption{Free parameters for each benchmark fit to LHC data.  Details
  on the model parameterizations are provided in
  Appendix~\ref{app:para}.  For all of the benchmarks we fix the light
  (or twin) Higgs mass to $m_{\hzero, \Hzero} = 126$~GeV.}
\end{table}

\subsection{Free Standard Model couplings}
\label{sec:fit_sm}

\begin{figure}[b!]
\begin{center}
 \includegraphics[width=0.5\textwidth]{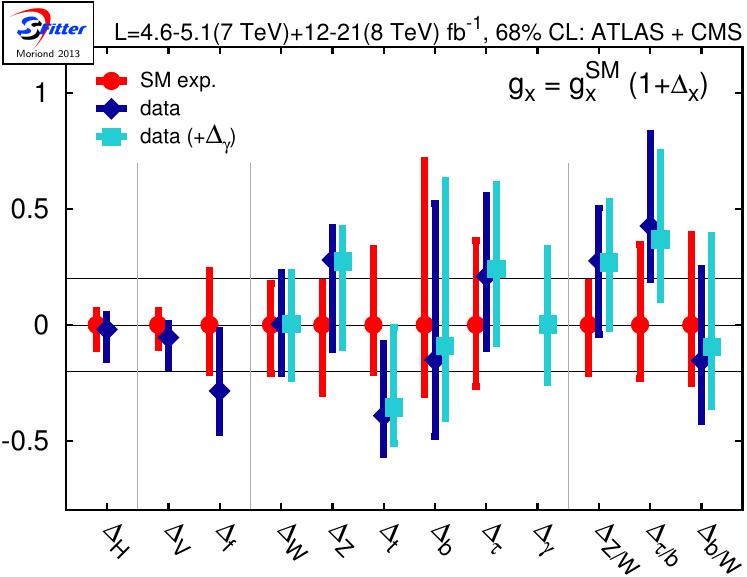}
\end{center}
\caption{Higgs coupling extraction based on all ATLAS and CMS results
  presented until the Winter conferences 2013 in Moriond and
  Aspen. The results shown are a yet unpublished update of the
  analyses published in Ref.~\cite{sfitter_higgs}.}
\label{fig:coup_sm}
\end{figure}

Before we study different extended Higgs sectors we review the Higgs
coupling extraction in the Standard Model. In Figure~\ref{fig:coup_sm}
we show all Higgs couplings to Standard Model particles with their
best--fit central values as well as their non--Gaussian error
bars. The red points indicate what we would expect using the current
data sets with the given experimental and theoretical errors, but with
all rate measurements fixed to the Standard Model predictions. The
dark blue points assume that the dimension-5 Higgs coupling to
photons, $g_\gamma$, is fully determined by Standard Model
loops. Finally, for the light blue dots we allow for additional states
contributing to $g_\gamma$. 
In this case, the quantity $\Delta_\gamma$ defined in Eq.~\eqref{eq:deltagamma}
enters the fit as an additional free parameter. 
The dimension-5 Higgs coupling to
gluons is identified with the Standard Model top contribution because
there is no independent measurement of the top Yukawa available yet.

We see that there is no visible tension between the Higgs measurements
and a purely Standard Model explanation. However, this outcome is not
at all unexpected.  The precision on the individual couplings still
ranges around 20 to 50\%, while typical models for physics beyond the
Standard Model predict significantly smaller deviations once all other
model constraints are taken into account~\cite{heidi}. Only strongly
constrained models, like a universal Higgs coupling modification
$\Delta_H$ or universal fermion and gauge coupling modifications
$\Delta_{V,f}$ are restricted at a level which significantly limits
their underlying toy models. As a matter of fact, for a single
modification $\Delta_H = 0 \pm \mathcal{O}(10\%)$ the LHC constraints
are in a similar range as electroweak precision data constraints on
the weakly interacting underlying models~\cite{heidi}.

In the following, we will discuss constraints on extended Higgs
sectors, which represent consistent perturbative new physics scenarios
in between the most general and extremely simplified scenarios shown
in Figure~\ref{fig:coup_sm}. Eventually, the question is how close the
results in the most general Yukawa--aligned 2HDM without loop
corrections will be to the general fit including all Standard Model
Higgs operators.

\subsection{Dark singlet}
\label{sec:fit_singlet}

\begin{figure}[t]
\begin{center}
 \includegraphics[width=0.50\textwidth]{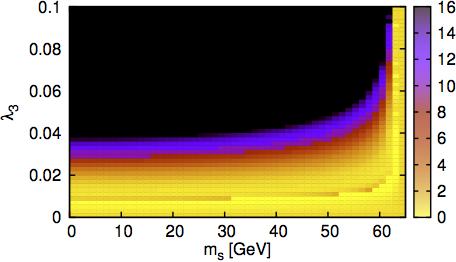} 
\end{center}
\vspace*{-2em}
\caption{Correlated relative log-likelihood $- 2 \Delta \log
  \mathcal{L}$ for the dark scalar mass $m_s$ vs the portal
  interaction $\lambda_3$ assuming an additional dark singlet.}
\label{fig:mc-darksinglet}
\end{figure} 

The lowest--dimension Lorentz and gauge invariant field
combination in the Standard Model is $\Phi^\dagger \Phi$. In a minimal
setup it can couple to an $SU(2)$ singlet from a hidden
sector~\cite{portal-orig}. In that case the Higgs self-interactions
form the only link, or portal, between the visible and the hidden
domains. For an additional singlet the corresponding potential is
given in Eq.\eqref{eq:singlet-potential}.  If it does not develop a
second VEV it does not contribute to electroweak symmetry breaking and
hence does not mix with the Higgs doublet. Nevertheless, the portal
interaction $\lag \supset \lambda_3\,(\Phi^\dagger\,\Phi)\,S^2$
defined in Eq.\eqref{eq:singlet-potential} triggers a new triple
scalar coupling. If kinematically allowed it gives rise to an
invisible decay mode
\begin{alignat}{5}
\Gamma_\text{inv}
\equiv 
\Gamma(h \to ss)
&= \frac{1}{32 \pi\,m_h}\; 
   \sqrt{1-\cfrac{4m_s^2}{m_h^2}} \; 
   \lambda^2_3 v^2 \; .
\label{eq:invwidth-text}
\end{alignat}
It depends only on two new model parameters, the dark singlet mass and
the strength of the portal interaction. In
Figure~\ref{fig:mc-darksinglet} we show the $m_s$ vs $\lambda_3$
correlation after confronting the model with the most recent LHC Higgs
measurements.  Because there are no significant LHC analyses directly
probing invisible Higgs decays, we only obtain an indirect limit from
the assumption that the sum of the visible and invisible partial
widths forms the total Higgs width. The total Higgs width can be
extracted as a common normalization factor to all predicted event
rates. As expected in the absence of a signal for invisible Higgs
decays, a light inert scalar is only allowed with weak portal
interactions.  A stronger singlet--doublet coupling $\lambda_3 \gtrsim
0.03$ is only compatible with LHC data when the invisible decay is
suppressed by phase space effects.

\subsection{Hierarchical 2HDM}
\label{sec:fit_hierarchical}

The simplest two Higgs doublet model includes one light Higgs scalar
and an approximately degenerate set of heavy states.  From
Eq.\eqref{eq:spectrum-custodial} we know that for $m_{\PHiggs^\pm}
\simeq m_{\Azero}$ the custodial symmetry is protected. In the absence
of significant non-decoupling effects a large mass hierarchy is
trivially in agreement with the stringent $B$-physics
bounds~\cite{bounds-flavor}, which disfavor charged Higgs masses
$m_{\PHiggs^\pm} \le 300$~GeV for a wide $\tan\beta$ range. For
extended Higgs sectors such indirect flavor physics constraints are
presently much stronger than constraints from direct
searches.\bigskip

\begin{figure}[t]
\hspace*{1.8cm} type-I 
\hspace*{2.7cm} type-II
\hspace*{2.1cm} lepton--specific
\hspace*{1.9cm} flipped \\[-1em]
\mbox{
\raisebox{-\height}{\includegraphics[width=0.23\textwidth]{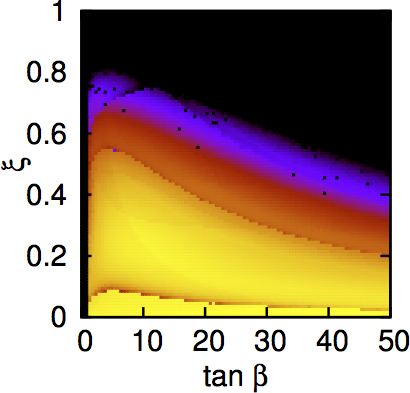}}
\raisebox{-\height}{\includegraphics[width=0.23\textwidth]{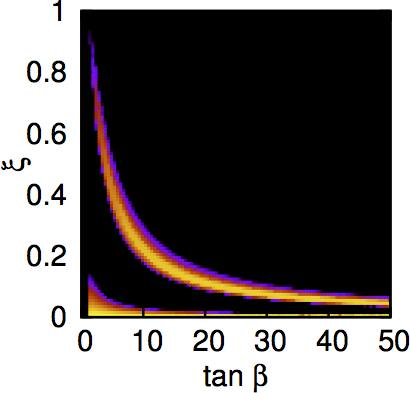}}
\raisebox{-\height}{\includegraphics[width=0.23\textwidth]{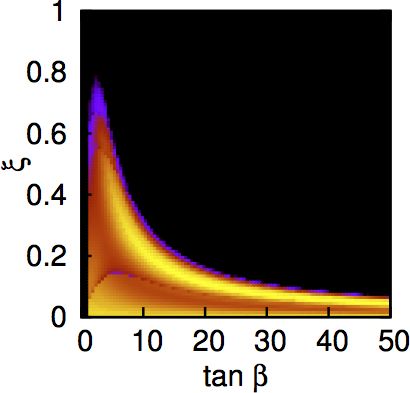}}
\raisebox{-\height}{\includegraphics[width=0.23\textwidth]{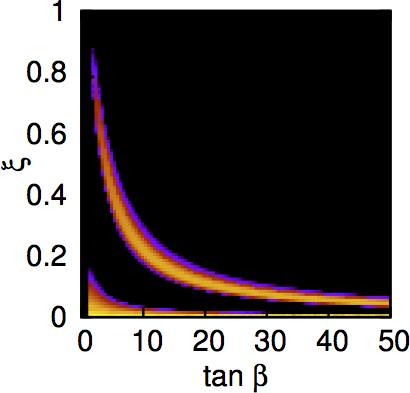}}
\raisebox{-\height}{\includegraphics[width=0.035\textwidth]{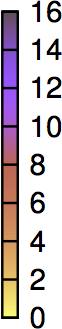}}}
\caption{Correlated relative log-likelihood $- 2 \Delta \log
  \mathcal{L}$ for $\tan\beta$ vs $\xi$, assuming a hierarchical 2HDM
  with natural flavor conservation.}
\label{fig:mc-hierarchical}
\end{figure}

A hierarchical 2HDM can be described by the effective field theory
described in Appendix~\ref{app:hierarchy}. Its expansion parameter is
$v/M_\text{heavy} \ll 1$ or equivalently $\xi \equiv
\cos(\beta-\alpha) \simeq v^2/M_\text{heavy}^2$.  The consistency
condition $\xi < 1$ needs be taken into account in the discussion of
the results. It immediately forbids sign changes in Higgs couplings
$\Delta_x \simeq -2$.  We assume natural flavor conservation and
separately examine the type-I, type-II, lepton--specific, and flipped
setups discussed in Section~\ref{sec:doublet}.  They are initially
characterized by four parameters, leading to coupling shifts
\begin{alignat}{5}
 \Delta_f (\tan\beta, \xi) \qqquad 
 \Delta_V (\tan\beta, \xi) \qqquad 
 \Delta_\gamma (\tan\beta, \xi, m^2_{\PHiggs^{\pm}}(\xi), \tilde{\lambda}(\xi)) \; .
\label{eq:dep2}
\end{alignat}
\noindent Only two of them turn out to be independent, which we take
to be the VEV ratio $\tan\beta$ and the decoupling parameter $\xi$.
By construction, the heavy Higgs mass and the self-coupling
$\tilde{\lambda}$ are fixed at 
\begin{alignat}{5}
M^2_\text{heavy} = \frac{v^2}{\xi} 
\qqquad \text{and} \qqquad 
\tilde{\lambda}\,v^2 & \simeq m^2_{\hzero} + 2M^2_\text{heavy} \; .
\label{eq:selfcoup-hierachical}
\end{alignat}
This guarantees that the light Higgs boson self-interactions to the
heavy fields exhibit a consistent $\mathcal{O}(\xi)$ behavior in the
decoupling limit, as discussed in Appendix~\ref{app:hierarchy}.  The
parameter ranges are defined in Table~\ref{tab:parameters}.\bigskip

In Figure~\ref{fig:mc-hierarchical} we show the likelihood correlations
for the hierarchical 2HDM fit to LHC data.  The favored regions are
particularly clear for type-II and flipped 2HDM models, where the top
and bottom Yukawas are largely independent when we vary $\tan\beta$
and $\xi$. This enhanced flexibility allows for an easier convergence
to parameter regions preferred by the LHC results.

As derived in Appendix~\ref{app:hierarchy} the correlation between
$\tan \beta$ and $\xi$ follows the line $\xi \sim \tan\beta / (1 +
\tan^2\beta)$ for $\tan\beta \gg 1$. For the Yukawa couplings in
type-II and flipped models large deviations in $\xi$ only occur at
relatively small $\tan \beta$, because simultaneously large $|\xi|$
and $\tan \beta$ values would lead to too large shifts in the bottom
and tau Yukawas.  The gauge boson couplings scale as $\Delta_V \sim -
\xi^2/2$, which means that typical allowed deviations $|\Delta_V|
\lesssim 0.2$ correspond to $\xi \lesssim 0.63$.

Type-I models are less efficient in profiling the LHC measurements, so
their parameters are less clearly constrained. We will come back to
this kind of effect when comparing the fits to the different models.
In all setups of the hierarchical 2HDM the LHC measurements broadly
prefer $\xi \lesssim 0.4$, possibly correlated with $\tan \beta$ and
entirely consistent with the original assumption $\xi <1$ underlying
our effective theory.

The sub-leading charged Higgs parameters can be tested by computing
the correlated $\Delta_\gamma$ and $\Delta_W$ ranges. We find that we
are limited to $\Delta_\gamma = 0...0.1$ with preferred $\Delta_W <
0$, except for type-I models. There, the scan over the model
parameters fills the generally allowed range $\Delta_\gamma =
-0.3...0.3$ shown in Figure~\ref{fig:coup_sm}.  This is in line with
the also broader $\Delta_\gamma^\text{SM}$ range which 
results from the wider shifts in the top Yukawas.\bigskip

In Table~\ref{tab:bestfit-h} we show the single best--fit parameter
points for the hierarchical 2HDM. We do not quote error bars on the
individual parameters because of the strong correlations shown in
Figure~\ref{fig:mc-hierarchical}. Not surprisingly, the best
log-likelihood is reached in the $\xi \to 0$ limit. The results
highlight a clear preference for a SM-like pattern with vanishing
coupling deviations within numerical precision. A slightly improved
log-likelihood value we can only achieve for the lepton--specific
setup. Here, the tau Yukawa can be adjusted independently of the quark
Yukawas, which allows us to accommodate the central values
$\Delta_\tau >0$ and $\Delta_{b,t} \lesssim 0$ in the general
couplings fit --- at the expense of a simultaneously larger deviation
from $\Delta_V \simeq -\xi^2/2 \to 0$. Nevertheless, the general
agreement of the LHC measurements with the hierarchical 2HDM is
satisfactory.

In practice, vacuum stability precludes $\tan\beta \gtrsim 15$ for a
hierarchical spectrum in the $\xi \to 0$ limit. Similar constraints
follow from self-coupling perturbativity.  This explains the downward
shift in the favored $\tan\beta$ values, once the model constraints
are included in the fit.  A consistent $\tan\beta \gg 1$ limit can
only be reached in this benchmark if $\lambda_{7} \neq
0$~\cite{Gunion:2002zf}.

\begin{table}[t]
\begin{center} \begin{small}
\begin{tabular}{c|r|r|r|r||r|r|r|r} \hline
 & \multicolumn{4}{c||}{unconstrained} & \multicolumn{4}{c}{constrained} \\ \hline
& type-I & type-II & lepton & flipped & type-I & type-II & lepton & flipped \\ \hline \hline
$\tan \beta$ & 24.0 & 19.9 & 16.4 & 36.9 &10.6 & 9.2 & 14.2 & 13.9  \\ 
$\xi$ & 0.000 & 0.000 & 0.134 & 0.000 & 0.000& 0.000& 0.002& 0.000   \\ 
$\Delta_V$ & 0.000 & 0.000 & -0.009 & 0.000  &0.000 &0.000 & 0.000 & 0.000 \\ 
$\Delta_t$ & 0.000 & 0.000 & -0.001 & 0.000  &0.000 & 0.000& 0.000&  0.000 \\ 
$\Delta_b$ & 0.000 & 0.000 & -0.001 & 0.000 & 0.000& 0.000&0.000 &  0.000\\ 
$\Delta_\tau$ & 0.000 & 0.000 & -2.208 & 0.000  & 0.000& 0.000& -0.030 &  0.000\\ 
$\Delta_\gamma$ & 0.000 & 0.000 & -0.055 & 0.000 & 0.000& 0.000&-0.001 &  0.000 \\ 
$\Delta_\gamma^\text{tot}$ & 0.000 & 0.000 & -0.074 & 0.000 &0.000 & 0.000& -0.001& 0.000   \\ \hline
$-2\log \mathcal{L}$ & 31.7 & 31.7 & 30.8 & 31.7 & 31.7& 31.7& 31.6& 31.7\\\hline 
\end{tabular}
\end{small} \end{center}
\caption{Best--fit parameter points for the hierarchical 2HDM. Separate results
are shown for a fully unconstrained fit (left columns) and after
we include the theoretical and experimental bounds as fit priors (right columns).}
\label{tab:bestfit-h}
\end{table}

\subsection{General 2HDM}
\label{sec:fit_general}

In the next step we release the assumption of decoupled heavy Higgs
states and instead vary their common single mass parameter as
$M_\text{heavy} =200...1000$~GeV.  By the same token, we no longer
impose the decoupling condition $\cos(\beta-\alpha) \to 0$ and allow
for an independent variation of both mixing angles.  The individual
coupling shifts are based on four independent model parameters,
\begin{alignat}{5}
\Delta_f (\tan\beta, \sin\alpha) \qquad
\Delta_V (\tan\beta, \sin\alpha) \qquad
\Delta_\gamma (\tan\beta, \sin\alpha, m^2_{\PHiggs^\pm},\tilde{\lambda}) \; .
\label{eq:dep3}
\end{alignat}
The $\sin\alpha$ and $\tan\beta$ ranges given in
Table~\ref{tab:parameters} cover the whole range of modified Yukawas,
including sign flips.  The range of $\tan\beta$ is severely
constrained by flavor physics, disfavoring $\tan \lesssim 1$ as well
as large $\tan\beta$ values.  Barring the type-I models, charged Higgs
bosons below $300$~GeV tend to be also ruled out. As for the
$\tilde{\lambda}$ range, large (negative) values are excluded by
unitarity, while the upper (positive) edge is constrained by vacuum
stability.\bigskip

\begin{figure}[t]
\hspace*{1.8cm} type-I 
\hspace*{2.7cm} type-II
\hspace*{2.1cm} lepton--specific
\hspace*{1.9cm} flipped \\[-1em]
\mbox{
 \raisebox{-\height}{\includegraphics[width=0.23\textwidth]{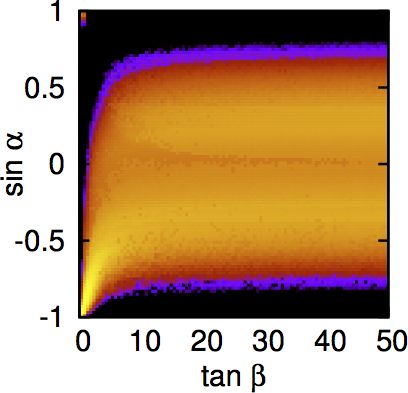}}
 \raisebox{-\height}{\includegraphics[width=0.23\textwidth]{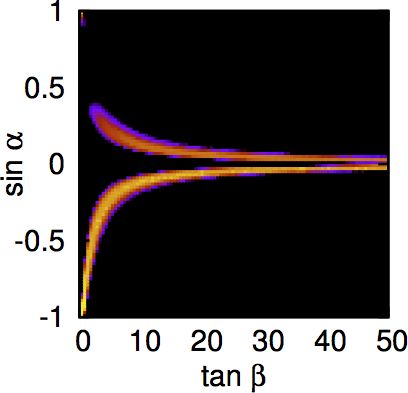}}
 \raisebox{-\height}{\includegraphics[width=0.23\textwidth]{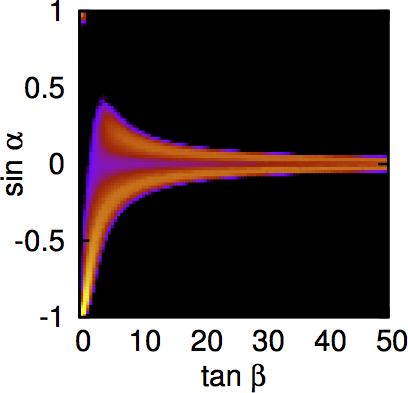}} 
 \raisebox{-\height}{\includegraphics[width=0.23\textwidth]{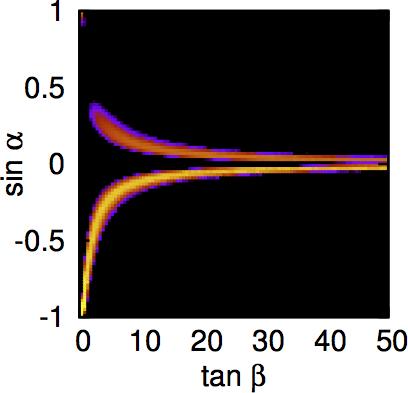}} 
 \raisebox{-\height}{\includegraphics[width=0.035\textwidth]{./figs_jpg/plot_colorbox.jpg}}}
\caption{Correlated relative log-likelihood $- 2 \Delta \log
  \mathcal{L}$ for $\tan\beta$ vs $\sin\alpha$ ranges assuming a
  general 2HDM with a common heavy Higgs mass scale and natural
  flavor conservation.}
\label{fig:mc-tanb-sina}
\end{figure}

In Figure~\ref{fig:mc-tanb-sina} we show how a 2HDM with a single
variable heavy Higgs mass describes the current LHC data.  The
preferred regions in the $\tan\beta$ vs $\sin\alpha$ plane are similar
to those for the hierarchical spectrum displayed in
Figure~\ref{fig:mc-hierarchical}.  They cover the entire $\tan\beta$
range with some preference for moderately small values.  The second
mixing angle $\sin\alpha$ spans a correlated symmetric range
$\sin^2\alpha \sim 1/(1+\tan^2\beta)$, reflecting the decoupling
condition $\cos(\beta-\alpha) \to 0$ discussed in
Appendix~\ref{app:hierarchy}.\bigskip

\begin{table}[t]
\begin{center} \begin{small}
\begin{tabular}{c|r|r|r|r||r|r|r|r} \hline
 & \multicolumn{4}{c||}{unconstrained} & \multicolumn{4}{c}{constrained} \\ \hline
& type-I & type-II & lepton & flipped &   type-I & type-II & lepton & flipped\\ \hline 
$\tan \beta < 1$ &  0.920 & 0.115 & 0.594 & 0.270 & 0.999 & 0.816 & 0.976 & 0.857 \\ 
$\sin \alpha$ & -0.891 & -0.996 & -0.950 & -0.975 & -0.859 & -0.826 & -0.866 & -0.814 \\ 
$M_\text{heavy}$ & 614.7 & 202.3 & 220.2 & 243.5 & 608.0 & 624.0 & 627.8 & 607.0  \\ 
$\tilde{\lambda}$ & -8.19 & 4.49 & -2.31 & 4.63 & 3.14 & 2.17 & 4.11 & 1.50 \\\hline 
$\Delta_V$ & -0.037 &  0.000 & -0.024 & -0.001 & -0.031 & -0.003 & -0.031 & -0.004  \\ 
$\Delta_t$ & -0.328 & -0.213 & -0.387 & -0.150  & -0.276 & -0.108 & -0.284 & -0.107  \\ 
$\Delta_b$ & -0.328 & 0.003 & -0.387 & 0.010 & -0.276 & 0.066 & -0.284 & 0.072 \\ 
$\Delta_\tau$ & -0.328 & 0.003 & 0.105 & -0.150 & -0.276 & 0.066 & 0.210 & -0.107\\ 
$\Delta_\gamma$ & -0.082 & 0.098 & -0.124 & 0.057 & -0.038 & -0.043 & -0.035 & -0.046 \\ 
$\Delta_\gamma^\text{tot}$ & -0.043 & 0.155 & -0.052 & 0.096 & -0.005 & -0.019 & 0.002 & -0.022 \\ 
\hline$-2\log \mathcal{L}$ & 28.8 & 28.7 & 26.6 & 29.2 & 28.9 & 30.5 & 27.1 & 30.8 \\\hline  \hline
$\tan \beta > 1$ & 1.000 & 1.051 & 1.000 & 1.000 & 1.324 & 1.491 & 1.093 & 2.918 \\ 
$\sin \alpha$ & -0.878 & -0.746 & -0.862 & -0.756 & -0.842 & -0.574 & -0.829 & -0.340  \\ 
$M_\text{heavy}$ & 783.4 & 215.2 & 959.1 & 234.7 & 412.1 & 356.4 & 565.4 & 378.5  \\ 
$\tilde{\lambda}$ & -8.49 & 5.00 & 5.00 & 4.91 & -0.65 & 4.11 & 3.03 & 4.68 \\\hline 
$\Delta_V$ & -0.045 & -0.003 & -0.032 & -0.003 & -0.062 & 0.000 & -0.028 & 0.000  \\ 
$\Delta_t$ & -0.322 & -0.080 & -0.283 & -0.075 & -0.324 & -0.014 & -0.242 & -0.006  \\ 
$\Delta_b$ & -0.322 & 0.082 & -0.283 & 0.069 & -0.324 & 0.031 & -0.242 & 0.049 \\ 
$\Delta_\tau$ & -0.322 & 0.082 & 0.219 & -0.075 & -0.324 & 0.031 & 0.228 & -0.006 \\ 
$\Delta_\gamma$ & -0.070 & 0.071 & -0.041 & 0.082 & -0.057 & -0.004 & -0.036 & -0.002 \\ 
$\Delta_\gamma^\text{tot}$ & -0.037 & 0.131 & -0.007 & 0.099 & -0.052 & 0.000 & -0.007 & -0.001 \\ \hline
$-2\log \mathcal{L}$ & 28.8 & 29.7 & 27.2 & 29.8 & 28.9 & 30.8 & 27.1 & 30.7 \\\hline 
\end{tabular}
\end{small} \end{center}
\caption{Best--fit parameter points for the general 2HDM with a
  variable single heavy Higgs mass $M_\text{heavy}$, computed for $\tan\beta <
  1$ (top panel) and $\tan\beta > 1$ (down panel). Separate results are shown for a fully unconstrained fit (left
  columns) and after we include the theoretical and experimental
  bounds as fit priors (right columns).}
\label{tab:bestfit-g}
\end{table}

The best--fit parameter points for $\tan\beta < 1$ and $\tan\beta > 1$
are separately quoted in Table~\ref{tab:bestfit-g}. First, we focus on
the results for the unconstrained fit.  The absolute best--fit point
lies at $\sin\alpha \simeq -1$ and, with the exception of the type-I
setup, $\tan\beta < 1$.  The sign of the mixing angle $\alpha$ is
linked to the sign of the Yukawa couplings, which for the bottom quark
and the tau lepton is not determined by LHC data. Only for the top
Yukawa the destructive interference in the effective Higgs--photon
coupling resolves this sign ambiguity. The larger best--fit value of
$\tan \beta$ in type-I models can be traced to competing effects in
this strongly constrained setup: in all fits data prefers tiny
deviations of $\Delta_V \ll 0.1$. In type-I models the central value
is pulled down to $\Delta_V = -0.037$.  At the same time the shift in
the top Yukawa, scaling like $\Delta_t \sim \cos
(\beta-\alpha)/\tan\beta$, prefers to be negative and small. This
moves the very small $\tan \beta$ values towards unity.  Conversely,
for the type-II 2HDM the value $\Delta_V \simeq 0$ pulls towards
$\tan\beta = \mathcal{O}(0.1)$. The preference for small finite
coupling shifts to the top-quark and the other heavy fermions also
explains the preference for $\tan\beta \simeq 1$ for all models when
we focus on the $\tan\beta > 1$ range.\bigskip

\begin{figure}[t]
\hspace*{1.8cm} type-I 
\hspace*{2.7cm} type-II
\hspace*{2.1cm} lepton--specific
\hspace*{1.9cm} flipped \\[-1em]
\mbox{
\raisebox{-\height}{\includegraphics[width=0.23\textwidth]{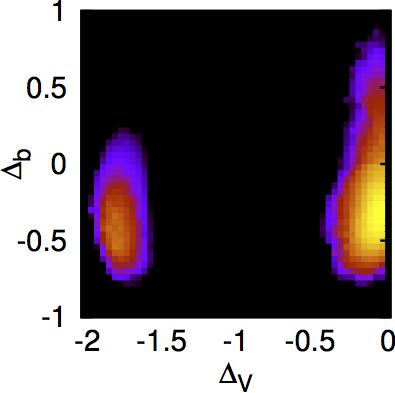}}
\raisebox{-\height}{\includegraphics[width=0.23\textwidth]{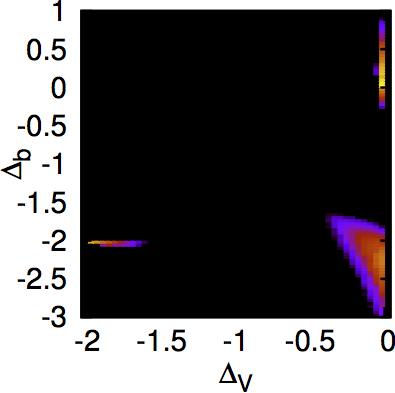}}
\raisebox{-\height}{\includegraphics[width=0.23\textwidth]{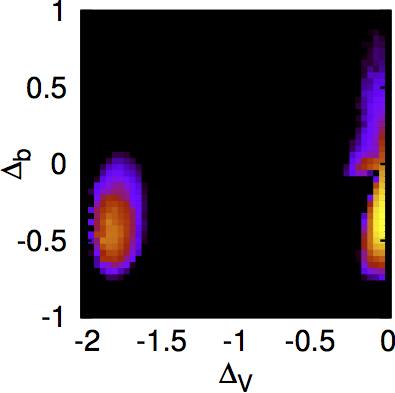}}
\raisebox{-\height}{\includegraphics[width=0.23\textwidth]{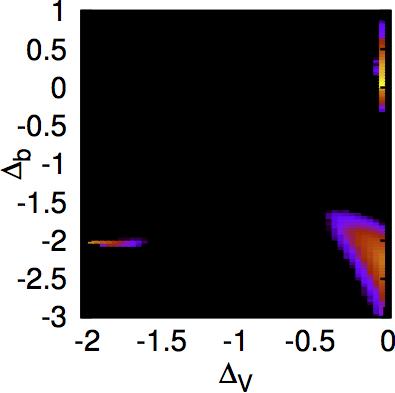}}
\raisebox{-\height}{\includegraphics[width=0.035\textwidth]{./figs_jpg/plot_colorbox.jpg}}}\\
\mbox{
\raisebox{-\height}{\includegraphics[width=0.23\textwidth]{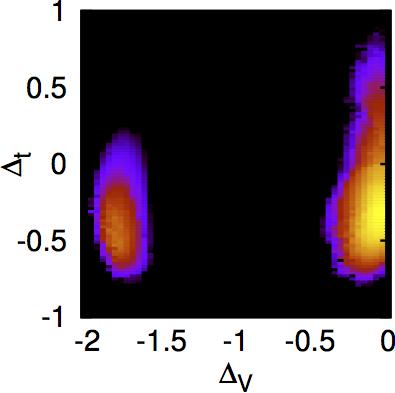}}
\raisebox{-\height}{\includegraphics[width=0.23\textwidth]{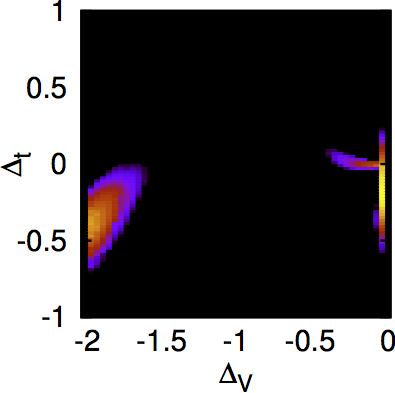}}
\raisebox{-\height}{\includegraphics[width=0.23\textwidth]{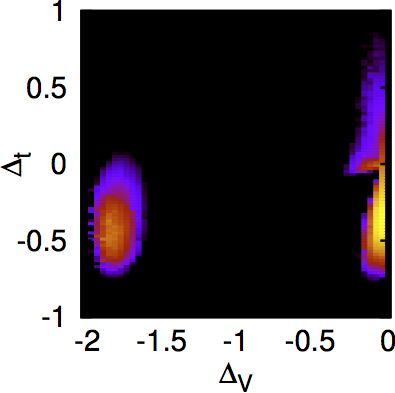}}
\raisebox{-\height}{\includegraphics[width=0.23\textwidth]{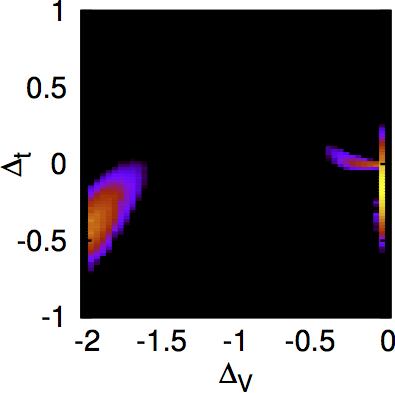}}
\raisebox{-\height}{\includegraphics[width=0.035\textwidth]{./figs_jpg/plot_colorbox.jpg}}}\\
\mbox{
\raisebox{-\height}{\includegraphics[width=0.23\textwidth]{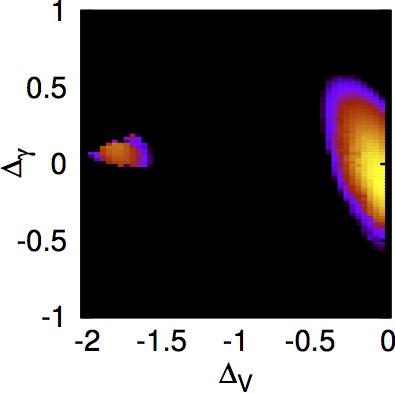}}
\raisebox{-\height}{\includegraphics[width=0.23\textwidth]{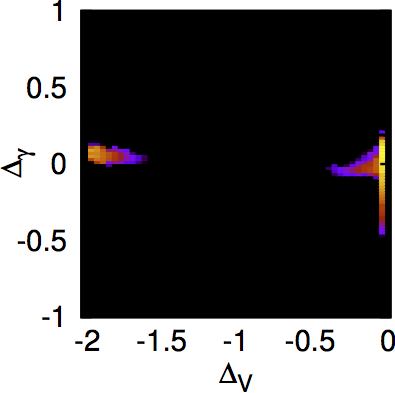}}
\raisebox{-\height}{\includegraphics[width=0.23\textwidth]{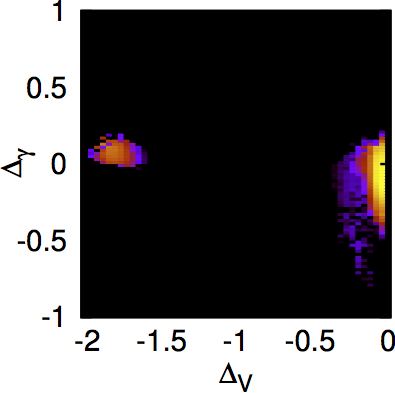}}
\raisebox{-\height}{\includegraphics[width=0.23\textwidth]{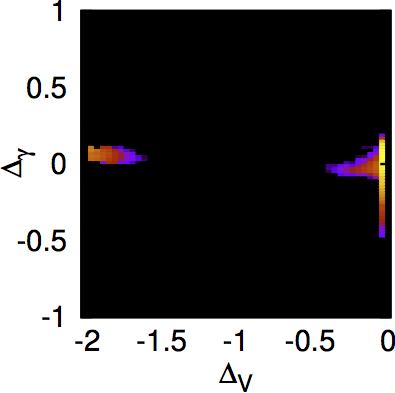}}
\raisebox{-\height}{\includegraphics[width=0.035\textwidth]{./figs_jpg/plot_colorbox.jpg}}}\\
\caption{Correlated relative log-likelihood $- 2 \Delta \log
  \mathcal{L}$ for $\Delta_{\PW}$ vs $\Delta_b$ (top row),
  $\Delta_{\PW}$ vs $\Delta_t$ (center row), and $\Delta_{\PW}$ vs
  $\Delta_\gamma$ (bottom row) assuming a general 2HDM with natural
  flavor conservation.}
\label{fig:mc-general-coup}
\end{figure}

It is useful to translate the favored 2HDM configurations into the
associated coupling strength deviations, illustrated in
Figure~\ref{fig:mc-general-coup}. We see that in the fermion sector
type-II and flipped realizations show a slight preference for enhanced
couplings to the bottom quarks, staying either way in the SM-like region $\Delta_b \sim
0$.  All of the remaining cases favor a mild suppression, and neatly
peak around the decoupling limit $\Delta_V \simeq 0$. The largest
deviation from the decoupling limit we find for type-I and
lepton--specific models. This reflects the more constrained nature of
their correlated top and bottom Yukawa couplings.  This is also the
reason for a tendency towards a slightly larger negative $\Delta_t$ in the same models.

When the tau Yukawa can be shifted independently, like in the
lepton--specific setup, the data prefers a slight positive deviation
$\Delta_\tau \gtrsim 0$.  This is in line with the general Standard
Model fit from Figure~\ref{fig:coup_sm} and stems from the mild CMS
excess in the $\tau\tau$ search channel. Correspondingly, an
independent tau Yukawa turns out to be key to improve the overall fit
quality. As seen in Table~\ref{tab:bestfit-g} the lepton--specific
2HDM yields a best--fit log--likelihood value $-2\log \mathcal{L}
\simeq 27.2$, which is roughly as good as the values for a general Yukawa
alignment, as will be discussed next.

In addition to the best fit regions centered around the SM-like
solution $\Delta_x \simeq 0$ our results also spotlight additional
solutions including a sign--flipped bottom Yukawa $\Delta_b \lesssim
-2$ at $\Delta_W > 0$ in Figure~\ref{fig:mc-general-coup} or
$\Delta_{b,W} \lesssim -2$ corresponding to SM-like coupling strengths
with opposite signs relative to the top Yukawa. All these observations
are in agreement with the literature~\cite{Chen:2013kt}.\bigskip

Aside from the two key parameters which affect all coupling shifts
we also examine the impact of the charged Higgs mass
$m_{\PHiggs^{\pm}} = M_\text{heavy}$ and the self-coupling
$\tilde{\lambda}$ corresponding to the PQ-breaking scale $m^2_{12}$,
Eq.\eqref{eq:ltilde}.  They determine the charged Higgs contribution
$\Delta_\gamma$, which can be sizable for $m_{\PHiggs^{\pm}}
\lesssim 300$~GeV and/or $|\tilde{\lambda}| \gtrsim \mathcal{O}(5)$.
The bulk of the scanned $m_{\PHiggs^{\pm}}$-$\tilde{\lambda}$ plane is
allowed and carries no significant log-likelihood variation.  Small
shifts of the tree-level couplings always provide excellent agreement
with the LHC results within given uncertainties.  In the bottom panels of 
Figure~\ref{fig:mc-general-coup} we examine the charged Higgs effect
on $\Delta_\gamma$ by correlating it with $\Delta_V$. While centered
at $\Delta_V = 0$, the statistically preferred region spans the range
$\Delta_\gamma = -0.5...0.1$ to $2\sigma$, in particular for the
type-I 2HDM.  We also find a secondary solution around $\Delta_V = -2$
related to the sign shift in the top Yukawa. This solution is by
definition absent in the hierarchical 2HDM.\bigskip

The theoretical and experimental constraints described in
Section~\ref{sec:doublet}, including electroweak precision and flavor observables,
Higgs mass exclusion limits, 
perturbative unitarity and vacuum stability are accounted for in the right columns of
Table~\ref{tab:bestfit-g}. As anticipated, values of $\tan\beta \ll 1$
are ruled out by flavor observables, like $B_d^0 -\overline{B}_d^0$
mixing. The charged Higgs boson contribution to this process scales
like $1/\tan^4\beta$, common to all of the natural flavor
conserving models. When we focus on the $\tan \beta <1$ region this
pushes the constrained best--fit solutions towards $\tan\beta \simeq
1$. Even in the $\tan\beta > 1$ region the preferred values are
slightly shifted and now fall in the $\tan\beta \simeq 1.5$
ball--park. This is below the large $\tan\beta$ range which would lead
to enhanced bottom and tau Yukawas in type-II, lepton--specific and
flipped models, and which is severely limited $\text{BR}(b \to
s\gamma)$.  Similarly, the unitarity and the vacuum stability
conditions do not feature.  These flavor constraints unify the heavy
Higgs mass $M_\text{heavy}$ and self-coupling $\tilde{\lambda}$ ranges
in the different setups. Electroweak precision constraints play no
role as long as the model setup essentially respects the custodial
symmetry.  As a further consequence, the compatibility with all these
additional constraints eliminates secondary solutions with
sign-flipped quark Yukawas.
 
\subsection{Yukawa--aligned 2HDM}
\label{sec:fit_aligned}

\begin{figure}[t]
\begin{center}
\mbox{
\raisebox{-\height}{\includegraphics[width=0.26\textwidth]{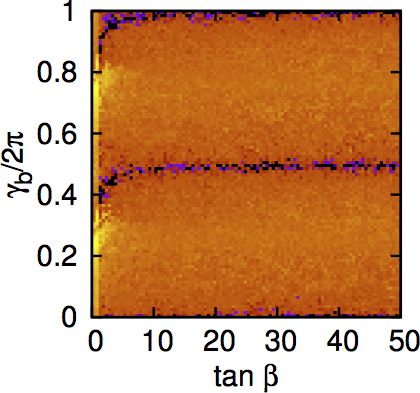}} \qquad
\raisebox{-\height}{\includegraphics[width=0.26\textwidth]{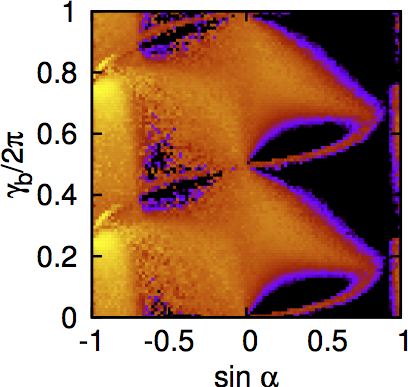}} \qquad
\raisebox{-\height}{\includegraphics[width=0.26\textwidth]{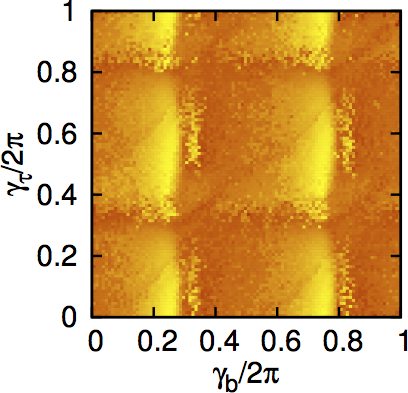}} 
\raisebox{-\height}{\includegraphics[width=0.040\textwidth]{./figs_jpg/plot_colorbox.jpg}}}\\
\mbox{
\raisebox{-\height}{\includegraphics[width=0.26\textwidth]{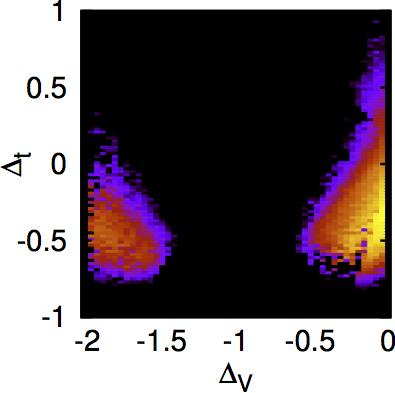}} \qquad
\raisebox{-\height}{\includegraphics[width=0.26\textwidth]{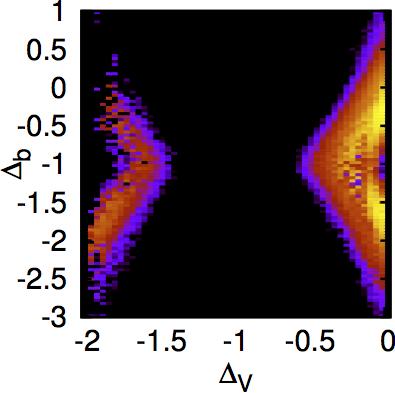}} \qquad
\raisebox{-\height}{\includegraphics[width=0.26\textwidth]{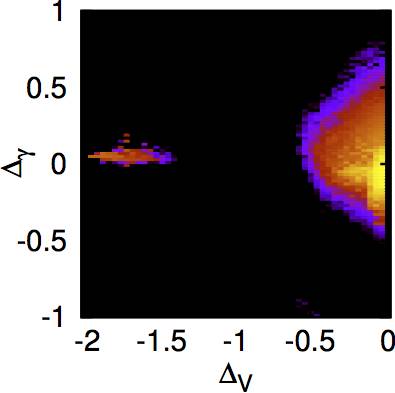}} 
\raisebox{-\height}{\includegraphics[width=0.041\textwidth]{./figs_jpg/plot_colorbox.jpg}}}
\end{center}
\caption{Correlated relative log-likelihood $- 2 \Delta \log
  \mathcal{L}$ for different pairs of parameters (top row) and
  coupling shifts (bottom row), assuming a 2HDM with Yukawa alignment.}
\label{fig:mc-aligned}
\end{figure}

The final step towards a fully independent Higgs couplings
determination in a consistent 2HDM framework replaces the natural
flavor conservation hypothesis with the more general Yukawa alignment
described in Section~\ref{sec:doublet}. The third-generation Yukawas can
now be shifted independently through the angles $\gamma_{b,\tau}$
defined in Table~\ref{tab:2hdmcouplings1}.  The different Higgs
couplings are functions of 6 independent model parameters, namely
\begin{alignat}{5}
&\Delta_t (\tan\beta, \sin\alpha) \qqquad
&&\Delta_b (\tan\beta, \sin\alpha, \gamma_b) \qqquad 
&&\Delta_\tau (\tan\beta, \sin\alpha, \gamma_{\tau}) \notag \\
&\Delta_V (\tan\beta, \sin\alpha) 
&&\Delta_\gamma (\tan\beta, \sin\alpha, m^2_{\PHiggs^{\pm}}, \tilde{\lambda},\gamma_b, \gamma_{\tau}) \; .
\label{eq:dep111}
\end{alignat}
The results of the corresponding fit to data are shown in
Figure~\ref{fig:mc-aligned}.  We find broad allowed regions in the
$\sin\alpha$ vs $\tan\beta$ plane, with a slight preference for low
$\tan\beta$ values and $\sin\alpha < 0$.  The log-likelihood profiles
are relatively smooth under variations of $\gamma_{b,\tau}$. In the
upper panels of Figure~\ref{fig:mc-aligned} the sign degeneracy in
$g_b$ and $g_\tau$ is clearly manifest as a twofold strip in the
$\gamma_{b,\tau}$ vs $\tan\beta$ plane and a double-sided lobular area
in the $\gamma_{b,\tau}$ vs $\sin\alpha$ plane.  By the same token, we
find corresponding excluded strips and oval--shaped areas for the
parameter space regions with overly suppressed Yukawa
interactions.

\begin{table}[b!]
\begin{center} \begin{small}
\begin{tabular}{c|r|r||r|r} \hline
 &  \multicolumn{2}{c||}{unconstrained} &  \multicolumn{2}{c}{constrained} \\ \hline
 & $\tan < 1$ & $\tan\beta > 1$ & $\tan\beta < 1$ & $\tan\beta > 1$ \\ \hline
$\tan \beta$               &  0.338  &  1.231 & 0.890 & 1.324 \\
$\sin \alpha$              & -0.977  & -0.871 &  -0.900 & -0.810 \\
$\gamma_b/(2\pi)$            &  0.744   &    0.261 & 0.732 &  0.267 \\
$\gamma_\tau/(2\pi)$         &  0.389   &  0.070 & 0.542 &  0.678 \\
$M_H$                      & 750.5  &     257.3 & 587.2 & 487.4  \\
$\tilde{\lambda}$          & -3.00  & 0.44  & 3.80  &  2.88\\  \hline   
$\Delta_V$                 & -0.006 &-0.069  & -0.038  &  -0.044\\
$\Delta_t$                 & -0.332 &-0.367  & -0.345 & -0.265\\
$\Delta_b$                 & -0.298 &-0.412  &-0.281 &  -0.318\\
$\Delta_\tau$              &  0.176 &0.107   &  0.099 & -0.102 \\
$\Delta_\gamma$            & -0.058  &-0.045 &  -0.034 & -0.031 \\
$\Delta_\gamma^\text{tot}$ &  0.023  &-0.036 &  0.009 & -0.017 \\\hline
$-2\log \mathcal{L}$& 26.6 & 26.9 & 27.1 & 28.5   \\ \hline
\end{tabular}
\end{small} \end{center}
\caption{Best--fit parameter points for the 2HDM with Yukawa
  alignment. Separate results are shown for a fully unconstrained fit
  (left columns) and after we include the theoretical and experimental
  bounds as fit priors (right columns).}
\label{tab:bestfit-aligned}
\end{table}

In the lower panels of Figure~\ref{fig:mc-aligned} we illustrate
multifarious deviations of the Yukawas with respect to the Standard
Model, including enhancements, suppressions, and sign flips. Unlike
for natural flavor conservation data now favors $\Delta_{b,t} \lesssim
0$, while the mild $\tau\tau$ excess points to $\Delta_\tau \gtrsim
0$.

For the effective Higgs coupling to photons our fit remains at
$|\Delta_\gamma| \lesssim 0.2$.  The correlation between
$\Delta_\gamma$ and $\Delta_V$ in Figure~\ref{fig:mc-aligned} follows
from an interplay of the interferences.  The experimentally favored
SM-like profile $\Delta^\text{tot}_\gamma \simeq 0$ can be realized in
the general 2HDM in different ways: first, there can be consistently
small deviations $\Delta_{f,V,\gamma} \simeq 0$, including a heavy
and/or weakly coupled charged Higgs.  Second, we can reduce $\Delta_W$
and $\Delta_t$ by similar finite amounts while keeping a negligible
charged Higgs contribution. Third, we can slightly reduce $g_t$,
preserve $\Delta_V \simeq 0$, and allow for a tempered charged Higgs
contribution.  Finally, we fix $\Delta_W \simeq -2$ together with a
sign-flipped top Yukawa.  These options are nicely visible in the
log-likelihood correlation in the right lower panel of
Figure~\ref{fig:mc-aligned}, including the strip $\Delta_\gamma =
-0.5...+0.5$ with $\Delta_W \simeq 0$.\bigskip

The best--fit central values for the coupling deviations in the
Yukawa--aligned 2HDM are shown in Table~\ref{tab:bestfit-aligned}.  In
line with the observations we have made for the natural flavor
conserving 2HDM benchmarks, the constrained fit solutions shift the
preferred $\tan\beta$ values upwards because the region $\tan\beta
\simeq 0.1$ is in tension with $B^0_d-\overline{B}^0_d$ mixing results. Once
all the theoretical and phenomenological restrictions are included as
fit priors, the preferred configurations feature $\tan\beta \simeq 1$,
$\sin \alpha \simeq -1$, a wide range of Yukawa alignment angles
$\gamma_{b,\tau}$, and heavy charged Higgs bosons, in particular for $\tan\beta < 1$.  None of the
coupling deviations is substantially altered with respect to the
unconstrained fit. This result starts looking much more like
a general coupling fit because the number of model parameters in the
aligned 2HDM is identical to a general fit of all Standard Model
coupling structures, with the exception of $\Delta_W = \Delta_Z$. 
We will show the direct comparison of the two approaches in the Summary.

\subsection{Degenerate spectrum}
\label{sec:fit_degenerate}

\begin{figure}[b!]
\begin{center}
\mbox{
\raisebox{-\height}{\includegraphics[width=0.26\textwidth]{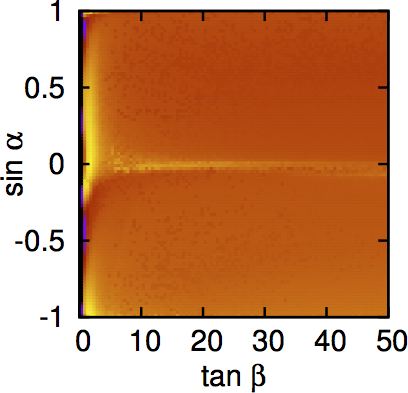}} \qqqquad
\raisebox{-\height}{\includegraphics[width=0.26\textwidth]{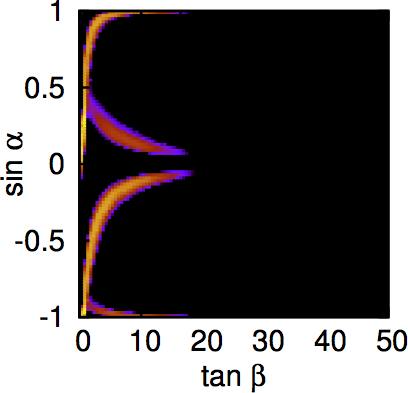}}
\raisebox{-\height}{\includegraphics[width=0.040\textwidth]{./figs_jpg/plot_colorbox.jpg}}}
\end{center}
\caption{Correlated relative log-likelihood $- 2 \Delta \log
  \mathcal{L}$ for $\tan\beta$ vs $\sin\alpha$ ranges for
a type-I 2HDM (left) and type-II 2HDM (right) with degenerate mass spectrum.}
\label{fig:degenerate-param}
\end{figure}

So far, we have interpreted the LHC measurements in terms of models
where only one state contributes to the observed Higgs resonance. We
can depart from this hypothesis and entertain the possibility of two
mass--degenerate \CP-even states $\hzero,\Hzero$, both contributing to
the same signal strength. Technically, we assume a very small
splitting $m_{\Hzero} = m_{\hzero} + \delta m$, to avoid the parameter
space singularity at $m_{\hzero} =
m_{\Hzero}$~\cite{Gunion:2002zf}. Away from this singular point, the
two neutral \CP-even mass-eigenstates and the respective mixing angle
$\alpha$ are well defined. On the other hand, if $\delta m$ is below
the experimental resolution, both states can in practice be viewed as
mass-degenerate.  We begin by focusing our degenerate spectrum fits to
the type-I and type-II natural flavor conserving 2HDMs, which we next
compare to the more flexible Yukawa--aligned setup. The type-I setup,
where the second Higgs scalar does not couple to fermions at all, can
be viewed as an effective low-energy realization of the GUT-based
extended Higgs sectors described in
Section~\ref{sec:degenerate-description}.  The parameter regimes
giving the best fit results are summarized in
Tables~\ref{tab:bestfit-degenerate}-\ref{tab:bestfit-degenerate-align}. The
relevant correlations among the model parameters are shown in
Figure~\ref{fig:degenerate-param}.

\begin{table}[t]
\begin{center} \begin{small}
\begin{tabular}{c||r|r||r|r||r|r||r|r} \hline 
&  \multicolumn{4}{c||}{type-I} &  \multicolumn{4}{c}{type-II} \\ \hline
 &  \multicolumn{2}{c||}{unconstrained} &  \multicolumn{2}{c||}{constrained} &  \multicolumn{2}{c||}{unconstrained} &  \multicolumn{2}{c}{constrained} \\ \hline
 & $\tan < 1$ & $\tan\beta > 1$ & $\tan\beta < 1$ & $\tan\beta > 1$  & $\tan < 1$ & $\tan\beta > 1$ & $\tan\beta < 1$ & $\tan\beta > 1$\\ \hline
$\tan \beta$                  &  0.987 & 1.646 & 0.990&  1.568 &  0.570 & 1.921 & 0.836 & 1.325 \\
$\sin \alpha$                 &  -0.139  & -1.000 & -0.783& 0.014 & 0.303 & -0.516 & -0.806 & 0.774\\
$M_H$                         & 364.4   & 545.7 & 646.8 & 622.5  & 200.0 & 226.4 & 608.2 & 570.2\\
$\tilde{\lambda}$             & 0.67   & 1.27 & 1.03&  0.26 & 4.96  & 4.86 & -1.86 & 0.81 \\\hline
$\Delta_V(h)$                 & -0.205 & -0.471& -0.006& -0.165 & -0.791 & -0.002 & -0.002 & -0.960  \\
$\Delta_t(h)$                 &  0.409  &  -0.987& -0.115& 0.186 & 0.924 & -0.034 & -0.077 & -0.207\\
$\Delta_b(h)$                 & 0.409  &  -0.987 & -0.115& 0.186 & -1.348 & 0.118 & 0.051 & -2.286\\
$\Delta_\tau(h)$              & 0.409  &  -0.987& -0.115& 0.186 & -1.348 & 0.118 & 0.051 & -2.286\\
$\Delta_\gamma(h)$            & -0.037  & -0.022& -0.048& -0.042 & 0.133 & 0.097 & -0.060 & 0.115\\
$\Delta_\gamma^\text{tot}(h)$ & -0.405  & -0.362& -0.025&  -0.300 & -0.871 & 0.105 & -0.041 & -0.818\\\hline
$\Delta_V(H)$                 & -0.393 & -1.849& -1.108&  -0.450 & -0.022 & -1.062 & -1.063 & -0.001\\
$\Delta_t(H)$                 & -1.198 &  -2.170& -2.112&  -0.983 & -0.389 & -1.582 & -2.257 & -0.031 \\
$\Delta_b(H)$                 & -1.198 &  -2.170 &-2.112 & -0.983 &  0.097 & 0.855  & -0.229 & 0.052 \\
$\Delta_\tau(H)$              & -1.198  & -2.170& -2.112&  -0.983 & 0.097 & 0.855 & -0.229 & 0.052\\
$\Delta_\gamma(H)$            & -0.028  & 0.040  &0.005& -0.028  & 0.112 & 0.089  & 0.006 & -0.048\\
$\Delta_\gamma^\text{tot}(H)$ & -0.214  & -0.261& -0.849& -0.343  & 0.190 & -0.828 & -0.739 & -0.041 \\\hline
$-2\log \mathcal{L}$         & 27.4  & 26.6& 29.1 &  26.9  & 27.7 & 29.6 & 29.8 & 30.3\\\hline 
\end{tabular}
\end{small} \end{center}
\caption{Best--fit parameter points for the 2HDM with degenerate
  spectrum, considering type-I (left columns) and type-II Yukawa structures (right columns). 
  Both sets of $\Delta_x$ are defined as deviations from the
  Standard Model couplings.}
\label{tab:bestfit-degenerate}
\end{table}

\paragraph{Type-I models} exhibit a rather featureless log-likelihood spanning the
entire $\tan\beta$ vs $\sin\alpha$ plane. Optimal agreement to data is
achieved for two separate strips. One of them corresponds to
$\sin\alpha \simeq 0$ and extends to very large $\tan\beta$ values. It
accommodates large variations in $\Delta_V(\hzero) \sim
\sin(\beta-\alpha)$. Large departures from the decoupling limit
$\sin(\beta-\alpha) \to 1 $ are protected by the unitarity sum rule
$g_{VV\hzero}^2 + g^2_{VV\Hzero} = (g_{VV\PHiggs}^\text{SM})^2$.  In
contrast, the Yukawas barely deviate from the Standard Model, because
the additional Higgs field is by construction fermiophobic,
$g_{ff\Hzero} \sim \sin\alpha/\sin\beta \sim 0$, as long as $\alpha$
is small.

A second favored configuration emerges at $\tan\beta \lesssim 1$ and
either $\sin\alpha \simeq -1$ or $\sin\alpha = 0...0.5$. The
absolute best fit point at maximal mixing $\sin\alpha = -1$ for $\tan\beta > 1$ is no
longer allowed after all the model bounds are imposed.  The
constrained fit instead favors the small mixing angle regions. In the
$\tan\beta < 1$ range the situation is the opposite; the best fit
point mildly shifts towards $\tan\beta \simeq 1$ with a stronger
mixing among the two Higgs doublets, $|\sin\alpha|$.\bigskip

\begin{figure}[t]
\begin{center}
\mbox{
\raisebox{-\height}{\includegraphics[width=0.26\textwidth]{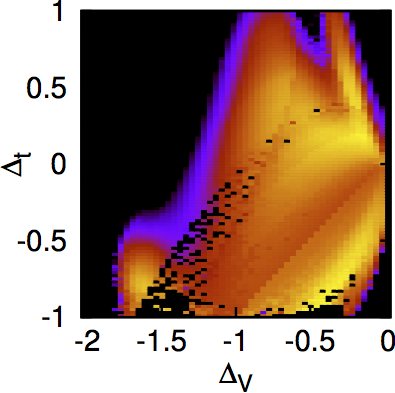}} \qquad
\raisebox{-\height}{\includegraphics[width=0.26\textwidth]{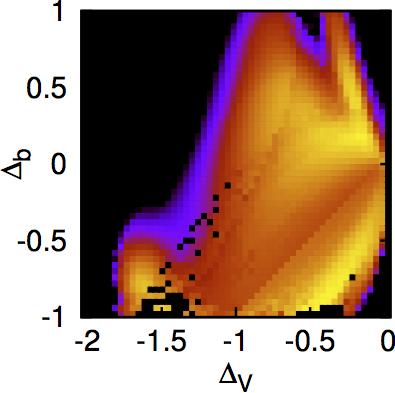}} \qquad
\raisebox{-\height}{\includegraphics[width=0.26\textwidth]{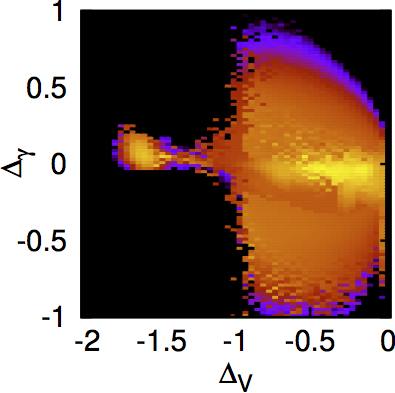}}
\raisebox{-\height}{\includegraphics[width=0.040\textwidth]{./figs_jpg/plot_colorbox.jpg}}}
\mbox{
\raisebox{-\height}{\includegraphics[width=0.26\textwidth]{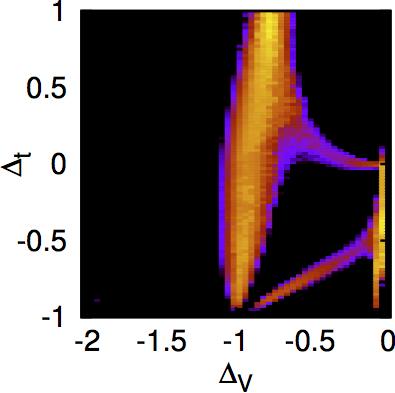}} \qquad
\raisebox{-\height}{\includegraphics[width=0.26\textwidth]{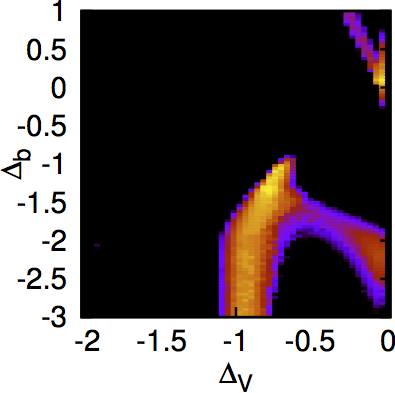}} \qquad
\raisebox{-\height}{\includegraphics[width=0.26\textwidth]{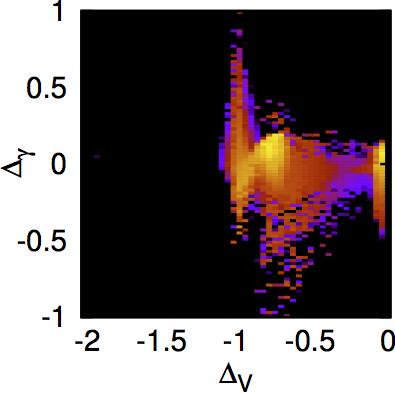}}
\raisebox{-\height}{\includegraphics[width=0.040\textwidth]{./figs_jpg/plot_colorbox.jpg}}}
\end{center}
\caption{Correlated relative log-likelihood $- 2 \Delta \log
  \mathcal{L}$ for different pairs of assuming a type-I 2HDM (top row)
  and type-II 2HDM (bottom row) with degenerate mass
  spectrum.}
\label{fig:degenerate-coup}
\end{figure} 

Again, we can switch from the analysis of the model parameters to the
coupling shifts.  In the upper panels of
Figure~\ref{fig:degenerate-coup} we display the correlated deviations
in $\Delta_V(\hzero)$ vs $\Delta_{t/b/\gamma}(\hzero)$. Systematic
suppressions in the $\hzero$ couplings are balanced by $\Hzero$, both
contributing to the 126~GeV signal. The strongly correlated
variations manifest as a thumb-print-shaped area, and can be tracked
analytically by
\begin{alignat}{5}
(1+\tan^2\beta)\, \Delta_V^2(\hzero) 
+ 2 \left( 1 - \tan^2\beta\,\Delta_{t/b}(\hzero) \right) \, \Delta_V(\hzero)
+ \tan^2\beta \; \Delta_{t/b}^2(\hzero) = 0 \; ,
\label{eq:bestfit-degenerate}
\end{alignat}
which follows from the parameter dependencies of the coupling shifts
$\Delta_{V,t/b}$ and the unitarity sum rules relating $\Delta(\hzero)$
and $\Delta(\Hzero)$~\eqref{eq:sumrules}.  Two limiting cases are
\begin{alignat}{5}
\Delta^2_{V}(\hzero) &\simeq - \frac{\Delta_f^2(\hzero)}{2} 
\qqquad &&\text{for} \; \tan\beta \simeq 1
\notag \\
\Delta_V(\hzero) &\simeq \Delta_f(\hzero)
\qqquad &&\text{for} \; \tan\beta \gg 1 
\; .
\label{eq:limiting-cases}
\end{alignat}
The first relation describes an ellipse in the $\tan\beta$ vs
$\sin\alpha$ plane, whereas the second relation corresponds to the
diagonal line crossing the best fit area. \bigskip

As for the Higgs--photon coupling we see charged Higgs--mediated
contributions extending over a rather symmetric range
$\Delta_\gamma(\hzero) \simeq \pm 0.5$, with best fit configurations
neatly peaking around $\Delta_\gamma(\hzero) \simeq 0$, and well
compatible with large departures from the decoupling limit.

\paragraph{Type-II models} allow for separate shifts of the
top and bottom Yukawa couplings to $\hzero$ and $\Hzero$.  This
typically leaves us with sharper log-likelihood profiles and more
focused statistically preferred regions.  Moreover, they may
give rise to enhanced Yukawas which are severely limited by the LHC
observation, as well as by the flavor constraints.\bigskip

In terms of the model parameters, the best-fit solutions split into
two separate classes, with interpolating areas in between. On the one
hand we identify the usual $\sin^2\alpha \simeq 1/(1+\tan^2\beta)$
pattern following the decoupling condition. The two-fold degeneracy
stems from the sign ambiguity in the bottom and tau Yukawas. The
narrow band around the no--mixing limit $\sin\alpha \simeq 0$ for
$\tan\beta > 1$ would lead to an overly enhanced bottom quark Yukawa
to the additional scalar field, $g_{bb\Hzero} \sim
\cos\alpha/\cos\beta$.  The $|\sin\alpha| \gtrsim 0.5$ areas for
$\tan\beta \gtrsim 5$ would yield exceedingly large rates due to the
combined contributions from both $\hzero,\Hzero$ states.  On the other
hand, we find additional solutions for sizeable neutral Higgs mixing
and away from the decoupling condition.  They correspond to
$|\sin\alpha| > 0.5$ and a more or less constrained $\tan\beta$ range
--- up to $\tan\beta \simeq 10$ for maximal mixing $\sin\alpha = \pm
1$.\bigskip

When viewed as correlated coupling shifts, we recognize the SM-like
profile $\Delta_V \simeq \Delta_f \simeq 0$ over the entire decoupling
regime $\sin^2\alpha \simeq 1/(1 + \tan^2\beta)$. In addition, novel
solutions departing from $\Delta_V \simeq 0$ lead to suppressed top
Yukawas, $\cos\alpha/\sin\beta < 1$.  For large $\tan\beta$ and
$\sin\alpha$, these variations are linearly correlated, as shown in
Eq.\eqref{eq:limiting-cases}.  Similarly, the different best fit
$\sin\alpha$ vs $\tan\beta$ configurations away from the decoupling
limit trigger the multifarious bottom and tau Yukawa patterns
displayed in Figure~\ref{fig:degenerate-coup}, which include enhanced,
suppressed and sign-flipped interactions.

Solutions with a vector-phobic $\hzero$ field $\Delta_V(\hzero) \simeq
-1$ or $\alpha \simeq \beta$ cover a wide range of coupling
variations. They include maximal mixing solutions $\sin\alpha \simeq
\pm 1$ for $\tan\beta \gg 1$ and no--mixing solutions $\sin\alpha
\simeq 0$ for $\tan\beta \ll 1$.  The former is characterized by
enhanced (sign-reverted) bottom Yukawas with suppressed top
interactions. Conversely, no--mixing solutions feature
$\Delta_b(\hzero) \simeq -1$ with enhanced top Yukawas.  All of them
are nicely visible in the lower panels of
Figure~\ref{fig:degenerate-coup}.

The black voids in the $\Delta_V$ vs $\Delta_t$ and $\Delta_V$ vs
$\Delta_b$ planes correspond to exceedingly large combined
$\hzero,\Hzero$ contributions to the relevant search channels.  This
is particularly apparent at moderate and large $\tan\beta$ for
$|\sin\alpha| \gtrsim 0.5$. This implies enhanced bottom and tau
Yukawas, a type-II structure hard to reconcile with the LHC
observation.\bigskip

Finally, for type-II models the best fit solutions at large
$\sin\alpha$ survive after imposing the theoretical and experimental
constraints. All we observe is a slight pull towards $\tan\beta \simeq
1$. The poor fit quality reflects the strong competition between the
model constraints and the compatibility with the LHC measurements.
The bulk of the phenomenologically viable parameter space fails to
accommodate two roughly 126~GeV mass-degenerate states with SM-like
couplings measurements.  For example, the best fit for the constrained
type-II model involves $\Delta_b \simeq -2.3 $, \ie a
$\mathcal{O}(30)\%$ enhancement in the Yukawas
$g_{bb\hzero/\tau\tau\hzero}$. This universal enhancement is
reinforced by the contribution from the $\Hzero$ state with SM-like
Yukawa interactions. In the view of LHC data, a type-I twin Higgs
model is clearly more likely.

\begin{table}[t]
\begin{center} \begin{small}
\begin{tabular}{c||r|r||r|r} \hline 
&  \multicolumn{4}{c}{Yukawa alignment} \\ \hline
 &  \multicolumn{2}{c||}{unconstrained} &  \multicolumn{2}{c}{constrained}  \\ \hline
 & $\tan < 1$ & $\tan\beta > 1$ & $\tan\beta < 1$ & $\tan\beta > 1$  \\ \hline
$\tan \beta$                  &  0.404 & 1.651 & 0.988&  1.509\\
$\sin \alpha$                 &  0.023  & 0.148 & -0.961& -0.986 \\
$\gamma_b$                    &  0.149  & 0.275 & 0.218 & 0.752 \\
$\gamma_\tau$                 &  0.959 & 0.137 & 0.180 & 0.120 \\
$M_H$                         & 425.6   & 390.2 & 661.7 & 397.0  \\
$\tilde{\lambda}$             & -1.35   & 3.79 & -0.60&   2.49 \\\hline
$\Delta_V(h)$                 & -0.647 & -0.231& -0.122&  -0.315  \\
$\Delta_t(h)$                 &  1.669  &  0.156& -0.606&  -0.797\\
$\Delta_b(h)$                 & -0.067  &  0.307 & -0.440&  -0.814\\
$\Delta_\tau(h)$              & -1.350 &  -0.336& -0.298&  -0.145\\
$\Delta_\gamma(h)$            & -0.039  & -0.023& -0.047&  -0.014\\
$\Delta_\gamma^\text{tot}(h)$ & -0.685  & -0.358& -0.042&  -0.203\\\hline
$\Delta_V(H)$                 & -0.064 & -0.361& -1.478&   -1.728\\
$\Delta_t(H)$                 & -0.938 &  -0.827& -2.367&  -2.182 \\
$\Delta_b(H)$                 & -0.283 &  -1.009 &-2.063  & -2.198 \\
$\Delta_\tau(H)$              & 0.201  & -0.235& -1.802&  -1.569\\
$\Delta_\gamma(H)$            & 0.115  & 0.004  &0.025&  0.027\\
$\Delta_\gamma^\text{tot}(H)$ & 0.023  & -0.236& -0.773&  -0.407  \\\hline
$-2\log \mathcal{L}$         & 25.7  & 26.0& 26.8 & 26.1   \\\hline 
\end{tabular}
\end{small} \end{center}
\caption{Best--fit parameter points for the 2HDM with degenerate
  spectrum, considering Yukawa-aligned Higgs couplings to fermions. 
  Both sets of $\Delta_x$ are defined as deviations from the
  Standard Model couplings.}
\label{tab:bestfit-degenerate-align}
\end{table}

\paragraph{Yukawa--alignment} completely disentangles the variations in the
heavy fermion Yukawas. The corresponding best fit results are given in
Table~\ref{tab:bestfit-degenerate-align}.  As expected for a 2HDM
setup with very large flexibility, we find wide ranges of
$\sin\alpha$, $\tan\beta$ and $\gamma_{b,\tau}$ to be compatible with
data.  The best agreement singles out $\tan\beta \simeq 1$ alongside
with large mixing $|\sin\alpha| \gtrsim 0.7$, Yukawa structures not
far from type-I $\gamma_b/2\pi = 0.25(0.75)$, and small charged Higgs
contributions.  The manifold coupling variations identified for type-I
and type-II models, including sizable departures from $\Delta_V \simeq
0$ are recovered in this more general setup. In addition, we find
broader parameter regimes with coexisting $\Delta_V(\hzero) < 0$ and
$\Delta_f(\hzero) > 0$. This reflects how Yukawa--aligned structures
unlink the fermion and the gauge boson coupling shifts.

\section{Summary}
\label{sec:summary}

In this paper we have studied extended Higgs sectors in the light of
the 2011--2012 ATLAS and CMS measurements. Starting from the simple
modifications of adding a dark singlet, not participating in
electroweak symmetry breaking, we have looked at increasingly complex
models.\bigskip

\begin{figure}[b!]
\mbox{
\raisebox{-\height}{\includegraphics[width=0.33\textwidth]{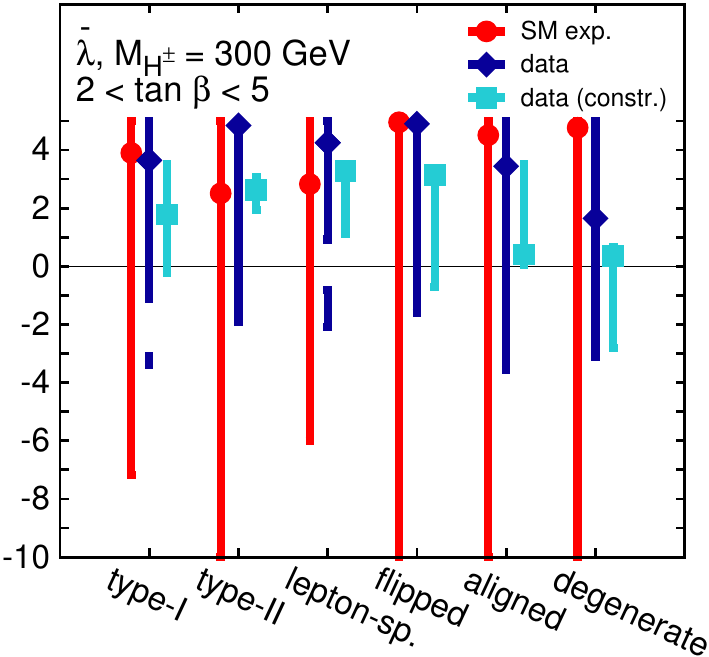}}
\raisebox{-\height}{\includegraphics[width=0.33\textwidth]{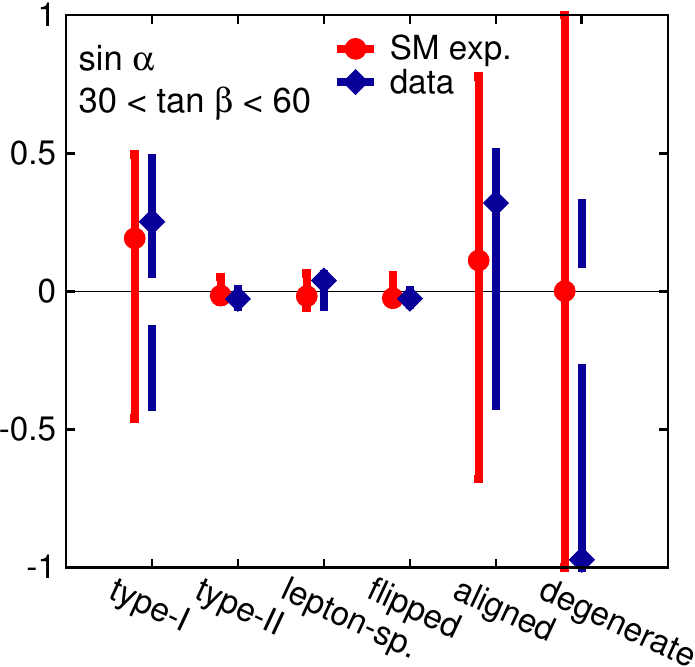}}
\raisebox{-1.01\height}{\includegraphics[width=0.335\textwidth]{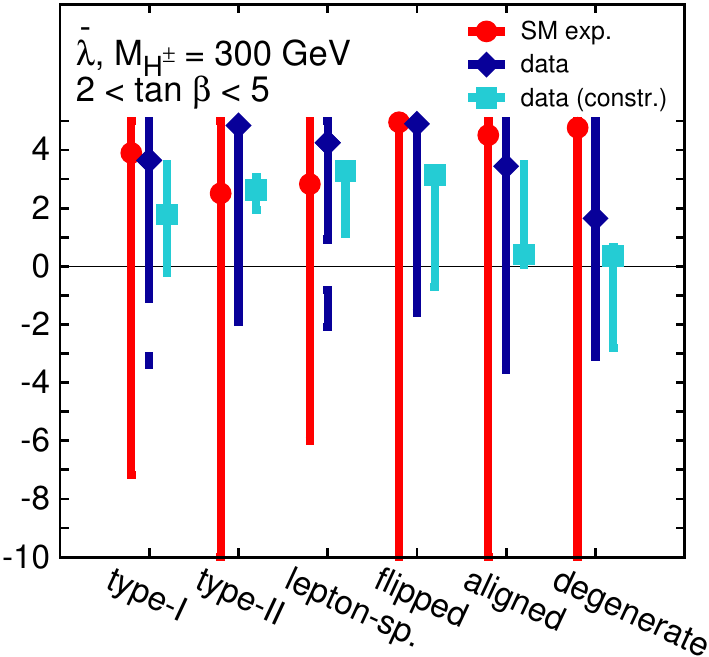}}}
\vspace*{-1ex}
\caption{Extracted $\sin\alpha$ and $\tilde{\lambda}$ values and error
  bands for the different 2HDM benchmarks.  The self couplings
  $\tilde{\lambda}$ is defined in Eq.\eqref{eq:ltilde}.  The variation
  of $\tan\beta$ is restricted to small (left and right panels) and
  large (center panel) values.}
\label{fig:higgserrors}
\end{figure} 

\begin{figure}[t]
\begin{center}
\includegraphics[width=0.50\textwidth]{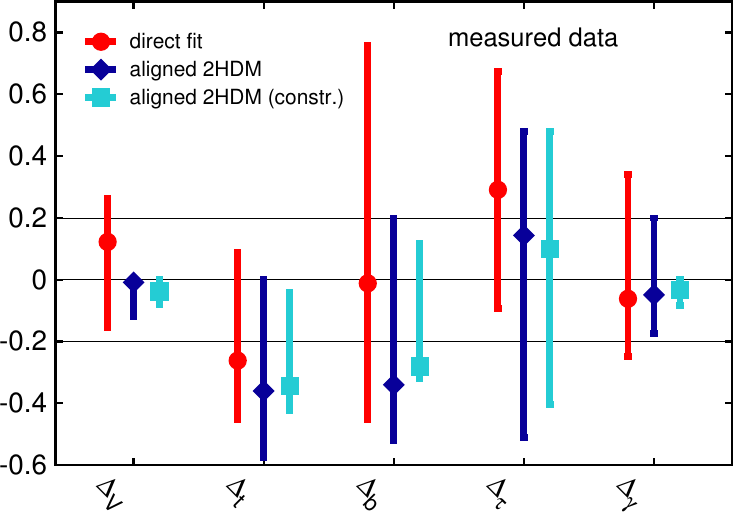} 
\end{center}
\vspace*{-5mm}
\caption{Extracted couplings and errors bars for the general fit of
  all Standard Model couplings and the Yukawa--aligned 2HDM fit. For
  the 2HDM we show the error bars without and with additional non-LHC
  constraints.}
\label{fig:deltas-aligned}
\end{figure}

As long as all our models include a decoupling limit we first confirm
that all models are consistent with the observed data. The preferred
parameter ranges depend on the model structure, in particular the
different Yukawa couplings in type-I, type-II, lepton--specific, and
flipped setups. In general, we find strong correlations between the
VEV ratio $\tan \beta$ and the Higgs mixing parameters $\xi$ or $\sin
\alpha$.  In Figure~\ref{fig:higgserrors} we show the corresponding
best--fit parameter points and error bars for some of the model
parameters. The red curves show the expected limits from current data,
assuming all LHC measurements right on the Standard Model parameter
choices. The dark blue results from the actual data are fully
consistent with the Standard Model limit, both in terms of the central
values and in terms of the error bars. The results shown in light blue
include the theoretical and phenomenological constraints on the
respective models. They can be significantly different and most
notably do not feature in the large-$\tan\beta$ panel, illustrating
the considerable problems this parameter range has for example with
flavor physics constraints.

The type-I 2HDM most naturally fits the observed Higgs rates, so it
has the widest error bands.  Segments close to the no-mixing point are
excluded for those models where they overly suppress the bottom or tau
Yukawas. For large $\tan\beta$ the decoupling condition
$\sin(\beta-\alpha) \simeq 1$ strongly limits the allowed amount of
mixing.  Sign shifts in couplings are only attainable for $\tan\beta <
1$. Particularly interesting parameter regimes open when we allow for
two mass-degenerate Higgs bosons to contribute to the observed
resonance. The results shown in Figure~\ref{fig:higgserrors} confirm
our general conclusion that extended Higgs sectors are significantly
constrained by current data, unless their structure is complex enough
to resemble the full Standard Model Higgs Lagrangian in terms of
degrees of freedom.\bigskip

In addition, we target the question if the most general,
Yukawa--aligned 2HDM can serve as a consistent model for a fully
flexible fit of all Standard Model Higgs couplings. To begin with,
deviations $\Delta_W \ne \Delta_Z$ and $\Delta_{W,Z} > 0$ can only be
realized when we include quantum effects, as described in
Section~\ref{sec:quantum_higgs} and \ref{sec:quantum_fermion}. In our
current LHC fit we omit quantum corrections, because they are
CPU-intensive and carry no numerical relevance; all we need to
remember is that $\Delta_W \ne \Delta_Z$ as well as $\Delta_{W,Z} >0$
can be achieved in principle.  In the custodial limit $\Delta_W =
\Delta_Z$ we compare the free SM-like couplings fit and the
Yukawa--aligned 2HDM fit in Figure~\ref{fig:deltas-aligned}. The error
bars of the actual Higgs couplings are indeed similar, as long as we
ignore the experimental constraints on the 2HDM setup.  The
tree--level assumption $\Delta_V < 1$ explains the comparably larger
downward shifts in the Higgs couplings to fermions, which compensate
for the suppressed observed di-boson rates.

This numerical outcome confirms that once we fold in quantum effects
we should indeed be able to structurally use an extended Higgs sector
as the ultraviolet completion of a SM-like Higgs sector with free
couplings. Obviously, this correspondence holds only when we ignore
the model-specific non--LHC constraints for the 2HDM structure.  In
such a model electroweak quantum corrections can be computed for LHC
and ILC observables and applied to the Higgs coupling measurement
based on the renormalizable dimension-4 Lagrangian.


\acknowledgments First of all, we would like to thank Berthold Stech
for suggesting the degenerate spectrum setup and for the fruitful and
enjoyable collaboration on this major part of our analysis.  It is a
pleasure to thank Dirk and Peter Zerwas for their help all along this
analysis. Moreover, we would like to thank our \textsc{SFitter}
co-authors Markus Klute and Remi Lafaye for a fun, fruitful, and
ongoing collaboration.  TP would like to thank the CCPP at New York
University for their hospitality while this paper was finalized.  MR
acknowledges partial support by the Deutsche Forschungsgemeinschaft
via the Sonderforschungsbereich/Transregio SFB/TR-9 ``Computational
Particle Physics'' and the Initiative and Networking Fund of the
Helmholtz Association, contract HA-101 (``Physics at the Terascale'').

\appendix

\section{Parameterizations}
\label{app:para}

\subsection{Additional singlet}

Starting from the scalar potential defined by
Eq.\eqref{eq:singlet-potential} and expanding the neutral component of
the doublet field around its VEV $\Phi = (v + h_1^0)/\sqrt{2}$ we
arrive at the mass-squared matrix
\begin{equation}
\left( \begin{array}{cc}  
      2\lambda_1\,v_1^2 & \lambda_3\,v_1\,v_2 \\
      \lambda_3\,v_1\,v_2 & 2\lambda_2\,v_2^2  
     \end{array} \right) \; .
\label{eq:nd-mass-matrix}
\end{equation}
Diagonalization in terms of $v_1 \equiv v = 246$~GeV and $v_2 =
v\tan\beta$ returns the physical masses
\begin{alignat}{5}
\frac{m_{\hzero,\Hzero}^2}{v^2} = \lambda_1 + \lambda_2\tan^2\beta \mp (\lambda_1 -
\lambda_2\tan^2\beta)\,\sqrt{1+\tan^2\,2\theta} \; ,
\label{eq:nd-spectrum}
\end{alignat}
and the mixing between the singlet and the doublet Higgs components
\begin{equation}
 R(\theta) = \left( \begin{array}{cc}  
      \cos\theta & \sin\theta \\
      -\sin\theta & \cos\theta  
     \end{array} \right) \qqquad 
\tan^2 (2\theta) = \cfrac{\lambda_3^2 v_1^2 v_2^2}{(\lambda_1 v_1^2-\lambda_2 v^2_2)^2} \; ,
\label{eq:nd-rotation}
\end{equation}
We can invert these relations for the coupling parameters
$\lambda_{1,2,3}$ as
\begin{equation}
 \lambda_1 = \cfrac{m^2_{\hzero}\,\cos^2\theta + m^2_{\Hzero}\,\sin^2\theta }{2v^2}; 
\quad \lambda_2 = \cfrac{m^2_{\hzero}\,\sin^2\theta + m^2_{\Hzero}\,\cos^2\theta }{2v^2\tan^2\beta};
\quad \lambda_3 = \cfrac{\sin 2\theta}{\tan\beta}\,\frac{(m^2_{\Hzero}-m^2_{\hzero})}{2v^2}\; .
\label{eq:nd-quartics}
\end{equation}
The singlet mixing reduces the coupling strength of the $h^0$ to all
the Standard Model fields by $\cos \theta$.  If kinematically allowed,
the heavier scalar $\Hzero$ may decay into $\hzero\hzero$ pairs at a
rate
\begin{equation}
 \Gamma(\Hzero \to \hzero\hzero) = \frac{|\lambda_{\Hzero\hzero\hzero}|^2}{32\pi m_{\Hzero}}\,\sqrt{1-\frac{4 m^2_{\hzero}}{m^2_{\Hzero}}}
\label{eq:singlet-width}.
\end{equation}
In the absence of a second VEV the dark singlet does not contribute to
electroweak symmetry breaking and hence does not mix with the SM-like
Higgs doublet. Notwithstanding, the portal interaction
\begin{alignat}{5}
 \lag \supset \lambda_3\,(\Phi^\dagger\,\Phi)\,S^2 \label{eq:portal-interaction},
\end{alignat}
triggers a new $hss$ triple coupling.  If kinematically allowed, the
corresponding decay mode
\begin{alignat}{5}
 \Gamma(h \to ss) &= \frac{\lambda^2_3\,v^2}{32\,\pi\,m_h}\,
  \sqrt{ 1-\cfrac{4m_s^2}{m_h^2}} 
\label{eq:invwidth-appendix}
\end{alignat}
will contribute to the invisible Higgs width.

\subsection{Additional doublet}

The 2HDM potential we refer to is given in
Eq.\eqref{eq:2hdmpotential}. The Higgs mass-eigenstates require a set
of rotations
\begin{equation}
\left(\begin{array}{c} \Hzero \\ \hzero \end{array} \right) = R(\alpha)\,\left(\begin{array}{c} h^0_1 \\ h^0_2 \end{array} \right)
\qqquad
\left(\begin{array}{c} G^0 \\ \Azero \end{array} \right) = R(\beta)\,\left(\begin{array}{c} a^0_1 \\ a^0_2 \end{array} \right)
\qqquad
\left(\begin{array}{c} G^\pm \\ \PHiggs^\pm \end{array} \right) = R(\beta)\,\left(\begin{array}{c} h^\pm_1 \\ h^\pm_2 \end{array} \right) \; ,
\end{equation}
with $R(\theta)$ defined as a rotation matrix with a mixing angle
$\theta$.  The VEVs $v_{1,2}$ associated with each of the doublets we
write as $v_1 \equiv v\,\cb, v_2 \equiv v\,\sb$, such that $\tan\beta
\equiv v_2/v_1$. Electroweak symmetry breaking requires $v_1^2 + v_2^2
= v^2$.\bigskip

The mass parameters in the minimized
Higgs potential Eq.\eqref{eq:2hdmpotential} are
\begin{alignat}{5}
 m^2_{11} &= m^2_{12}\,\tanb - \frac{v^2}{2}\,(\lambda_1\cbd + (\lambda_3+\lambda_4+\lambda_5)\,\sbd 
+ 3\lambda_6\,\sin\beta\cos\beta + \lambda_7\sin^2\beta\tan\beta)
 \notag \\
 m^2_{22} &= m^2_{12}\,\cotb - \frac{v^2}{2}\,(\lambda_2\sbd + (\lambda_3+\lambda_4+\lambda_5)\,\cbd
+ 3\lambda_7\,\sin\beta\cos\beta + \lambda_6\cos^3\beta\sin\beta) \; .
\label{eq:massterms}
\end{alignat}
The corresponding self-couplings read
\begin{alignat}{5}
 \lambda_1 &= \frac{1}{v^2\,\cos^2\beta}\,\left[- m_{12}^2 \frac{\sb}{\cb} + \mhhd\,\cad + \mlhd\sad \right] \notag \\
 \lambda_2 &= \frac{1}{v^2\,\sin^2\beta}\,\left[- m_{12}^2 \frac{\cb}{\sb} + \mhhd\,\sad + \mlhd\cad \right] \notag \\
\lambda_3 &= \frac{1}{v^2}\,\left[-\frac{m_{12}^2}{\sb\cb} + 2\mhpd + (\mhhd-\mlhd)\,\frac{\sin2\alpha}{\sin2\beta}\right] \notag \\
\lambda_4 &= \frac{1}{v^2}\,\left[\frac{m_{12}^2}{\sb\cb} + \mad - 2\,\mhpd \right]; \notag \\
\lambda_5 &= \frac{1}{v^2}\,\left(\frac{m_{12}^2}{\sb\cb}-\mad \right) - \lambda_6\,\cot\beta - \lambda_7\,\tan\beta \; .
\label{eq:lambdas}
\end{alignat}
For the physical masses of the five Higgs states we find
\begin{alignat}{5}
 \mhpd &= \frac{m_{12}^2}{\sb\cb} - \frac{v^2}{2}\,\left[\lambda_4 + \lambda_5 + \lambda_6\,\cotb + \lambda_7\tanb \right] \notag \\
\mad &= \frac{m_{12}^2}{\sb\cb}-\frac{v^2}{2}\,\left[2\lambda_5 + \lambda_6\,\cotb + \lambda_7\,\tanb \right] \notag \\
m^2_{\hzero,\Hzero} &= \frac{1}{2}\,\left[A^2 + B^2 \mp \sqrt{(A^2-B^2)^2 + 4C^4}\right] \; ,
\label{eq:2hdm-masses}
\end{alignat}
with $A,B,C$ defined as 
\begin{alignat}{5}
 A^2 &= v^2\,\left[ \lambda_1\,\cbd + \frac{3}{2}\,\tanb\,\cbd\,\lambda_6- \,\tan\beta\sin^2\beta\,\lambda_7\right]
+ m_{12}^2 \; \frac{\sb}{\cb} \notag \\
B^2 &= v^2\,\left[ \lambda_2\,\sbd + \frac{3}{2}\,\cotb\,\sin^2\beta \lambda_7- \,\cot\beta\cos^2\beta\,\lambda_6\right]
+ m_{12}^2 \; \frac{\cb}{\sb} \notag \\ 
C^2 &= v^2\,\left[\sb\cb\,(\lambda_3+\lambda_4+\lambda_5)\, + \frac{3}{2}\,(\lambda_6\,\cbd + \lambda_7\,\sbd) \right] - m_{12}^2 \; .
\label{eq:auxmasses}
\end{alignat}
\bigskip

\begin{table}[t]
\begin{center} \begin{small}
\renewcommand{\arraystretch}{2.0}
\begin{tabular}{l|c|c|c} \hline
 & \multicolumn{1}{c|}{$\hzero$} & \multicolumn{1}{c|}{$\Hzero$}& \multicolumn{1}{c}{$\Azero$} \\ \hline
$1 + \Delta_{\PW}$ & $\sin(\beta-\alpha)$  & $\cos(\beta-\alpha)$ & 0 \\
$1 + \Delta_{\PZ}$ & $\sin(\beta-\alpha)$ & $\cos(\beta-\alpha)$  & 0  \\ \hline
$1 + \Delta_t$ & $\dfrac{\ca}{\sb}$ &  $\dfrac{\sa}{\sb}$ & $\dfrac{1}{\tan\beta}$ \\
$1 + \Delta_b$ & $-\dfrac{\sin(\alpha-\gamma_b)}{\cos(\beta-\gamma_b)}$ & $\dfrac{\cos(\alpha-\gamma_b)}{\cos(\beta-\gamma_b)}$ & $\tan(\beta-\gamma_b)$ \\
$1 + \Delta_{\tau}$ & $-\dfrac{\sin(\alpha-\gamma_\tau)}{\cos(\beta-\gamma_\tau)}$ & $\dfrac{\cos(\alpha-\gamma_\tau)}{\cos(\beta-\gamma_\tau)}$ & $\tan(\beta-\gamma_\tau)$ \\
\hline 
\end{tabular}
\end{small} \end{center}
\caption{Neutral Higgs boson couplings to fermions and gauge bosons
  within a generic 2HDM. The flavor sector herewith we define
  according to the Yukawa alignment hypothesis.}
\label{tab:2hdmcouplings1}
\end{table}

The general form of the couplings quoted in
Table~\ref{tab:2hdmcouplings1} with aligned Yukawa structures may be
regarded as an interpolation between the canonical realizations of the
2HDM, \ie featuring the natural flavor conservation hypothesis with
the couplings modifications quoted in
Table~\ref{tab:2hdmcouplings2}. The couplings and expressed in terms
of $\Delta_x$, defined by Eq.\eqref{eq:delta}.\bigskip

\begin{table}[t]
\begin{center} \begin{small}
\renewcommand{\arraystretch}{2.0}
\begin{tabular}{l|rrrr}
\hline
 & type-I & type-II & lepton--specific & flipped \\[-1em]
 & $\gamma_b  = \pi/2$ & $\gamma_b = 0$ & $\gamma_b  = \pi/2$ & $\gamma_b = 0$  \\[-1em] 
 & $\gamma_\tau = \pi/2$ & $\gamma_\tau = 0$ & $\gamma_\tau =0$ & $\gamma_\tau = \pi/2$  \\  \hline 
$1 + \Delta_t(\hzero)$ & $\dfrac{\cos\alpha}{\sin\beta}$& $\dfrac{\cos\alpha}{\sin\beta}$ & $\dfrac{\cos\alpha}{\sin\beta}$ & $\dfrac{\cos\alpha}{\sin\beta}$  \\ 
$1 + \Delta_b(\hzero)$ & $\dfrac{\cos\alpha}{\sin\beta}$ &  $-\dfrac{\sin\alpha}{\cos\beta}$ & $\dfrac{\cos\alpha}{\sin\beta}$ & $-\dfrac{\sin\alpha}{\cos\beta}$ 
           \\ 
$1 + \Delta_\tau(\hzero)$ & $\dfrac{\cos\alpha}{\sin\beta}$ &  $-\dfrac{\sin\alpha}{\cos\beta}$ & $-\dfrac{\sin\alpha}{\cos\beta}$ & $\dfrac{\cos\alpha}{\sin\beta}$ 
          \\ \hline
$1 + \Delta_t(\Hzero)$ & $\dfrac{\sin\alpha}{\sin\beta}$& $\dfrac{\sin\alpha}{\sin\beta}$ & $\dfrac{\sin\alpha}{\sin\beta}$ & $\dfrac{\sin\alpha}{\sin\beta}$  \\ 
$1 + \Delta_b(\Hzero)$ & $\dfrac{\sin\alpha}{\sin\beta}$ & $\dfrac{\cos\alpha}{\cos\beta}$ & $\dfrac{\sin\alpha}{\sin\beta}$ & $\dfrac{\cos\alpha}{\cos\beta}$ 
          \\ 
$1 + \Delta_\tau(\Hzero)$ & $\dfrac{\sin\alpha}{\sin\beta}$ &  $\dfrac{\cos\alpha}{\cos\beta}$ & $\dfrac{\cos\alpha}{\cos\beta}$ & $\dfrac{\sin\alpha}{\sin\beta}$ 
          \\ \hline
$1 + \Delta_t(\Azero)$ & $\cot\beta$ & $\cot\beta$& $\cot\beta$& $\cot\beta$\\[-1em]
$1 + \Delta_b(\Azero)$ & $-\cot\beta$ & $\tan\beta$ & $-\cot\beta$ & $\tan\beta$ \\[-1em]
$1 + \Delta_\tau(\Azero)$ & $-\cot\beta$ & $\tan\beta$ & $\tan\beta$&$-\cot\beta$ \\ \hline
\end{tabular}
\end{small} \end{center}
\caption{Neutral Higgs boson couplings to fermions for the canonical realizations of the 2HDM
featuring Natural Flavor Conservation.}
\label{tab:2hdmcouplings2}
\end{table}

Finally, we need analytic expressions for the effective dimension 5
interactions. The light Higgs boson coupling to photons can be cast
into
\begin{alignat}{5}
\Delta_\gamma = 
 - \dfrac{m_W\,\sw}{e\,m^2_{\PHiggs^{\pm}}}\,
   \lambda_{h\PHiggs^+\PHiggs^-} \; A_s(\tau_{\PHiggs^{\pm}})
\label{eq:phot1};
\end{alignat}
The usual loop function for a scalar contributing to this coupling is
\begin{alignat}{5}
A^H_s(x) & = \frac{\arcsin^2 \sqrt{x} -x}{x^2}  
\label{eq:loopf}
\end{alignat}
in the low energy limit.  The relevant trilinear interaction involving
the light Higgs boson and the charged Higgs fields renders
\begin{alignat}{5}
 \lambda_{\hzero\PHiggs^+\PHiggs^-} &= \dfrac{e}{2m_W\sw}\,\left[
\sin(\beta-\alpha)(m^2_{\hzero}-2m^2_{\PHiggs^{\pm}}) - \cfrac{\cos(\alpha+\beta)}{\sin (2\beta)}\,
(2m_{\hzero}^2 - \tilde{\lambda} v^2)
\right] \; ,
\label{eq:3h}
\end{alignat}
with $\tilde{\lambda} = 2m^2_{12}/(v^2\sin\beta\cos\beta)$.
If we also entertain the possibility of the heavy neutral
Higgs contributing to the overall signal strength we need
to extract the corresponding coupling strength including
the trilinear coupling 
\begin{alignat}{5}
 \lambda_{\Hzero\PHiggs^+\PHiggs^-} &= \dfrac{e}{2m_W\sw}\,\left[
\cos(\beta-\alpha)(m^2_{\Hzero} - 2m^2_{\PHiggs^{\pm}})
-\cfrac{\sin(\alpha+\beta)}{\sin (2\beta)}\,(2m_{\Hzero}^2- \tilde{\lambda} \,v^2)
\right]
\label{eq:3hheavy-full}. 
\end{alignat}
\bigskip
%

The MSSM Higgs sector is a constrained 2HDM setup. Its main feature is
the link between the Higgs masses and the Higgs couplings through the
mixing angle $\alpha$. The quartic couplings from the superpotential
$D$-terms are gauge couplings, giving us
\begin{equation}
\lambda_1 = \lambda_2 = \cfrac{\pi\alpha_{em}}{\sw\cw}
\qqquad 
\lambda_3 = \cfrac{\pi\alpha_{em}(\cwd-\swd)}{\swd\cwd}
\qqquad 
\lambda_4 = -\cfrac{2\pi\alpha_{em}}{\swd}
\qqquad  
\lambda_5 = 0 \; .
\label{eq:susylimit}
\end{equation}
At tree level the entire Higgs sector is determined by two parameters,
often chosen as $m_{\Azero}$ and $\tan\beta$. Typically, the MSSM
Higgs sector shows a mass hierarchy $m_{\hzero} \ll
m_{\Hzero,\Azero,\PHiggs^\pm}$, in line with the generic hierarchical
benchmark.  Custodial symmetry and tree-level FCNC
suppression are guaranteed.  Dominant quantum corrections are due to
third generation quarks and squarks. They rely on the corresponding
soft-SUSY breaking parameters ($A_{t,b}, \mu, M_\text{SUSY}$) and have
a sizable quantitative impact into the resulting Higgs mass spectrum
and coupling pattern~\cite{Heinemeyer:2004ms}.

\section{Hierarchical 2HDM}
\label{app:hierarchy}

The low-energy realization of the 2HDM with a large mass hierarchy
corresponds to the case with one SM-like, weak-scale Higgs doublet,
and a second heavier field with $M_\text{heavy} \gg v$.  The
underlying EW symmetry requires that scalar mass splittings within
each of the doublets cannot be larger than $\mathcal{O}(v)$ for a
weakly-coupled theory, which gives rise to the purported mass
hierarchy between the light Higgs $m_{\hzero} = \mathcal{O}(v)$ and
its heavier companions $m_{\Hzero,\PHiggs^{\pm},\Azero} \simeq
\mathcal{O}(M_\text{heavy})$.  If these are integrated out, we are
left with an effective field theory description in terms of a single,
SM-like Higgs field.\bigskip

To construct this effective theory~\cite{Gunion:2005ja,Randall:2007as}
we start from the generic relations between the Higgs self-couplings,
masses and mixing angles Eqs.\eqref{eq:massterms} and
\eqref{eq:lambdas},
\begin{alignat}{5}
(\lambda_1\cos^2\beta - \lambda_2\sin^2\beta)\,v^2 
=& M_\text{heavy}^2\cos\, (2\beta) + (m^2_{\Hzero} - m^2_{\hzero})\,\cos (2\alpha) \notag \\
(\lambda_3+\lambda_4 + \lambda_5)\,v^2 
=& M_\text{heavy}^2 
  + (m^2_{\Hzero} - m^2_{\hzero})\, \dfrac{\sin (2\alpha)}{\sin (2\beta)} 
  -\lambda_6\cot\beta - \lambda_7 \tan\beta \; .
\label{eq:comb2}
\end{alignat}
For simplicity, we impose $\lambda_{6,7} = 0$ and define
\begin{alignat}{5}
\hat\lambda v^2 & = (m^2_{\Hzero} - m^2_{\hzero})\sin(\beta - \alpha)\,\cos(\beta - \alpha) \notag  \\
\hat{\lambda} &= \sin\beta\,\cos\beta\,(\lambda_1\,\cos^2\beta - \lambda_2\,\sin^2\beta - \lambda_{345}\,\cos (2\beta)) + \lambda_6\,\cos\beta\cos 3\beta
+ \lambda_7\,\sin\beta\sin 3\beta \; ,
\label{eq:comb3}
\end{alignat}
with $\lambda_{345} \equiv \lambda_3 + \lambda_4 + \lambda_5$. 
The small parameter in the scale separation
$m^2_{\Hzero,\Azero,\PHiggs^{\pm}} \simeq M_\text{heavy}^2 \gg m^2_{\hzero}$ can
therefore be defined as 
\begin{alignat}{5}
\xi = \hat{\lambda} \; \frac{v^2}{M_\text{heavy}^2} \ll 1 \; .
\end{alignat}
The fact that $\hat{\lambda}$ remains small rests on the assumption
that the different Higgs self-couplings remain perturbative
$\mathcal{O}(1)$ quantities.  We are then left with the condition
\begin{alignat}{5}
 \sin(\beta - \alpha)\,\cos(\beta - \alpha) \simeq \xi \ll 1 \; .
\label{eq:comb4}
\end{alignat}
By construction, the heavy Higgs mass follows $M_\text{heavy} \sim
v^2/\xi$. For a hierarchical spectrum we need to assume a
non-vanishing PQ soft-breaking term $m^2_{12} \neq 0$ for a stable
Higgs potential if $\lambda_{6,7} = 0$~\cite{Gunion:2002zf}.\bigskip

If the low-energy effective field theory description of the 2HDM
should approach the SM-like limit $\sin(\beta-\alpha) \to 1$,
%
%
this limit requires
\begin{equation}
 \cos(\beta-\alpha)\simeq \xi \ll 1 
\qquad \Leftrightarrow \qquad 
\cos\beta \sim \sin\alpha \; .
\label{eq:correl}
 \end{equation}
The correlation of the mixing angle is a key property of the
hierarchical 2HDM.  In the limit of large $\tan \beta$ we can derive
the leading correlation between $\xi$ and $\tan \beta$,
\begin{alignat}{5}
\sin^2 \alpha = \frac{1}{1+ \tan^2 \beta} 
\qqquad \text{or} \qqquad
\xi = \dfrac{2 \tan \beta}{1 + \tan^2 \beta} \; .
\label{eq:xi_tgb}
\end{alignat}
\bigskip

The relation $\cos(\beta - \alpha) = \xi \ll 1$ allows us to write the
scale separation $v^2 \ll M_\text{heavy}^2$ in terms of 2HDM
parameters.  The effective theory can be implemented by imposing
Eq.\eqref{eq:correl} on the conventional analytic expressions for the
2HDM couplings.  For example, the type-II bottom Yukawa can be written
as
\begin{alignat}{5}
g_b 
&= -\dfrac{\sin\alpha}{\cos\beta}
 = \sin(\beta-\alpha) - \tan\beta\cos(\beta-\alpha) \notag \\
&= 1-\tan\beta\,\cos(\beta-\alpha) - \dfrac{1}{2}\,\cos^2(\beta-\alpha) + \mathcal{O}(\cos^3(\beta-\alpha)) \notag  \\
& = 1 - \tan\beta\xi - \dfrac{1}{2}\,\xi^2 + \mathcal{O}(\xi^3) \; .
\label{eq:cbb}
\end{alignat}
The complete set of effective light Higgs boson interactions to
leading order in $\xi$ (up to $\mathcal{O}(\xi)^3$ corrections we
present in Table~\ref{tab:eff-coup1}, again written in terms of the
coupling shifts $\Delta_x$ defined in Eq.\eqref{eq:delta}.\bigskip

\begin{table}[t]
\begin{center} \begin{small}
\renewcommand{\arraystretch}{2.0}
\begin{tabular}{l|l|l} \hline
 & \multicolumn{1}{c|}{$\hzero$} & \multicolumn{1}{c}{$\Hzero$} \\ \hline
$1 + \Delta_W$ & $1-\dfrac{\xi^2}{2}$  & $\xi$ \\
$1 + \Delta_Z$ & $1-\dfrac{\xi^2}{2}$ & $\xi$ \\ \hline
$1 + \Delta_t$ & $ 1 + \cot\beta\xi - \dfrac{\xi^2}{2} + \mathcal{O}(\xi^3)$ & $-\cot\beta + \xi + \cot\beta \dfrac{\xi^2}{2}+ \mathcal{O}(\xi^3)$ \\
 $1 + \Delta_b$ & $ 1 - \tan(\beta-\gamma_b)\xi - \dfrac{\xi^2}{2} + \mathcal{O}(\xi^3)$ 
& $\tan(\beta-\gamma_b) + \xi - \tan(\beta-\gamma_b) \dfrac{\xi^2}{2} + \mathcal{O}(\xi^3)$ \\
 $1 + \Delta_{\tau}$ & $ 1 - \tan(\beta-\gamma_\tau)\xi - \dfrac{\xi^2}{2} + \mathcal{O}(\xi^3)$ 
& $\tan(\beta-\gamma_\tau) + \xi - \tan(\beta-\gamma_\tau) \dfrac{\xi^2}{2} + \mathcal{O}(\xi^3)$ \\
\hline
\end{tabular}
\end{small} \end{center}
\caption{Effective light Higgs boson couplings to fermions and gauge
  bosons up to $\mathcal{O}(\xi^3)$ within the 2HDM with a large mass
  hierarchy. The flavor sector features a general Yukawa alignment
  structure. The different interactions are normalized to the Standard
  Model and expressed in terms of $\Delta_x$ as defined in
  Eq.\eqref{eq:delta}.}
\label{tab:eff-coup1}
\end{table}

For the loop-induced Higgs interaction to photons we need to
consistently expand the corresponding trilinear Higgs coupling to the
charged Higgs bosons up to $\mathcal{O}(\xi^3)$ terms, reading
\begin{alignat}{5}
 \lambda_{\hzero\PHiggs^+\PHiggs^-} &=
\dfrac{-e}{2\,m_W\,\sw}\,\left[\left(1-\dfrac{\xi^2}{2}\right)\,(m^2_{\hzero} + 2m^{\pm\,2}_{\PHiggs}
-\lambda\,v^2) + \dfrac{\xi}{2}\,(2m^2_{\hzero} - \tilde{\lambda}\,v^2)\,(\cot\beta-\tan\beta) \right]
\label{eq:triple-hier}.
\end{alignat}
%
%
The leading dependence in the decoupling limit,
\begin{alignat}{5}
\lambda_{\hzero\PHiggs^+\PHiggs^-} &=  \lambda_{\hzero\Hzero\Hzero} = \lambda_{\hzero\Azero\Azero}
= m^2_{\hzero} + 2M^2_\text{heavy} - \tilde{\lambda} v^2 + \mathcal{O}(\xi) 
\label{eq:leading-decoupling}
\end{alignat}
is common to all trilinear self-interactions involving the light
Higgs boson and two heavy companions. Similarly,
in this limit we find $\lambda_{\hzero\hzero\Hzero} = \mathcal{O}(\xi)$ and 
$\lambda_{\hzero\hzero\hzero} = \lambda_{\hzero\hzero\hzero}^\text{SM} = -3im^2_{\hzero}/v^2$.
Therefore, the condition 
\begin{alignat}{5}
m_{\hzero}^2 + 2M_\text{heavy}^2 - \tilde{\lambda}\,v^2 \simeq 0 
\label{eq:decoup-selfcouplings}
\end{alignat}
ensures the consistent decoupling of the heavy fields 
in the $\xi \to 0$ limit, with dominant non-decoupling effects
reading
\begin{alignat}{5}
\lambda_{\hzero HH}^\text{non-dec} \simeq \dfrac{e}{4m_W\,\sw}\,\xi\,\tilde{\lambda}v^2\,(\cot\beta-\tan\beta)
&\simeq  \dfrac{e}{2m_W\,\sw}\,\xi\,M^2_\text{heavy} \,(\cot\beta-\tan\beta) \notag \\
 & \simeq \dfrac{e v^2}{2 m_W\,\sw}\,(\cot\beta-\tan\beta)
 \label{eq:nondec-3h}.
\end{alignat}
%


\end{document}